\documentclass[
    twocolumn,
    superscriptaddress,
    amsmath,amstex,amssymb,
    aps, prl,
    floatfix]{revtex4-2}

\usepackage{graphicx}
\usepackage{dcolumn}
\usepackage{bm}
\usepackage{lipsum}
\usepackage{gensymb}
\usepackage{physics}

\usepackage[overload]{empheq}
\usepackage{diagbox}
\usepackage{tabularx,booktabs}
\newcolumntype{Y}{>{\centering\arraybackslash}X}

\usepackage{hyperref}
\hypersetup{
  colorlinks   = true, 
  urlcolor     = blue, 
  linkcolor    = red, 
  citecolor   = red 
}

\begin{document}

\title{Probing Site-Resolved Current in Strongly Interacting\\
Superconducting Circuit Lattices}

\author{Botao Du}%
\author{Ramya Suresh}%
\author{Santiago L\'opez}
\author{Jeremy Cadiente}
\author{Ruichao Ma}%
 \email{maruichao@purdue.edu}

\affiliation{%
Department of Physics and Astronomy, Purdue University, West Lafayette, IN 47907, USA }%

\date{July 8, 2024} 

\begin{abstract}

Transport measurements are fundamental for understanding condensed matter phenomena, from superconductivity to the fractional quantum Hall effect. 
Analogously, they can be powerful tools for probing synthetic quantum matter in quantum simulators. Here we demonstrate the measurement of in-situ particle current in a superconducting circuit lattice and apply it to study transport in both coherent and bath-coupled lattices. Our method utilizes controlled tunneling in a double-well potential to map current to on-site density, revealing site-resolved current and current statistics. We prepare a strongly interacting Bose-Hubbard lattice at different lattice fillings, and observe the change in current statistics as the many-body states transition from superfluid to Mott insulator. Furthermore, we explore non-equilibrium current dynamics by coupling the lattice to engineered driven-dissipative baths that serve as tunable particle source and drain. We observe steady-state current in discrete conduction channels and interaction-assisted transport. These results establish a versatile platform to investigate microscopic quantum transport in superconducting circuits.

\end{abstract}

\maketitle

Charge transport plays a crucial role in the exploration of quantum phases and phase transitions in condensed matter physics \cite{Sachdev2011-mn}. Electrical conductivity measurements probe the intrinsic properties of charge carriers, while current fluctuations from shot-noise measurements reveal the quantum dynamics and correlations of charge carriers in strongly interacting systems\,\cite{Blanter2000-ah}. Meanwhile, synthetic quantum matter serve as emerging platforms for quantum simulation of condensed matter models, providing pristine many-body systems with exquisite control\,\cite{Georgescu2014-mt, Altman2021-yw, Daley2022-xm}. 
For instance, transport properties in ultracold atomic gases can be extracted from the time evolution of particle density either in response to an engineered non-uniform initial density\,\cite{Choi2016-ej, Brown2019-nx} or to applied external forces\,\cite{Anderson2019-xy}. In another example, the particle current through an atomic quantum point contact was measured by monitoring the particle number change in the attached finite-size reservoirs\,\cite{Brantut2012-gz, Hausler2017-ag}.

In superconducting (SC) quantum circuits, arrays of coupled SC qubits and resonators realize lattice models to study synthetic quantum matter comprised of interacting microwave photons \cite{Noh2017-ow, Hartmann2016-tl, Gu2017-iu, Carusotto2020-ct}. Recent experiments have explored quantum states and dynamics in Bose-Hubbard lattices \cite{Hacohen-Gourgy2015-zc, Collodo2019-oa, Ma2019-ye, Saxberg2022-tt, Morvan2022-am, Zhang2023-fg}, many-body localization\,\cite{Roushan2017-hh, Guo2021-jy, Guo2020-eo}, entanglement generation and characterization \cite{Braumuller2021-nj, Zhang2022-ts,  Karamlou2023HCB}, and flat-band physics\,\cite{Kollar2019-vn, Martinez2023-fh}. Coherent transport of microwave photons has been probed via density measurements in a disordered lattice \cite{Guo2020-eo}, and chiral current was measured in a triangular unit-cell with synthetic magnetic field \cite{Roushan2016-nr}. SC circuits also provide an ideal playground to investigate non-equilibrium transport in driven-dissipative lattices \cite{Leib2010-ph,Biella2015-xg,Mertz2016-ad,Bychek2020-bl}, with recent experiments performing microwave transmission spectroscopy of driven-dissipative phases in 1D lattices \cite{Fitzpatrick2017-sy,Fedorov2021-sr, vrajitoarea2024ultrastrong}. Nevertheless, in-situ control and measurement of transport dynamics remain less explored.

\begin{figure}
    \includegraphics[width=0.95\columnwidth]{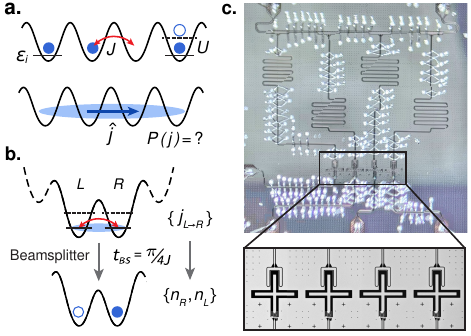}
    \caption{Measurement scheme. (a) Illustration of a strongly interacting Bose-Hubbard lattice, and the particle current between two neighboring lattice sites. (b) Controlled tunneling in the double-well potential realizes a beamsplitter operation that maps the nearest-neighbor current to on-site density. (c) Image of the superconducting circuit device with a zoomed-in view of the four-site transmon qubit lattice.
    }
    \label{fig:fig1} 
\end{figure}

In this Letter, we demonstrate the direct measurement of site-resolved current and current statistics in a SC circuit lattice, and apply it to probe the change in current fluctuations across a superfluid to Mott insulator transition. Furthermore, we couple the lattice to engineered particle baths to induce and study non-equilibrium transport through discrete conduction channels of the interacting 1D lattice.

\textbf{\textit{Measurement Scheme.---}}
The scheme is based on Ref.\,\cite{Kesler2014-gv} which we adapt to strongly interacting systems. We consider a Bose-Hubbard lattice (Fig.~\ref{fig:fig1}a) described by the Hamiltonian:
\begin{equation*}
\mathcal{H}_{\text{BH}}/\hbar = \sum_{<ij>}{J a_i^\dagger a_j}+\frac{U}{2}\sum_i{n_i(n_i-1)} + \sum_i \epsilon_i n_i
\end{equation*}
where $a_i^\dagger$ is the bosonic creation operator for a particle on site $i$, $J$ is the nearest neighbor tunneling rate, $U$ is the on-site interaction, $n_i=a_i^\dagger a_i$ is the on-site occupancy, $\epsilon_i$ is the local on-site energy, and $\hbar$ is the reduced Planck constant.
The current operator $\hat{j}_{l\rightarrow r}$ for particles flowing from site $l$ to site $r$ is defined, from the continuity equation for local particle density, as \cite{Kesler2014-gv}:
\begin{equation*}
\hat{j}_{l\rightarrow r} = i J ( a_l^\dagger a_r -  a_r^\dagger a_l )
\end{equation*}

For non-interacting particles ($U\approx 0$), the current can be measured using an effective beamsplitter operation (BS) implemented as controlled resonant tunneling between the two sites~\cite{Kesler2014-gv}, illustrated in Fig.~\ref{fig:fig1}b. After the evolution of an initial state in the isolated two-site system for a duration of $t_\text{BS}=\pi/4J$, the initial current is mapped onto density imbalance according to $\hat{j}_{l\rightarrow r} \xrightarrow{\text{BS}} J (n_r - n_l)$.
As indicated from this mapping, the current operator $\hat{j}$ has discrete eigenvalues $j \in \{-nJ,-(n-1)J,\dots, +nJ\}$, where $n = n_l+n_r$ is the total number of particles on the two sites. 
The current expectation value $\ev{j}$ and current statistics $P(j)$ are then extracted from a particle-number-resolved density measurement after the beamsplitter operation. 
This current measurement method, applicable to non- or weakly- interacting particles, has been
implemented in ultracold atoms to probe chiral currents albeit without spatial resolution \cite{Atala2014-tw} and recently in a quantum gas microscope to measure local current and current correlations \cite{Impertro2023current}.

Here, we extend this method to strongly interacting lattices in the hard-core boson limit ($U\gg J$). The on-site occupancy cannot exceed $n_i=1$ under the hard-core condition, thereby limiting the current eigenvalues to $j \in \{-2J, -J,0,+J, +2J\}$. We denote the on-site Fock states as $\ket{0}$ and $\ket{1}$. With the same resonant tunneling of duration $t_\text{BS}$ between two sites, now in the presence of large $U$, the current eigenstates corresponding to $j=\{-J,0,+J\}$ map uniquely to Fock states $\{\ket{10}, \ket{00}, \ket{01}\}$ after the beamsplitter, i.e. $P(j=-J) \xrightarrow{\text{BS}} P(\ket{10})$ etc. The Fock state $\ket{11}$ does not evolve in density during the beamsplitter operation in the hard-core limit, and can be written as an equal superposition of current eigenstates for $j=\pm 2J$. Therefore we obtain the probability of the remaining current components as $P(j=\pm 2J) \xrightarrow{\text{BS}} \frac{1}{2} P(\ket{11})$. 
From the current statistics, we calculate the current expectation value $\ev{j} = \sum_{j=-2}^{+2} jP(j)$.
This protocol for measuring current is tailored for analog quantum simulation experiments in SC circuits, as it only requires density readout and does not rely on measuring phase-sensitive correlation functions. See Supplemental Material (SM) Sec.\,D for a detailed derivation of the current measurement \footnote{See Supplemental Material [URL] for device characterization, measurement setup, numerical modeling, and additional analysis of the data, which includes Refs.\,\cite{Rol2020-mr, Johnson2011-jg, Geerlings2013-kl, Pedregosa2011-rf, roberts2023manybody}.}.

\textbf{\textit{SC circuit Bose-Hubbard lattice.---}}
In our SC circuit device, four transmon qubits \cite{Koch2007-cz} constitute the lattice sites of the 1D Bose-Hubbard lattice (Fig.~\ref{fig:fig1}c). The capacitive coupling between neighboring transmons results in tunneling $J \approx 2\pi\times 6$\,MHz, while the transmon anharmonicity provides the effective on-site interaction $U \approx - 2\pi\times 246$\,MHz.
The on-site energies $\epsilon_i$ are given by the frequency of the transmon $n=0\rightarrow 1$ transition $\omega_{\text{q}}$ and dynamically tunable via individual on-chip flux lines. We typically operate the lattice near $\omega_{\text{q}} \approx 2\pi\times 4.5$\,GHz. Microwave photons in the lattice (transmon excitations) have a relaxation rate of $\Gamma_1 = 1/T_1 \approx 2\pi\times 5$\,kHz and a dephasing rate of $\Gamma_\phi = 1/T_2^* \approx 2\pi\times 60$\,kHz. 
\begin{figure}
    \includegraphics[width=0.95\columnwidth]{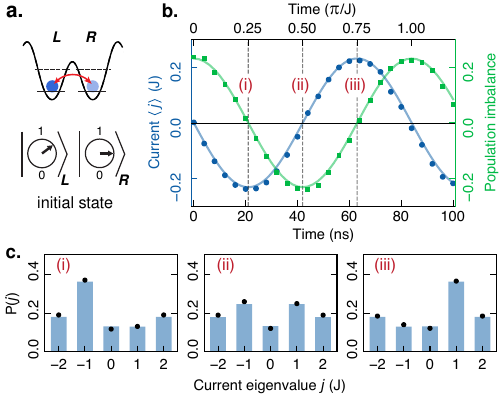}
    \caption{Current dynamics in a resonant double-well. (a) We start in a product state of two sites with unequal population. (b) Time evolution of current and population imbalance in the strongly interacting two-site system. (c) Current statistics at three different evolution times. Solid lines in (b) and bars in (c) are results from numerical simulation. Data shown in all figures are typically averaged over 40,000-100,000 experimental runs, with standard error of the mean smaller than the size of the data points. Other systematic uncertainties in the extracted density or current are below $\pm1\%$, see SM Sec.\,C.}
    \label{fig:fig2} 
\end{figure}
Each transmon lattice site is capacitively coupled to an individual coplanar waveguide resonator used for dispersive readout of the on-site occupancy. We simultaneously measure all lattice sites by performing frequency-multiplexed readout via a common readout transmission line. See SM Sec.\,A-C for details on device parameters, measurement setup, and lattice control and characterization.

\begin{figure*}
    \includegraphics[width=2.05\columnwidth]{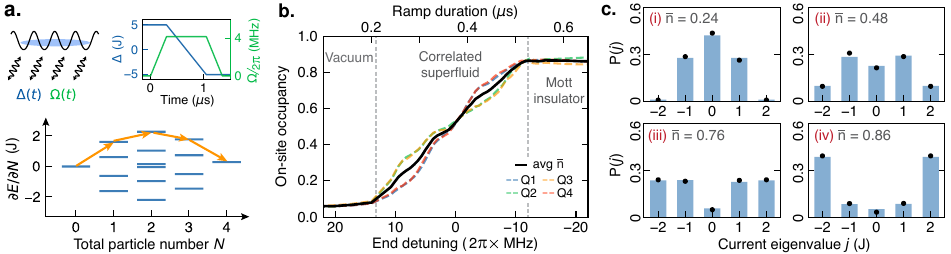}
    \caption{Current statistics in a coherent Bose-Hubbard lattice. (a) Many-body states with different lattice fillings are prepared using a global coherent drive with time-varying amplitude and detuning. As the drive detuning varies, the lattice is adiabatically filled through a sequence of transitions in the many-body spectrum (orange arrows). (b) The measured on-site occupancy and average filling as a function of the drive end detuning. (c) Current statistics at four different lattice fillings, showing the change in the probability distribution. Bars show results from numerical simulation.}
    \label{fig:fig3} 
\end{figure*}

\textbf{\textit{Current dynamics in a double-well.---}}
To illustrate our protocol, we measure current in a resonant double-well formed by two neighboring transmon lattice sites (Fig.~\ref{fig:fig2}a).
For any product state of the left and right sites with an initial population imbalance, we expect the density to start oscillating as a result of resonant tunneling and an oscillating current to develop between the two sites. Here we choose a product state with each site initialized in a superposition of $\ket{1}$ and $\ket{0}$. At the beginning of the experiment, the two sites start far-detuned in frequency in their equilibrium ground states, which have approximately 6\% thermal population in $\ket{1}$ due to the finite effective temperature of the device. 
We prepare superpositions with $P(\ket{1})$ of 76\% and 50\% in the left and right sites, by applying two resonant microwave pulses that correspond to $X$-rotations of $127^\circ$ and $90^\circ$ on the Bloch sphere of the transmon qubits.
We then rapidly bring the two sites into resonance and evolve for a variable time before applying the current measurement protocol. The data is shown in Fig.~\ref{fig:fig2}b where we observe the coherent oscillation of the current $\ev{j_{L\rightarrow R}}$ with a period of $\pi/J$. Separately, we measure the on-site density in the double-well and observe the population imbalance $\left\langle n_\text{R}-n_\text{L}\right\rangle$ oscillating $90\degree$ out of phase with the current, as expected from the relation between current and density under resonant tunneling discussed above. In this experiment, the density and current measurements share the same pulse sequence with the latter evolving for an extra duration of $t_\text{BS}$ in the double-well. The measured current statistics $P(j)$ at three different times are plotted in Fig.~\ref{fig:fig2}c, revealing the origin of the observed current expectation value $\ev{j}$ and its quantum fluctuations.
Due to the hard-core condition, the $j=\pm 2J$ components remain constant in time and contribute no net current. The oscillating non-zero current $\ev{j}$ comes from the $j=\pm J$ components as a result of single-particle tunneling in the double-well.
See SM Sec.\,D for calibration of the beamsplitter operation.

\textbf{\textit{Current statistics across the superfluid to Mott transition.---}}
We investigate current statistics in the hard-core Bose-Hubbard lattice at different average fillings $0<\bar{n}\leq 1$. At unit filling, the state corresponds to the Mott insulator phase with suppressed density fluctuations \cite{Bakr2010-jm}. At partial filling, the states are correlated superfluids where strong on-site interaction induces repulsive density correlations in the limit of 1D Tonks–Girardeau gases \cite{Saxberg2022-tt}.
SC circuits provide a platform for new approaches to creating strongly correlated quantum states. Many-body states can be prepared spectroscopically via direct driving \cite{Umucalilar2017-ak,Karamlou2023HCB}, or with disorder-assisted local adiabatic control \cite{Saxberg2022-tt}. Alternatively, engineered dissipation can serve as an effective chemical potential for microwave photons to stabilize many-body phases in SC circuits~\cite{Hafezi2015-mw, Lebreuilly2018-bs, Ma2017-vw, Mi2024-in}.

\begin{figure*}
    \includegraphics[width=2.05\columnwidth]{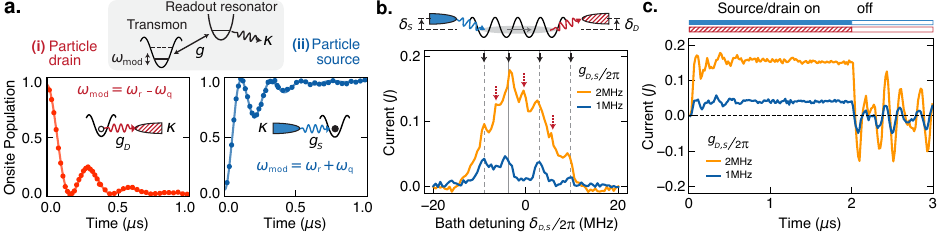}
    \caption{Transport in a bath-coupled lattice. (a) We engineer local particle drain (i) and source (ii) by parametric coupling of the transmon and the lossy readout resonator.
    Plots show the dynamics of a single site coupled to the drain or source. Dots are experimental data; lines are numerical simulations. 
    (b) Steady-state current through the lattice with end-coupled source and drain, measured as a function of bath detuning for two bath coupling rates. Black arrows on the current plot indicate frequencies of the single-particle eigenstates, and red dotted arrows indicate processes involving two-particle eigenstates.
    (c) Time dynamics of the current with the baths turned on at $t=0$, then turned off at $t=2\,\mu$s after reaching steady state. The bath detuning in (c) is indicated by the solid line in (b).}
    \label{fig:fig4} 
\end{figure*}

Here we prepare the Bose-Hubbard lattice at different fillings via adiabatic many-body Landau-Zener transitions using coherent external driving, similar to methods employed in recent Rydberg atom experiments \cite{De_Leseleuc2019-ux}. As illustrated in Fig.~\ref{fig:fig3}a, we apply a global microwave drive to all lattice sites via the readout transmission line. Starting with an empty lattice, we turn on the coherent drive to a Rabi amplitude $\Omega\approx 2\pi\times 4.2$\,MHz in 300\,ns. The initial drive frequency is detuned by $\Delta = 2\pi\times 30$\,MHz $\approx 5J$ above the lattice frequency, away from all single-particle resonances which are within $\pm 2J$ of the lattice frequency. We then ramp the drive frequency at a constant rate of $-2\pi\times 80$\,MHz/$\mu$s to populate the lattice through a sequence of avoided crossings between states of different total particle number $N$. 
The drive detuning at the end of the ramp is varied and determines the final lattice filling $\bar{n}$. 
Due to the relatively uniform drive amplitude and phase across the lattice sites, we prepare the highest energy states at different $\bar{n}$. Finally, we ramp down the drive amplitude to zero in 300\,ns and then measure the resulting many-body state. 
Figure~\ref{fig:fig3}b shows the measured on-site occupancy and average lattice filling as a function of the drive end detuning. The final state transitions from the empty vacuum, to strongly-interacting superfluids at partial filling, to the Mott insulator at unit filling. To verify the adiabaticity of the preparation, we perform the coherent drive twice with a second reversed detuning ramp and measure the final lattice density. Due to finite thermal population, the lattice starts with $P(\ket{0000})\approx 80$\%, and we extract a return probability of $P(\ket{0000})\approx 60$\% after the double ramp by comparing density data to numerical simulations. The preparation fidelity is consistent with theory and primarily limited by decoherence in the lattice. See SM Sec.\,E for the design and characterization of the drive parameters and additional analysis on adiabaticity.

The current statistics at different lattice fillings, measured between the two middle lattice sites, are shown in Fig.~\ref{fig:fig3}c. For low filling $\bar{n}=0.24$ ($N\approx 1$), a single microwave photon is delocalized over the lattice, with strong fluctuations of the on-site occupancy. The current statistics is expected to follow a Skellam distribution~\cite{Kesler2014-gv} with a broad distribution peaked at $j=0$. As $\bar{n}$ increases, both the density and current fluctuations are increasingly suppressed due to the strong interaction $U$. The observed $P(j)$ gradually becomes peaked at two eigenvalues $j=\pm 2J$, as a result of the Mott insulator at $\bar{n}=1$ being the equal superposition of current eigenstates $j =\pm 2J$. The adiabatically prepared stationary states all have a vanishing mean current $\ev{j}$, as reflected in the symmetric distribution of the measured current statistics. The current statistics vary slightly at different locations in the finite lattice due to boundary effects but have the same qualitative features.

\textbf{\textit{Non-equilibrium transport in bath coupled lattice.---}}
Finally, we explore particle transport in open lattices that are difficult to access in closed quantum systems or by local density measurements alone.
We couple the two ends of the Bose-Hubbard lattice to particle source and drain, and measure current through the interacting 1D system. Such boundary-driven transport has been the subject of extensive theoretical investigations \cite{ Landi2022-ga, Bertini2021-lo} and recently studied in a digital SC circuit experiment \cite{Mi2024-in}.

We utilize the transmon readout resonators as the source of dissipation to engineer local particle baths.
The far-detuned resonators, with frequency $\omega_\mathrm{r}\sim 2\pi\times (6.1-6.3)$\,GHz and linewidth $\kappa\approx 2\pi\times 1.5$\,MHz, are coupled to each transmon with $g \approx 2\pi\times 65$\,MHz (Fig.\,\ref{fig:fig4}a). We generate resonant interactions between the transmon site and the resonator by parametric driving of the transmon flux line at frequency $\omega_\mathrm{mod}$.
When $\omega_\mathrm{mod} \approx \omega_\mathrm{r}-\omega_\mathrm{q}$, the flux driving modulates the on-site frequency and induces a coherent modulation-assisted tunneling between the transmon site and the resonator \cite{Beaudoin2012-xh, Strand2013-wt}. Since the resonator is lossy, this leads to an effective drain for microwave photons in the lattice. Alternatively when $\omega_\mathrm{mod} \approx \omega_\mathrm{r}+\omega_\mathrm{q}$, the coherent parametric drive creates a pair of excitations, one to the transmon and one to the resonator \cite{Lu2017-uq}. One of the two excitations is lost via the resonator, leaving the other incoherently added to the transmon site -- this leads to a particle source. We hence realize a hardware-efficient implementation of local baths anywhere in the lattice. The drain (D) and source (S) both have a narrow energy bandwidth of $\kappa \ll J$.  The effective bath-lattice coupling rates $g_\text{D,S}$ and effective bath-lattice detunings $\delta_\text{D,S} = (|\omega_\mathrm{mod} - \omega_\mathrm{r}| - \omega_\mathrm{q})$ are dynamically tunable by controlling the amplitude and frequency of the parametric drive respectively. 
Similar parametric driving processes are widely used in SC circuits for engineering tunable interactions \cite{Beaudoin2012-xh, Strand2013-wt, Macklin2015-pj,  Naik2017-pi, Lu2017-uq, Reagor2018-we}.

In Fig.\,\ref{fig:fig4}a\,(i), we plot the measured dynamics of an isolated site coupled resonantly to the drain with $\delta_\text{D} =0$ and $g_\text{D} = 2\pi\times 1.75$\,MHz. The site is initialized with one photon and relaxes via its coupling with the drain to a steady-state population of $P(\ket{1})=0.02$, limited by the resonator thermal population. In (ii), an initially empty site is coupled to the source with $\delta_\text{S} =0$ and $g_\text{S} = 2\pi\times 2.4$\,MHz where the on-site population reaches a steady-state value of $P(\ket{1})=0.98$ in about 1\,$\mu$s. In this device, we achieve maximum coupling rates of $g_\text{D} \sim 2\pi\times 12$\,MHz limited by the flux tuning range of the transmon, and $g_\text{S} \sim 2\pi\times 4$\,MHz limited by drive-induced heating.

To probe energy-dependent transport through the Bose-Hubbard lattice, we couple the source and drain to the ends of our 1D lattice with the same coupling rates and same detunings from the lattice (Fig.\,\ref{fig:fig4}b). After turning on the baths for 2\,$\mu$s, we apply the current measurement protocol to the middle two lattice sites to measure the steady-state current $\ev{j}$ as a function of the bath detuning. At a relatively weak bath coupling of $g_\text{D,S}=2\pi\times 1$\,MHz, we observe four distinct current peaks when the narrow-band baths are tuned near the frequencies of the four single-particle eigenstates (i.e. $N$$=$$1$ states in Fig.\,\ref{fig:fig3}a). We note that a finite current $\ev{j}$ requires coherent superposition between many-body eigenstates with the same particle number $N$. This follows from the observation that each energy eigenstate is stationary with vanishing $\ev{j}$, and the current operator commutes with particle number so eigenstates with different $N$ contribute to the current independently.
For weak bath coupling and low lattice filling, the measured steady-state current near each single-particle eigenfrequency originates from coherent admixtures between the resonantly-driven single-particle eigenstate and other single-particle eigenstates that are off-resonantly excited by the finite-bandwidth bath.
At a stronger bath coupling of $g_\text{D,S}= 2\pi\times 2$\,MHz, we observe higher steady-state currents over a broader range of bath detuning. In addition to peaks at the single-particle eigenfrequencies, more peaks appear at the frequencies of $N=2$ eigenstates which correspond to non-linear transport processes where the baths add and remove two particles. Such processes are driven by the larger bath-lattice coupling and assisted by the effective interaction between many-body eigenstates that results from the on-site $U$. When eigenstates with more particles participate in the transport, the steady state current has contributions from superposition in different particle number manifolds.
See SM Sec.\,G for detailed modeling of the bath-coupled steady-state current and current dynamics.

In Fig.\,\ref{fig:fig4}c, we measure the time evolution of the current at the specific bath detuning of $\delta_\text{D,S} = -2\pi\times4$\,MHz. After the source and drain are turned on at $t=0$, we observe currents with initial oscillations that settle to the steady state value within $1\,\mu$s. At $t = 2\,\mu$s, we turn off both baths and observe the subsequent current dynamics in the isolated lattice.
The different eigenstates in the coherent superposition responsible for the steady-state current now evolve coherently at different eigenfrequencies. Therefore the current undergoes coherent oscillations with a time-averaged value of zero. Alternatively, this can be viewed as the particles reflecting elastically at the edges of the lattice, moving back and forth with an oscillatory current.
In bath-coupled lattices, the non-equilibrium steady state depends sensitively on the interplay between the coherent lattice interactions and parameters of the driven-dissipative baths \cite{Kordas2015-ec, Sieberer2016-rh, Dutta2021-pr}. Our future work will explore the detailed dependence of the steady-state current and current dynamics on the bath's spectral properties, lattice disorder and decoherence, and many-body interactions.

\textit{\textbf{Conclusion.---}}
In this Letter, we present the first measurement of site-resolved current \textit{and} current statistics in an analog quantum simulator using SC circuits. We leveraged both coherent control and bath engineering to generate and probe non-equilibrium quantum transport in the Bose-Hubbard lattice, establishing a versatile setup for future exploration of open quantum systems and quantum thermodynamics \cite{Binder2019-qo, Deffner2019-ps}.
Energy-dependent transport using tunable baths can be applied to probe critical behavior masked by transient finite-size effects in small closed systems\,\cite{Lenarcic2020-hh}. Measuring current in the quantum critical regime, where the quasiparticle picture breaks down, will provide insights into the nature of collective excitations, how dissipation influences quantum critical dynamics, and how quantum entanglement evolves. For example, current statistics can reveal correlations and entanglement in many-body systems \cite{Klich2009-id}. Furthermore, complex current correlation functions can be measured by simultaneous readout of multiple pairs of sites to investigate topological properties, e.g., in 2D lattices with chiral edge states \cite{Anderson2016-sp, Owens2022-ua}.

We would like to thank Qi Zhou for stimulating discussions, and  Srivatsan Chakram and Andrei Vrajitoarea for a careful reading of the manuscript. 
This work was supported by the National Science Foundation under award number DMR-2145323, and the Air Force Office of Scientific Research under award number FA9550-23-1-0491.

\let\oldaddcontentsline\addcontentsline
\renewcommand{\addcontentsline}[3]{}


%

\let\addcontentsline\oldaddcontentsline



\renewcommand{\thetable}{S\arabic{table}}
\renewcommand{\theequation}{S\arabic{equation}}
\renewcommand{\thefigure}{S\arabic{figure}}
\setcounter{equation}{0}
\setcounter{figure}{0}
\setcounter{secnumdepth}{3}

\newpage

\onecolumngrid

\renewcommand{\appendix}{\par
  \setcounter{section}{0}
  \setcounter{subsection}{0}
  \setcounter{subsubsection}{0}
  \gdef\thesection{\Alph{section}}
  \gdef\thesubsection{\Alph{section}.\arabic{subsection}}
  \gdef\thesubsubsection{\Alph{section}.\arabic{subsection}.\roman{subsubsection}}
}

\appendix

\clearpage
\begin{center} 
    \large {\textbf {Probing Site-Resolved Current in Strongly Interacting\\Superconducting Circuit Lattices\\}}
    \vspace{0.2in}
    \uppercase{\textbf {Supplemental Material}}
    \vspace{-0.1in}
\end{center}

\tableofcontents

\section{Device Fabrication and Parameters}

The superconducting circuit device used in this work consists of four frequency tunable transmon qubits \cite{Koch2007-cz} in a linear array with fixed nearest-neighbor capacitive coupling, and individual readout resonators. The transmons have cross-shaped capacitors with $E_C \approx 2\pi\times 246$\,MHz and flux tunable $E_J \approx 2\pi\times (9-18)$\,GHz, corresponding to a tunable frequency of $\omega_{01} \approx 2\pi\times (4.0-5.7)$\,GHz. The capacitance between the neighboring transmons $c \approx 0.2$\,fF results in tunneling $J \approx 6$\,MHz. The qubits are coupled to $\lambda/2$ coplanar-waveguide readout resonators with a coupling strength $g\approx 65$\,MHz. The resonators are staggered in a frequency range of $6.18-6.32$\,GHz, with a typical linewidth of $1.5$\,MHz. The flux bias lines are galvanically coupled to the SQUID loops of the transmon qubit with a mutual inductance $M \approx 0.86$\,pH. 

The device is fabricated using standard two-step lithography, at Purdue University's Birck Nanotechnology Center. The base layer is $150$\,nm of Niobium deposited using an e-beam evaporator (Plassys MEB550) on $c$-plane sapphire substrate. Optical lithography is performed with a direct pattern writer, followed by fluorine dry etching, to define all features on the chip except for the Josephson junctions. The $\textrm{Al}/\textrm{AlO}_x/\textrm{Al}$ Josephson junctions are defined with e-beam lithography on MMA-PMMA bilayer resist, and deposited in the same e-beam evaporator using Manhattan style angled-evaporation. Ion milling prior to deposition ensures electrical contact between the Nb base layer and the Al junction traces. The bottom/top layers of Al are $50$\,nm/$100$\,nm thick, deposited at $1$\,nm/s at an angle of $30^{\circ}$ to normal. The junction oxidation is performed in 85\%/15\% $A_r$/$O_2$ for $10$ minutes at $1$\,mBar. The frequency tunable transmons have two asymmetric junctions of sizes $130\,\text{nm}\times 130\,\text{nm}$ and $220\,\text{nm}\times 220\,\text{nm}$. The junction SQUID loops have a dimension of $20\,\mu\text{m}\times 20\,\mu$m. 

The device is mounted to a multi-layer copper printed circuit board, and enclosed in a pocketed OFHC copper sample holder. Aluminum wire bonds are used to suppress stray on-chip modes, reduce signal crosstalk between control lines, and connect the chip to the printed circuit board. The packaged device is mounted inside a dual-layer mu-metal shield, and anchored to the base of a dilution refrigerator at 10\,mK.

\section{Measurement Setup}

\begin{figure}
    \includegraphics[width=0.95\columnwidth]{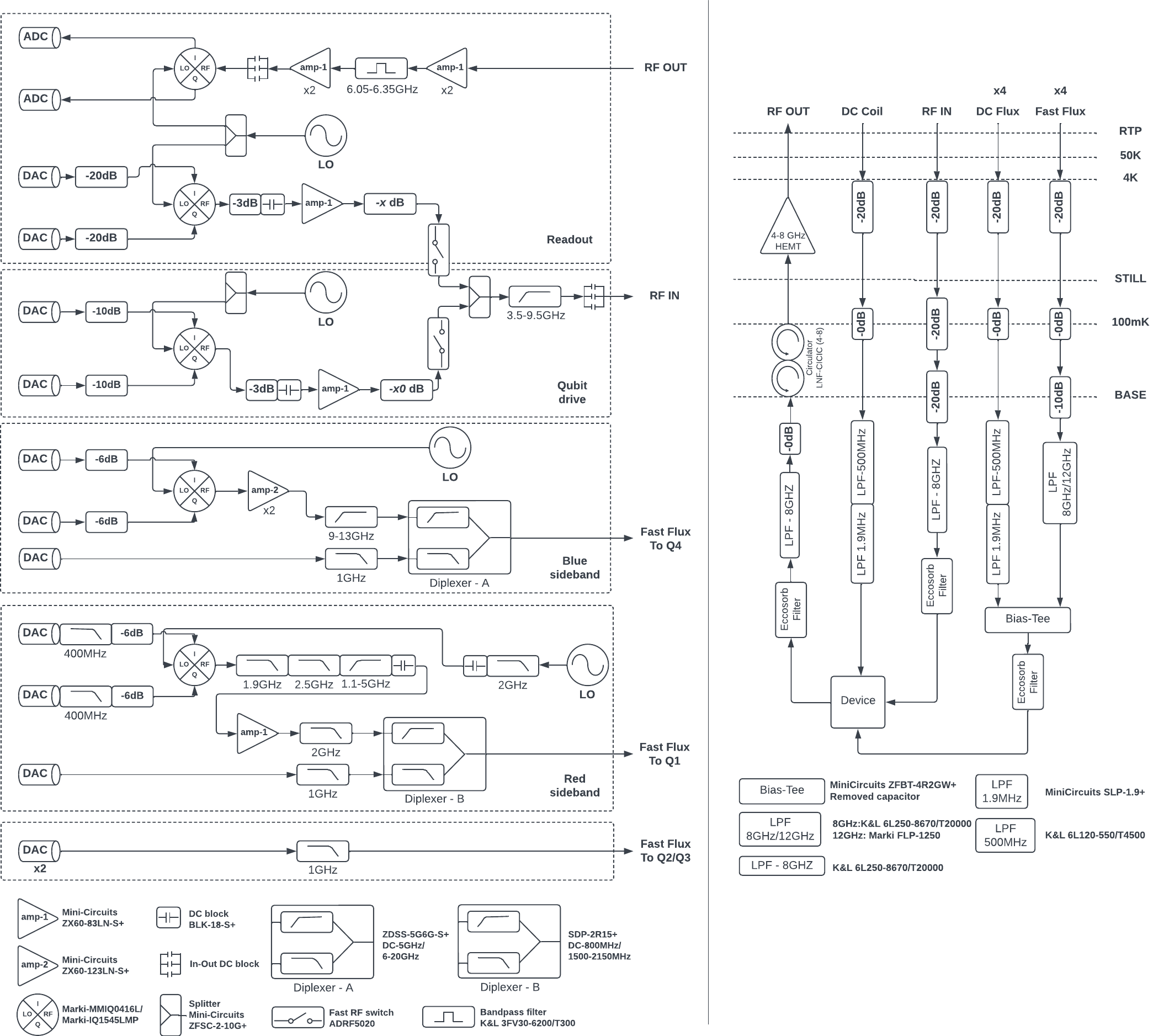}
    \caption{Fridge wiring and measurement setup.
    }
    \label{fig:SI_wiring}
\end{figure}

The cryogenic setup is shown in Fig.\,\ref{fig:SI_wiring}(right). Low pass filters (K\&L, Marki, Mini-circuit) and IR absorbers (Eccosorb CR-110, homemade) are used on all input and output coaxial lines to suppress noise and heating from high-frequency radiation. We use DC flux bias to control the idle frequency of the qubits using both on-chip flux lines, and an off-chip DC coil. The fast AC flux pulses, used to dynamically change the qubit frequencies during the experimental sequence, are combined with the DC flux on microwave bias tees at the mixing chamber. The bias-tees (Mini-circuit ZFBT-4R2GW+) have the capacitor on the rf port replaced by a short to allow the fast AC flux pulses to work down to DC. After passing through cryogenic circulators (Low Noise Factory), the readout output signal is amplified by a cryogenic HEMT amplifier (Low Noise Factory) and subsequently by additional room-temperature amplifiers.

The room temp control and measurement setups are shown in Fig.\,\ref{fig:SI_wiring}(left). A Quantum Machines OPX+ system provides multi-channel DC-coupled DAC outputs at 1GS/s, which are used directly for the fast AC flux pulses. Additional DAC outputs are used as I/Q signals and upconverted using IQ mixers (Marki) to generate qubit drive and readout pulses. Sideband flux control pulses used for bath engineering are generated similarly using IQ mixers near the required LO frequencies. After room temperature amplification and filtering, the frequency multiplexed readout signal from the fridge is downconverted with an I/Q mixer and recorded by the OPX+'s ADC at 1GS/s. This multiplexed heterodyne signal (centered around -120 MHz) is then digitally demodulated by the OPX+ to produce the I/Q signals for each qubit. Additionally, we generate digital pulses from the OPX+ to control fast microwave switches (Analog Devices ADRF5020) to avoid LO leakage when the pulses are idle. We use both low noise voltage sources (Analog Devices AD5780) and current sources (Yokogawa GS200) for applying the qubit DC flux bias.

\section{Experimental Control and Characterization}

\subsection{Flux control and calibration }

\subsubsection{DC flux crosstalk}

We observe relatively large crosstalk between our device's different flux control lines. A flux bias current $I_j$ applied to the flux line for qubit $j$ induces non-zero magnetic flux $\phi_i$ in the SQUID loops of all qubits $i$, due to the residual mutual inductance. To realize independent frequency control of each qubit, we measure the flux crosstalk matrix $M_{ij} = \partial\phi_i/\partial I_j$, defined from the flux response $d\phi_i = \sum_{j=1}^4 \frac{\partial\phi_i}{\partial I_j} dI_j $.
The eigenvectors of the inverted crosstalk matrix $M_{ij}^{-1}$ then provide the linear combinations of currents to independently tune each qubit to any desired flux value $\phi_i$. The conversion between qubit frequency $\omega_i$ and flux $\phi_i$ is obtained by fitting the measured flux-dependent qubit spectra to a Jaynes-Cummings model.

The measured static dc flux crosstalk matrix is shown in Table~\ref{tab:SI_flux-crosstalk}, normalized row-wise to the diagonal elements. 
While the dc flux signals tune the qubits to their initial idle frequency, the fast-flux signals provide additional dynamic tuning with nanosecond resolution. In Table~\ref{tab:SI_flux-crosstalk2}, we show the fast ac flux crosstalk matrix, measured from the time-averaged step-response to $3\,\mu$s-long square fast-flux pulses. Improved flux line design and on-chip air bridges can reduce the overall flux crosstalk in future devices.
We observe a frequency-dependent crosstalk of the fast flux control, which is a direct result of the different time-domain responses of the different flux lines. This suggests that part of the flux pulse distortion comes from the on-chip flux lines, which vary in length for each qubit. The time-domain characterization and compensation of the fast-flux pulses and crosstalk are detailed below.

\begin{table}
    \centering
    \begin{minipage}{0.5\textwidth}
        \centering
        {\renewcommand{\arraystretch}{1.4}%
    	\begin{tabularx}{0.8\textwidth}{ |Y||*{4}{Y|} }
    	    \multicolumn{5}{c}{$\partial \phi_i / \partial I_j \hspace{0.1in}$} \vspace{0.05in} \\  \hline
    		\diagbox[height=0.2in]{i  }{  j} & 1 & 2 & 3 & 4  \\ \hline\hline
    		1 & 1.0 & -0.189 & 0.104 & 0.206     \\ \hline
                2 & -0.317 & 1.0 & 0.059 & 0.261     \\ \hline
                3 & -0.247 & -0.267 & 1.0 & 0.287   \\ \hline
                4 & -0.153 & -0.153 & -0.101 & 1.0   \\ \hline
    	\end{tabularx}
    	}
        \caption{Measured static flux (dc) crosstalk matrix.}
    	\label{tab:SI_flux-crosstalk}
    \end{minipage}%
    \begin{minipage}{0.5\textwidth}
        \centering
        {\renewcommand{\arraystretch}{1.4}%
    	\begin{tabularx}{0.8\textwidth}{ |Y||*{4}{Y|} }
    	    \multicolumn{5}{c}{$\partial \phi_i / \partial I_j \hspace{0.1in}$} \vspace{0.05in} \\  \hline
    		\diagbox[height=0.2in]{i  }{  j} & 1 & 2 & 3 & 4  \\ \hline\hline
    		1 & 1.0 & -0.028 & 0.044 & 0.031     \\ \hline
                2 & -0.082 & 1.0   &  0.028 & 0.036     \\ \hline
                3 & -0.056 & -0.084 & 1.0  &   0.057   \\ \hline
                4 & -0.031 & -0.039 & -0.083 & 1.0   \\ \hline
    	\end{tabularx}
    	}
        \caption{Measured fast-flux (ac) crosstalk matrix.}
        \label{tab:SI_flux-crosstalk2}
    \end{minipage}
\end{table}

\subsubsection{Time domain characterization of flux control pulses}

\textbf{Measuring the flux response:} 
The fast-flux pulses generated by the room-temperature arbitrary waveform generators (QM OPX+, 1GS/s) can experience significant distortion when arriving at the qubit after passing through various lines, filters, and rf components. To characterize the distortion, we apply a square fast-flux pulse to a flux line and measure the frequency response of the qubit. The qubit frequency is measured using both Ramsey-type ``CryoScope'' experiments \cite{Rol2020-mr} and pump-probe-type qubit spectroscopy experiments. The former provides time resolution at the pulse resolution of 1\,ns but is limited to times shorter than the qubit dephasing time. The latter allows us to measure flux response at longer times, but the resolution is limited by the duration of the qubit excitation pulse to tens of ns. 

\begin{figure}
    \includegraphics[width=0.95\columnwidth]{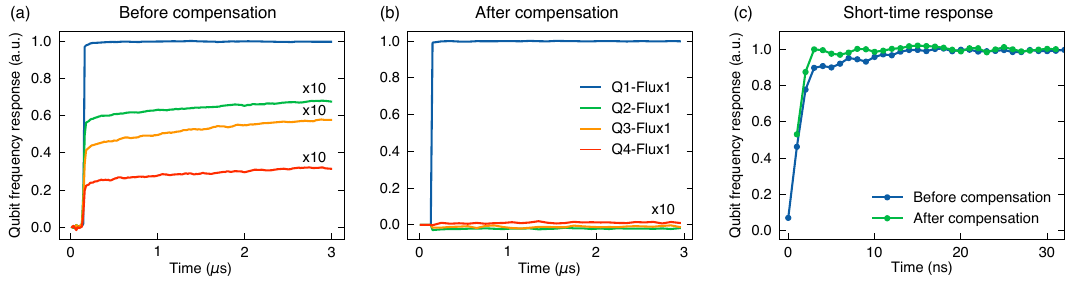}
    \caption{Fast-flux pulse calibration and compensation. (a-b) Step responses before and after the time-dependent crosstalk compensation. The off-diagonal responses are plotted with same signs and on a $\times10$ scale for easier visualization. (c) Typical short-time flux response before and after compensation.
    }
    \label{fig:SI_flux-comp}
\end{figure}

The measured fast-flux response is shown in Fig.\,\ref{fig:SI_flux-comp}. The typical distortion shows a slow exponential rise with a decay constant of a few hundred ns, and fast oscillations with a period of few ns that decay in the first tens of ns. We attribute the former to the effective low-pass filtering effect of the filters and lines, and the latter to impedance mismatches near the on-chip flux lines. We measure an initial ramp time of $\approx 2$\,ns for all flux lines, see Fig.\,\ref{fig:SI_flux-comp}(c). 

\vspace{10pt}

\textbf{Flux pulse distortion compensation:} 
We follow a procedure similar to that described in \cite{Johnson2011-jg}. 
We calculate the frequency response of the flux line from the measured time-domain step response, then take the Fourier transform of the inverted frequency response to construct a time-domain compensation kernel. To pre-compensate distortions in the flux line, the kernel is convoluted with the target flux pulse to generate the desired fast-flux waveform to be played on the pulse generators. All compensated pulses are computed off-line and then uploaded to the OPX+. We only compensate for distortions with time scales $>$ 10\,ns to reduce sensitivity to high-frequency noise in the step-response measurement and to retain the bandwidth of the fast flux pulse. 
The remaining fast oscillations at short times do not significantly affect the dynamics of the qubits in our experiments due to the relatively long tunneling time $2\pi/J \approx 160$\,ns. 

\vspace{10pt}

\textbf{Time-dependent crosstalk:} With relatively large flux crosstalk in the present device, we observe different step responses for all $4\times 4$ qubit-to-flux-line pairs. This indicates non-negligible flux pulse distortion from on-chip structures. To achieve independent dynamical control of all qubits, we must compensate for a \textit{time-dependent} crosstalk or, equivalently, a \textit{frequency-dependent} crosstalk. From all $4\times 4$ step responses, we calculate a frequency response matrix for all qubit-to-flux-line pairs and construct the corresponding time-domain kernel matrix to pre-compensate all fast-flux waveforms. In Fig.\,\ref{fig:SI_flux-comp}(a), we show the step responses of all qubits to the same flux line 1 with different time constants. The step responses after the time-dependent crosstalk compensation are shown in (b).

\vspace{10pt}

\textbf{Net-zero flux pulses:} We use net zero flux pulses for all experiment sequences to avoid the slow drifting of qubit frequencies due to the low-frequency response of the flux lines. This is realized by applying a flux-balancing pulse at the end of each experimental run so that the total area of the fast-flux pulse remains zero.

\subsection{Typical experimental sequence}

We start with all qubits tuned to the same frequency using dc flux bias, forming a degenerate lattice. The multi-qubit state in the lattice is then prepared in several ways: (1) Detune all qubits into a staggered frequency configuration using fast-flux pulses and apply resonant microwave pulses sequentially on each qubit via the readout transmission line; this is used for preparing the states in Fig.\,2 of the main text and for preparing Fock states for readout calibration. The microwave pulses have a typical duration of 40\,ns with a Gaussian width of $\sigma \approx 8$\,ns.
(2) Keep the qubits at the lattice frequency and directly apply a common microwave pulse that drives all qubits with time-varying amplitude and detuning; this is used in Fig.\,3. (3) Apply fast-flux modulation pulses to select qubits to enable resonant coupling to the readout resonator, to realize a lattice with locally-coupled driven-dissipation baths. The bath populates the lattice with excitations. This is used in Fig.\,4. After preparation, the many-body state can undergo an additional free evolution in the lattice with variable time to study their dynamics.

To probe the resulting state, we measure either the lattice density (on-site population of each qubit) or the current across two neighboring lattice sites (by performing the beamsplitter operation and then measuring the density). When measuring current, the two sites of interest are rapidly tuned into resonance and undergo a controlled resonant tunneling, while all other qubits are detuned. All qubits are then moved rapidly to the staggered readout configuration. The frequency-multiplexed readout microwave pulse is applied for a typical duration of 1.5\,$\mu$s. After the readout, the flux balancing pulses are applied to null the net current flowing into each flux line during the sequence. 

The experiment is repeated with a cycle period of 300\,$\mu$s, leaving enough idle time between sequences for the qubits to relax to their equilibrium ground states. Each sequence is run $\sim$40,000--100,000 times to accumulate statistics.

\begin{table}
	{\renewcommand{\arraystretch}{1.4}
	\begin{tabularx}{0.6 \textwidth}{| Y  Y | Y | Y | Y | Y |} 
      \hline
       ~ & ~ & $Q_1$ & $Q_2$ & $Q_3$ & $Q_4$   \\ \hline\hline
       \multicolumn{2}{|c|}{$T_1$ ($\mu$s)} & 29 & 26 & 25 & 29  \\ \hline
      \multicolumn{2}{|c|}{$T_2^*$ ($\mu$s)} & 2.7 - 3.0 & 2.2 - 2.7 & 1.0 - 2.2 & 1.5 - 2.5 \\ \hline

      \multicolumn{2}{|c|}{$U/2\pi$ (MHz)} & -246.5 & -246.6 & -246.4 & -246.7  \\ \hline
      \multicolumn{2}{|c|}{$J_{i,i+1}/2\pi$ (MHz)} & 5.78 & 5.80 & 5.86 & n/a  \\ \hline 
      \multicolumn{2}{|c|}{$n_{\textrm{th}}$ } & 0.064 & 0.059 & 0.056 & 0.052 \\ \hline 
\multicolumn{2}{|c|}{$\omega_{\textrm{read}}/2\pi$ (GHz)} & 6.323 & 6.276 & 6.229 & 6.178 \\ \hline
      \multicolumn{2}{|c|}{$g_{\textrm{read}}/2\pi$ (MHz)} & 67 & 65 & 64 & 65 \\ \hline
      \multicolumn{2}{|c|}{$\kappa_{\textrm{read}}/2\pi$ (MHz)} & 1.5 & 1.5 & 1.5 &  1.6  \\ \hline
    \end{tabularx}
    }
\caption{Measured device parameters.}
\label{tab:SI_qubit-param}
\end{table}

\subsection{Qubit and Lattice Characterization}

Table \ref{tab:SI_qubit-param} lists the measured device parameters when the qubits are near the typical lattice frequency of 4.55\,GHz. For performing multiplexed readout, the qubits are moved to staggered frequencies around (4.85, 4.40, 4.75, 4.30) GHz. 
The qubit relaxation times $T_1$ exhibit fluctuations up to a factor of 2 over time, while the dephasing times $T_2^*$ are limited by the noise in the fast-flux signals from room temperature electronics. The tunneling rates $J$ are measured from the oscillation of an initial excitation between two neighboring resonant sites. The effective interaction $U$, i.e. the anharmonicity of the transmon, is extracted from Ramsey experiments. The equilibrium thermal population of the qubits is measured from the amplitude of Rabi oscillation between the transmon excited states\,\cite{Geerlings2013-kl}. We observe typical $n_{th} \approx$ 5.0-6.5\,\%, corresponding to the effective qubit temperature of 70-80\,mK. The thermal population can be reduced in future experiments with better microwave line filtering and improved sample package thermalization.

\subsection{Readout characterization}

\begin{figure}
    \includegraphics[width=0.7\columnwidth]{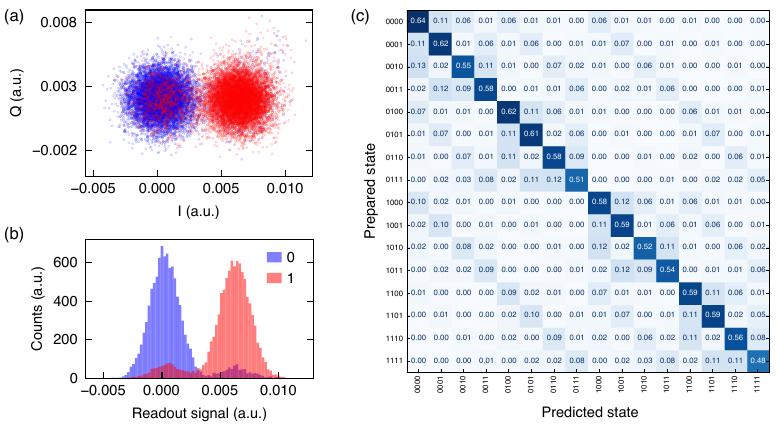}
    \caption{Qubit readout. (a) Typical single-shot readout signal showing the distinguishability between qubit states in the I/Q space. In the 1D projection (b), we see $\sim$10\,\% in the opposite state due to contributions from the initial thermal population and measurement-induced qubit transitions.
    (c) Assignment fidelity matrix for 4-qubit multiplexed readout using linear SVM.
    }
    \label{fig:SI_readout}
\end{figure}

We use a frequency multiplexed square pulse of length 1.5\,$\mu$s for readout. The readout output signal, after hardware down-conversion, is digitally demodulated and integrated with constant weights to obtain the quadrature (I/Q) signals for each qubit. For simultaneous single-shot readout of all 4 qubits, the 4x2 I/Q signal is fed to a linear support vector machine (SVM) for state discrimination. The SVM \cite{Pedregosa2011-rf} is trained using calibration data by preparing the qubits in the basis states using resonant $\pi$ pulses. Figure \ref{fig:SI_readout} shows the typical readout signal visualized in the I/Q plane of one readout resonator, and the assignment probability matrix for single-shot 4-qubit multiplexed readout.

Population in the transmon second excited state $n=2$ ($f$-level) is negligibly small in the thermal ground state of the qubit. During the lattice experiments, any $n=2$ population in the final state comes from Landau-Zener transfer between neighboring qubits as they are ramped from the lattice location to the readout location. We experimentally measure and numerically verify that this population transfer contributes to $\leq 1\%$ of readout error in our experiments. Hence we only include Fock states $n=0,1$ in the multiplexed readout.

We calculate the Fock state representation of a particular measured state by multiplying the SVM output probabilities with the inverted assignment probability matrix. Additionally, the prepared basis states include finite thermal populations. We use a linear transformation to map the state vector in the prepared basis state back to bare Fock states.

\textbf{Readout error estimation:}
(1) The basis states for readout calibration are prepared using sequential resonant $\pi$-pulses at the readout location where all qubits are far detuned from each other. Typical measured $\pi$-pulse fidelities are 99.8\,\%.
(2) Thermal population calibration. Throughout all experiments, we observed no significant change in the thermal population ($\lesssim 0.5$\,\%). The procedure we used above to extract the actual qubit state from readout calibration signal at finite temperature do not introduce any significant systematic readout errors.
(3) Systematic uncertainties on the measured density and current due to Landau Zener transition are less than $1$\% \cite{Ma2019-ye}.
(4) With each experiment repeated for typically 40,000--100,000 times, the uncertainty from statistical error (standard error of the mean) is small compared to the contributions above.

\section{Beamsplitter Operation for Measuring Current} 

\subsection{Non-interacting bosons}

We first consider the time evolution of non-interacting bosons in a resonant double-well potential, with Hamiltonian $\mathcal{H}_{\text{dw}}/\hbar = J ( a_l^\dagger a_r + a_r^\dagger a_l )$. The Heisenberg equations of motion reads
\begin{equation*}
    \dv{t} a_{l}(t) = - \frac{i}{\hbar}  \comm{a_l(t)}{\mathcal{H}_{\text{dw}}} =  - i J a_{r}(t), \quad \textrm{similarly} \quad  \dv{t} a_{r}(t) = - i J a_{l}(t).
\end{equation*}
These two differential equations can be solved, e.g. by a change of basis to the symmetric and anti-symmetric superpositions $(a_l\pm a_r)$ which are the eigenbasis of $\mathcal{H}_{\text{dw}}$. The results are
\begin{align*}[left=\empheqlbrace]
  a_{r}(t) &= \cos(Jt) a_r(0) - i \sin(Jt) a_l(0) \\
  a_{l}(t) &= -i\sin(Jt) a_r(0) + \cos(Jt) a_l(0)
\end{align*}
which simply describes the tunneling of bosons in the double-well potential. The time evolution of the density difference in the double-well can then be evaluated as
\begin{equation*}
n_r(t) - n_l(t) = a_r^\dagger(t) a_r(t) - a_l^\dagger(t) a_l(t) = \cos(2Jt) \left[n_r(0) - n_l(0)\right] + i \sin(2Jt) \left[ a_l^\dagger(0) a_r(0) -  a_r^\dagger(0) a_l(0) \right]
\end{equation*}
The second term on the right can be identified as the current
$\hat{j}_{l\rightarrow r} = i J ( a_l^\dagger a_r -  a_r^\dagger a_l )$, to write
\begin{equation*}
n_r(t) - n_l(t) =  \cos(2Jt) \left[n_r(0) - n_l(0)\right] + \sin(2Jt) \frac{\hat{j}_{l\rightarrow r}(0)}{J}
\end{equation*}
For an evolution time of $t_\text{BS} = \pi/(4J)$, the first term proportional to the initial density difference vanishes. Therefore, we arrive at the conclusion that a non-interacting double-well at the chosen beamsplitter duration $t_\text{BS}$  maps the current of the initial state to density difference after the evolution:
\begin{equation*}
n_r(t_\text{BS}) - n_l(t_\text{BS}) = \frac{\hat{j}_{l\rightarrow r}(0)}{J}
\end{equation*}
From here we see that the eigenvalues of $\hat{j}_{l\rightarrow r}$ are discrete and have values $j \in \{-nJ,-(n-1)J,\dots, +nJ\}$ where $n = n_l+n_r$. Measuring the distribution of density difference after the beamsplitter operation reveals the statistics of the current $P(j)$.

\subsection{Measuring current in the hardcore limit} 

Our strongly interacting SC circuit lattice has fixed $|U/J| \gtrsim 40$ in the so-called hard-core boson limit. The mapping from current before the beamsplitter to on-site occupancy after the beamsplitter using the controlled tunneling in the hard-core boson limit is summarized as follows:
\begin{equation*}
    \begin{cases}
        P(j=+J) &\xrightarrow{\text{BS}} \quad P(\ket{01})\\
        P(j=0) &\xrightarrow{\text{BS}} \quad P(\ket{00})\\
        P(j=-J) &\xrightarrow{\text{BS}} \quad P(\ket{10})\\
        P(j=-2J) &\xrightarrow{\text{BS}} \quad \frac{1}{2} P(\ket{11})\\
        P(j=+2J) &\xrightarrow{\text{BS}} \quad \frac{1}{2} P(\ket{11})
    \end{cases}
\end{equation*}

For the hardcore Bose-Hubbard lattice with onsite occupancy limited to 0 or 1, the current can be written in terms of the Pauli operators:
\begin{equation*}
\hat{j}_{l\rightarrow r} = i J ( \sigma_l^+ \sigma_r^- - \sigma_r^+ \sigma_l^- ) = 2 J ( \sigma_l^x \sigma_r^y - \sigma_r^x \sigma_l^y )
\end{equation*}
This amounts to measuring two-qubit correlation functions, requiring phase-sensitive single-qubit rotations before readout as in \cite{Roushan2016-nr}. This seemingly simple approach turns out challenging for \textit{analog} simulation experiments, where the qubits accumulate different phases during the continuous many-body evolution, and fully calibrating the phases for each qubit at each time step is extremely difficult. In contrast, our method is particularly suited for analog simulation experiments with frequency tunable lattice sites, eliminating the need to compensate for the phase accumulated during the evolution when applying the tomography pulses. We only measure onsite density after the controlled tunneling, which is phase-insensitive.

\subsection{Calibrating the beamsplitter}

\begin{figure}
    \includegraphics[width=0.7\columnwidth]{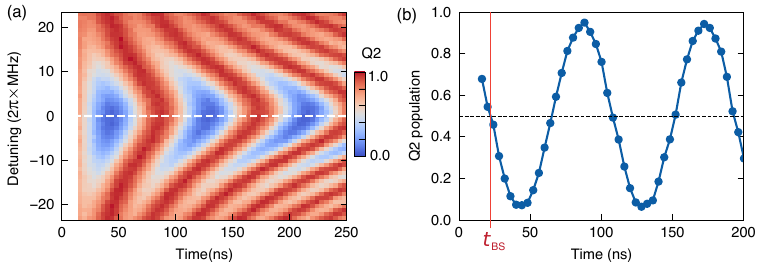}
    \caption{Beamsplitter operation via resonant tunneling between two sites. (a) Population in one site as a function of time for different detuning between sites, (b) Resonant tunneling when the detuning is zero.
    }
    \label{fig:SI_BS}
\end{figure}

The beamsplitter operation is a resonant tunneling between two neighboring sites for a duration of $\pi/4J$. This duration corresponds to a quarter period of the density oscillation for a single excitation initially localized in one site. We prepare the initial state at a detuning of $\approx 2\pi\times 150$\,MHz, then rapidly tune the two sites near-resonance and measure the density after a variable time, as shown in Fig.\,\ref{fig:SI_BS}. The vertical line marks the beamsplitter time $t_\text{BS}$. We use this same experiment to calibrate the lattice tunneling rate $J$.

\section{Coherent Preparation in Bose-Hubbard Lattice}

\subsection{Manybody spectra in the presence of driving}

Here we discuss the details of the state preparation scheme using the global coherent drive applied to all lattice sites. The driven Hamiltonian reads:
\begin{equation*}
\mathcal{H}_d/\hbar = \mathcal{H}_{\text{BH}}/\hbar + \sum_{i} \frac{\Omega_d^i}{2} [a_i^\dagger e^{i\phi_i} +h.c.] - \Delta_d \sum_i n_i
\end{equation*}
where the Rabi rate $\Omega_d^i$ and phase $\phi_i$ can differ for different lattice sites. In our experiments, the microwave drive is applied through the common readout transmission line and goes through separate readout resonators to reach individual qubits.
The detuning of the drive from the lattice frequency $\Delta_d$ resembles an effective chemical potential for the microwave photons.

Starting with an empty lattice, we would like to turn on the drive and sweep its detuning to go through avoided crossings that populate the lattice successively to different filling. To ensure adiabaticity, the rate of change of the drive frequency and amplitude must be slow compared to the instantaneous many-body gap at all times. We illustrate the requirements for adiabaticity by considering how the many-body gap depends on drive detuning, amplitude, and phase.

\vspace{10pt}

\textbf{Effect of drive phase:} 
The many-body gap reflects the effective coupling between many-body states under driving. It depends critically on the symmetry of the many-body states and the drive phase. The relative phases of the drive on different sites of the lattice determine the plausible adiabatic paths through the driven many-body spectra.

\begin{figure}
    \includegraphics[width=0.7\columnwidth]{ 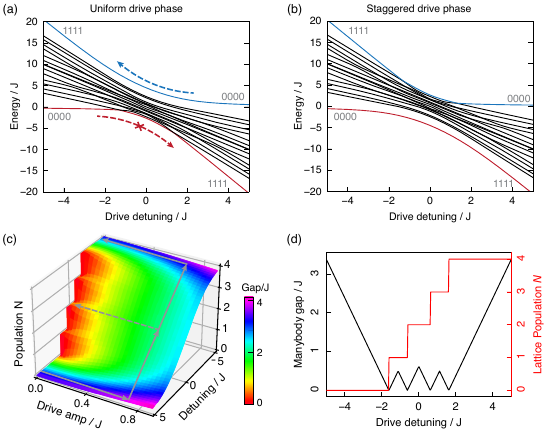}
    \caption{ Many-body spectra in the presence of coherent driving, for uniform (a) and staggered driving phases (b). The driving amplitude is $0.7 J$ for all sites, matching the value used in our experiment. The drive phase determines whether the lowest (red) or highest (blue) states are gapped. (c) The paths taken by the coherent-driven preparation, where the end detuning determines the final lattice population. The many-body gap is shown in color to illustrate the adiabaticity requirements. (d) Many-body gap and lattice filling as a function of drive detuning at zero driving amplitude, i.e., the vertical cut in (c) at zero drive amplitude. 
    }
    \label{fig:SI_driven-mb-gap}
\end{figure}

Consider the part of the adiabatic sweep where the lattice goes from empty to having one particle. As shown in the following sections, the global microwave drive has a near-uniform amplitude and phase when it reaches the different qubits in our present device. In addition, our capacitively coupled grounded transmon lattice has an alternating phase for the microwave field in the lowest energy single-particle eigenstate.
Due to this different symmetry between the single particle wavefunction (with phase pattern $+-+-$) and the drive (with phase pattern $++++$), the effective rate to go from $n=0$ to lowest energy $n=1$ state will remain zero (gapless) even at finite driving amplitude. This is true for all transitions between lowest-energy states with different particle numbers. In Fig.\,\ref{fig:SI_driven-mb-gap}(a) we show the many-body spectra when driving with uniform phase across the lattice. As the detuning is varied, the groundstate (red) transitions are all gapless, making it impossible to remain adiabatic when sweeping the drive detuning. 

For the coherent filling of the lattice, we instead adiabatically follow the \textit{highest} energy states of the system. These states in blue in Fig.\,\ref{fig:SI_driven-mb-gap}(a) have uniform phase patterns ($++++$), opposite to the lowest energy states, making them always gapped under uniform phase driving. This drive sequence, used in the experiments in Fig.\,3, is illustrated in Fig.\,\ref{fig:SI_driven-mb-gap}(c): The solid path shows the drive used to prepare a fully filled $\bar{n}=1$ lattice, while the dashed path indicates a drive sweep that ends at zero detuning and prepares the lattice at half filling $\bar{n}=0.5$.
For comparison, we plot in Fig.\,\ref{fig:SI_driven-mb-gap}(b) the many-body spectrum in the case of uniform driving amplitude but staggered driving phases (i.e. $\pi$ phase shift between neighbors). Now the lowest energy path is fully gapped and can be adiabatically prepared, while the highest energy states have the gap vanishing at a few detunings.

While we used a global drive with fixed relative amplitude and phase in the present experiments, we can readily implement individual qubit drives with independent amplitude and phase, by combining the charge driving signal with the flux bias signal\,\cite{Karamlou2023HCB}.
Generally, we expect local phase and amplitude-controlled driving to be a powerful tool for coherently preparing many-body states by selectively closing and opening level crossings in the many-body spectra. 

\vspace{10pt}

\textbf{Final states with vanishing many-body gaps:} 
The adiabatic requirement cannot be met for final undriven states with a vanishing gap, which happens when the lowest energy states in two adjacent particle number manifolds are resonant in the rotating frame of the drive. This is shown in Fig.\,\ref{fig:SI_driven-mb-gap}(d) where we plot the many-body gap at zero driving amplitude. When the end detuning of the drive is close to those values with a vanishing gap, we expect the state we prepare to be a superposition of two (nearly-) degenerate many-body eigenstates that differ in particle number by one. The exact details of the final state in our experiments are subject to further investigation. However, the current and current statistics we show in the main text at different lattice fillings are insensitive to the details of this superposition.

\vspace{10pt}

\textbf{Normal modes in transmon lattices:} 
We show that in our transmon array, when considering the phase of electric fields oscillating at the capacitor of each transmon qubit, the lowest-energy single-particle eigenstate in the lattice has alternating signs, i.e., electric field phase different of $\pi$ between nearest neighbors). Meanwhile, the highest-energy single-particle eigenstate has uniform signs, i.e., the electric fields oscillate in phase on all qubits. 

\begin{figure}
    \includegraphics[width=0.95\columnwidth]{ 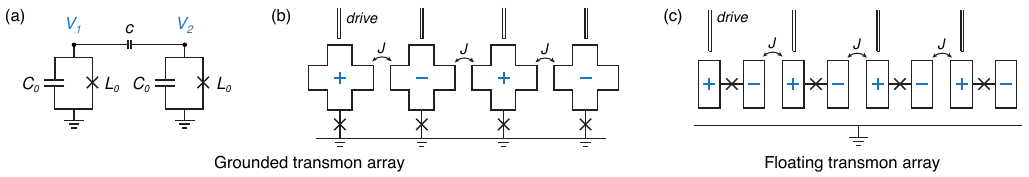}
    \caption{Normal modes in transmon lattices. (a) Circuit diagram for coupled grounded transmons. (b) Relative signs of electric fields on the qubit capacitors for the lowest-energy single-excitation state.
    }
    \label{fig:SI_transmon-J-config}
\end{figure}

We use grounded transmons with direct capacitive coupling, represented by the effective circuit in Fig.\,\ref{fig:SI_transmon-J-config}(a). Focusing on single-particle states, we have ignored non-linearity from the Josephson junction and keep only the transmon capacitance $C_0$, inductance $L_0$, and coupling capacitance $c$. Solving Kirchoff's rules, the normal modes and frequencies for two resonantly coupled qubits are:
\begin{equation*}
    \begin{cases}
    \omega_{+} &= \frac{1}{\sqrt{L_0 C_0}} = \omega_0 \quad \textrm{with} \quad V_1 = V_2 \\
    \omega_{-} &= \frac{1}{\sqrt{L_0 (C_0 + 2c)}} \approx \omega_0 (1-\frac{c}{C_0}) = \omega_0 - 2J \quad  \textrm{with} \quad V_1 = - V_2 
    \end{cases}
\end{equation*}
where $V$ is the voltage on the capacitor for each qubit. The mode `$+$' with $V_1=V_2$ has no electric field in the coupling capacitor, hence the frequency is unchanged from the single qubit. The mode `$-$' with $V_1=-V_2$ has a finite electric field in the coupling capacitance, lowering the curvature of the fields and hence lowering the energy. By the same argument, we see that the lowest-energy single particle eigenstate in the lattice will have the electric fields out of phase on neighboring sites, see Fig.\,\ref{fig:SI_transmon-J-config}(b). This corresponds to alternating signs relative to the individual qubit drive lines, which are also capacitively coupled to the qubit capacitors either directly or via the readout resonators.

The situation is different for a floating transmon array, depicted in Fig.\,\ref{fig:SI_transmon-J-config}(c). While the lowest, the spit capacitors for each qubit act as electric dipoles, making the electric fields to oscillate with the same phase for all qubits. The relative phase to the individual drive lines is now uniform.

Note that regardless of the circuit implementation, the node theorem always holds in 1D such that the lowest-energy state has no node in the electric field. The difference between the two configurations illustrated here can be alternatively viewed as having different signs for the effective tunneling $J$.

\subsection{Calibrating drive amplitude and phase}

We measure the drive amplitudes $\Omega_d^i$ via Rabi oscillations of each qubit at the lattice frequency while other qubits are detuned away. The measured amplitudes are uniform across the lattice with less than 5\% variation. 

To estimate the difference in drive phases between neighboring sites, we measure the driven time evolution when the two sites are on-resonance. The one-excitation eigenmodes of the two-site system are $\frac{\ket{01} \pm \ket{10}}{\sqrt{2}}$, which are driven at effective rates that depend on the drive phase: 
$\Omega_d^\pm = (\Omega_d^L e^{i\phi_L} \pm \Omega_d^R e^{i\phi_R})/\sqrt{2}$. 
Here the label $\pm$ on $\Omega$ follows the sign of the superposition, and according to the reasoning in the previous section, the `$+$' state has higher energy than the `$-$' state with a splitting of $2J$. In the limit where $\Omega_d^L \approx \Omega_d^R = \Omega_d$, the effective rates simplify to $|\Omega_d^-| = \sqrt{2} \Omega_d \cos{(\theta/2)}$ and $|\Omega_d^+| = \sqrt{2} \Omega_d \sin{(\theta/2)}$, where $\theta = |\phi_R-\phi_L|$. So by observing the Rabi frequencies of the two modes, we can extract the drive phase difference.

\begin{figure}
    \includegraphics[width=0.7\columnwidth]{ 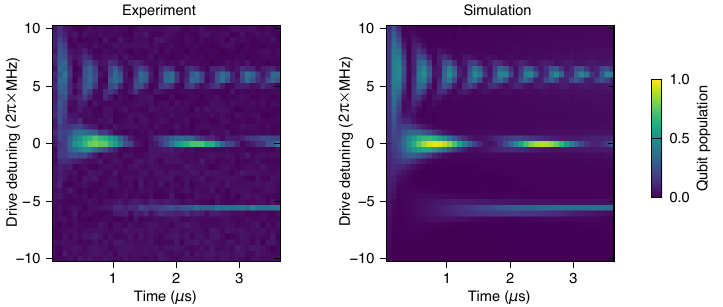}
    \caption{Calibrate drive amplitude and phase. Population in left site after driving two resonantly coupled neighboring sites at different drive detuning and drive pulse duration. 
    }
    \label{fig:SI_drive-cal}
\end{figure}

The data for the middle two sites of the lattice are shown in Fig.\,\ref{fig:SI_drive-cal}, where we plot the qubit population on one of the sites after applying a Gaussian pulse of varying length at different drive detuning. We observe the two single excitation resonances at drive detuning of $\pm J\approx 2\pi\times 6$\,MHz. The highly suppressed Rabi rate $\Omega_d^-$ compared to $\Omega_d^+$ indicates that the drive phase is almost identical for the two qubits. We fit the data to numerical simulations to extract $\theta \approx 7^{\circ}$. The additional resonance at zero detuning is a two-photon process oscillating between $\ket{01}$ and $\ket{11}$, enabled by a virtual transition through the detuned $\ket{02}$ and $\ket{20}$ states. We measure similar phase differences of $7^{\circ}$ between all pairs of sites in our device. Overall we have a relatively uniform drive phase, approximately corresponding to the case in Fig.\,\ref{fig:SI_driven-mb-gap}(a). The calibration method above does not reveal the sign of the phase difference. This sign can be extracted from an interferometer-type experiment between the two neighboring qubits\,\cite{Karamlou2023HCB}.

\subsection{Adiabaticity and many-body decoherence}

\begin{figure}
    \includegraphics[width=1\columnwidth]{ 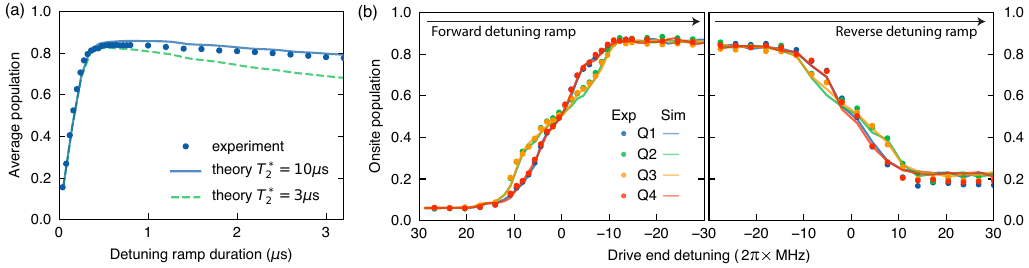}
    \caption{Adiabaticity of the coherent preparation. Dots are experimental data, and lines are numerical simulations.
    }
    \label{fig:SI_adiabaticity}
\end{figure}

When maximizing the fidelity of the adiabatic preparation, the optimal rate of change for the drive detuning results from the competition between diabatic transitions when the rate is faster (shorter ramp duration), and decoherence when the rate is slower (longer ramp duration). 
We perform the state preparation by fixing the start and end frequency of the drive detuning ramp at $+2\pi\times 30$\,MHz and $-2\pi\times 30$\,MHz, and varying the duration of the linear ramp. In Fig.\,\ref{fig:SI_adiabaticity}(a), we measure the final state average population $\bar{n}$ as a function of the ramp duration. For ramp duration less than $500$\,ns, we see effects of diabatic transitions and a reduced fidelity for reaching the $\bar{n}=1$ state. At durations larger than 1\,$\mu$s, we observe reduced filling due to lattice relaxation and dephasing. We choose an optimal duration of 750\,ns for the adiabatic preparation used in the main text, corresponding to a detuning ramp rate of $-2\pi\times 80$\,MHz/$\mu$s.

We show numerical simulations in good agreement with the observed data. The numerics are implemented in QuTiP by solving the Lindblad master equation. We use experimentally calibrated values for all coherent parameters, and the measured average single qubit lifetime at the lattice location $T_1 \approx 25\mu s$ for relaxation rate.
When we include a white noise dephasing rate corresponding to the measured average single qubit $T_2^* \sim 3\mu s$, we see a significant over-estimation of decoherence in the numerical results, shown in Fig.\,\ref{fig:SI_adiabaticity}(a). To match the adiabaticity data (and all other experimental data in the manuscript), we need to use a dephasing rate of $T_2^* \approx 10\mu s$ in the numerical calculations.
Such suppressed effective many-body dephasing rate $T_2^*$ has been observed in recent transmon lattice experiments \cite{roberts2023manybody}. For the numerical simulation in Fig.\,\ref{fig:SI_drive-cal}(b), we used the same effective dephasing time of $10\,\mu$s.

Using the optimal ramp rate, we perform a ``forward and back'' experiment where we coherently drive from $\bar{n}=0$ to $\bar{n}=1$ then apply a second drive with a reversed detuning ramp to coherently bring the many-body system back to the initial state $\bar{n}=0$. We vary the end detuning during the forward and back ramp and measure the onsite population after turning off the drive amplitude in 300\,ns. The data is shown in Fig.\,\ref{fig:SI_adiabaticity}(b)(right), where we have included the forward ramp data in Fig.\,3 of the main text for comparison (left). Here the drive amplitude is turned to zero at the minimum detuning of $-2\pi\times 30 $\,MHz, then turned back on before the reverse detuning ramp starts. Starting with an initial filling of 6\% due to thermal population, we observe a final lattice filling of $\approx 20$\% after the forward and back ramps. The numerical simulation, using the above mentioned many-body decoherence rates, show excellent agreement with the observed data. The initial state has $\approx 80$\% probability in $\ket{0000}$, the final state after the forward ramp has $\approx 70$\% probability in $\ket{1111}$, while the state after the forward-and-back ramp has $\approx 60$\% probability in $\ket{0000}$.

\subsection{Effect of initial thermal population}

\begin{figure}
    \includegraphics[width=1.0\columnwidth]{ 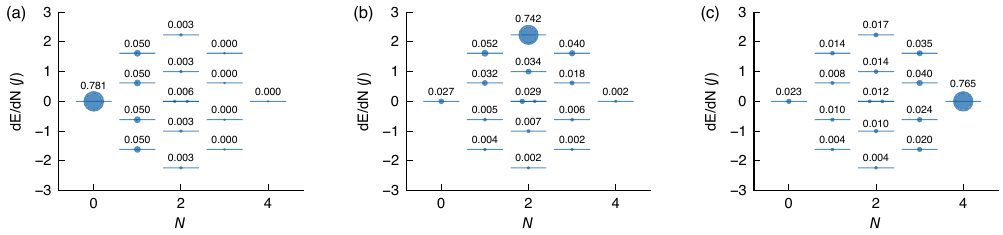}
    \caption{Thermal population after the coherent preparation, from numerical simulations without decoherence. (a) The initial ground state. The numbers indicate probability in each many-body eigenstate. (b) Final state at half-filling. (c) Final state at unit-filling. Note there are two near-degenerate states at zero detuning for $N=2$, where we show the total probability in both states.
    }
    \label{fig:SI_thermal-pop}
\end{figure}

The lattice starts in an equilibrium ground state with a finite thermal population of approximately 6\% on each lattice site. The thermal population can be coherently driven into different many-body eigenstates during the many-body preparation. In Fig.\,\ref{fig:SI_thermal-pop}, we show numerical results to illustrate the distribution of the thermal components after the coherent preparation. We intentionally excluded decoherence effects in this calculation to focus on visualizing the coherent transfer of the initial thermal population across the many-body spectra. All numerical simulations in the manuscript include the effect of thermal population, in good agreement with experimental data.

\section{Bath-Lattice Coupling Rates}

The parametric modulation that enables the transmon-resonator coupling is applied via the local flux line of each transmon lattice site and induces a flux modulation of
$\phi(t) = \phi_\text{mod}\cos(\omega_\text{mod}t) + \phi_0$ in the SQUID loop of the transmon. For relatively small modulation amplitude $\phi_\text{mod}$ around the lattice location we use, the transmon frequency is linearly dependent on the flux: $\omega_q(t) = \omega_0 + A_\text{mod}\cos(\omega_\text{mod} t)$. The Hamiltonian describing the capacitively coupled resonator and transmon system is:
\begin{equation*}
H = \omega_q(t)a^{\dag}a + \omega_r b^{\dag}b + g(a^{\dag} + a)(b^{\dag}+ b) + \frac{U}{2} a^{\dag}a^{\dag}aa
\end{equation*}
where $b^\dag$ is the creation operator for microwave photons in the resonator. The modulation frequencies that lead to the realization of effective particle drain (D) and particle source (S) are $\omega_\text{mod}^D\approx (\omega_r-\omega_q)$ (``red sideband'') and $\omega_\text{mod}^S\approx (\omega_r+\omega_q)$ (``blue sideband''). The red sideband modulation leads to an effective Hamiltonian with resonant tunneling between the qubit and resonator, written in the rotating frame of the qubit with the rotating wave approximation as:
\begin{equation*}
H_\text{red} = g_D (a^{\dag}b + b^{\dag}a) +  \frac{U}{2} a^{\dag}a^{\dag}aa
\end{equation*}
with effective red sideband coupling rate $g_D = J_1(\frac{A_\text{mod}}{|\omega_r-\omega_q|})\times g$, where $J_1$ is the 1st order Bessel function of the first kind.
On the other hand, the blue sideband modulation leads to an effective Hamiltonian:
\begin{equation*}
H_\text{blue} = g_S (a^{\dag}b^{\dag} + ba)  +  \frac{U}{2} a^{\dag}a^{\dag}aa
\end{equation*}
with effective blue sideband coupling rate $g_S = J_1(\frac{A_\text{mod}}{|\omega_r+\omega_q|})\times g$. The readout resonators in our device are dissipative with linewidth $\kappa$ through their coupling to the common readout transmission line. This turns the above coherent sideband couplings into incoherent processes which we utilize as narrow-band particle drain and source. 

In our experiments, to achieve a red sideband coupling rate of $g_D=2 \pi\times2$\,MHz, we used a sideband power of approximately -11\,dBm before entering the fridge. This is consistent with analytical models and numerical simulations which predict a required flux modulation amplitude of $\phi_\text{mod} \approx 0.04\Phi_0$ at the lattice location, where $\Phi_0$ is the magnetic flux quantum. This corresponds to a frequency modulation amplitude of $A_\text{mod} \approx 2 \pi\times 115$\,MHz. At the typical lattice location, the highest rate we can achieve in the present device is $g_D \approx 2 \pi\times 12$\,MHz, limited by the finite qubit frequency tuning range of approximately $2 \pi\times(4-5.8)$\,GHz.

To achieve a blue sideband coupling rate of up to $g_S=2 \pi\times 2$\,MHz, we used modulation power up to 0.7\,dBm before entering the fridge, in agreement with numerical modeling for $A_\text{mod} \approx 2 \pi\times 640$\,MHz. Due to the finite qubit frequency tuning range, the numerical model indicates a maximum $g_S \approx 2 \pi\times 2.4$\,MHz. Experimentally, we observed $g_S=2 \pi\times 4$\,MHz with a sideband power of 7\,dBm. We attribute this additional coupling to a three-wave mixing process enabled by a weak charge coupling between the flux line and the transmon. At even higher modulation amplitudes, we start to observe drive-induced heating of the transmon to higher excited states.

During the flux modulation, the average qubit frequency will experience a shift coming from both the non-linear dependence of the qubit frequency on flux and drive-induced ac-Stark shifts. In our experiments, these shifts are on the order of a few MHz. To ensure the transmon sites remain at the lattice frequency when the baths are turned on, these shifts are measured and compensated using the dc component of the fast flux control.

\section{Extended Analysis on Current in Bath-Coupled Lattice}

Here we present modeling of the bath-coupled transport experiments. In the main text, we made this statement: \textbf{A finite current $\ev{j}$ requires coherent superposition between many-body eigenstates with the same particle number.} 
For this discussion, we label the many-body eigenstates in the lattice as $\ket{\psi_{N,k}}$, where $N$ is the number of particles and $k$ labels the states within each manifold $N$. For instance, there are 4 states ($k=$1...4) for $N=1$ and 6 states for $N=2$ in our 4-site hard-core Bose-Hubbard lattice (see the many-body spectrum shown in Fig.\,\ref{fig:fig3}a, replotted here in Fig.\,\ref{fig:SI_bath-spec} and \ref{fig:SI_bath-curr}).
The statement above in bold can be understood from the following two properties of the current operator:
\begin{itemize}
    
    \item Each many-body energy eigenstate is a stationary state with zero current expectation value $\ev{j}$ anywhere in the lattice, i.e. $\expval{j}{\psi_{N,k}} = 0$. It follows that a system in an incoherent admixture of different many-body eigenstates would also have $\ev{j}=0$. Therefore, having a non-zero current in a many-body state requires coherent superposition between different many-body eigenstates.
    
    \item The current operator and the total particle number operator commute, $ \comm*{\hat{N}}{\hat j} = 0$. Intuitively, the current describes the flow of particles which does not change the total particle number in the system. Mathematically, this implies that $\mel{\psi_{N,k}}{j}{\psi_{M,l}} = 0$ for any $N\ne M$. Therefore, the total current is the sum of independent contributions from different particle number manifolds $\ev*{\hat j} = \text{Tr} (\rho \hat j) = \sum_N \ev*{\hat j_N}$. Here $\rho$ is the density matrix of the system, and the current in particle number manifold $N$ is $\ev*{\hat j_N} = \sum_k \expval{\rho \hat j}{\psi_{N,k}}$.
\end{itemize}

\subsection{Energy-dependent transport in Fig.\,\ref{fig:fig4}b}

\begin{figure}
    \includegraphics[width=0.7\columnwidth]{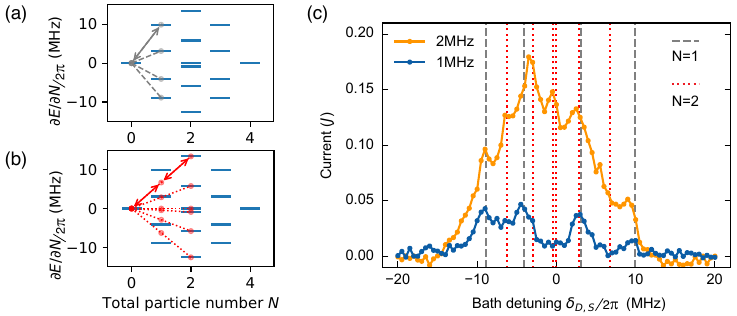}
    \caption{Identification of observed features in the bath-coupled steady-state current. (a, b) Resonant single-particle and two-particle processes starting from the ground state. (c) The same data from Fig.\,\ref{fig:fig4}b, with lines indicating the frequencies of the single-particle transitions (grey dashed) and the two-particle transitions (red dotted).}
    \label{fig:SI_bath-spec}
\end{figure}

\begin{figure}
    \includegraphics[width=1.0\columnwidth]{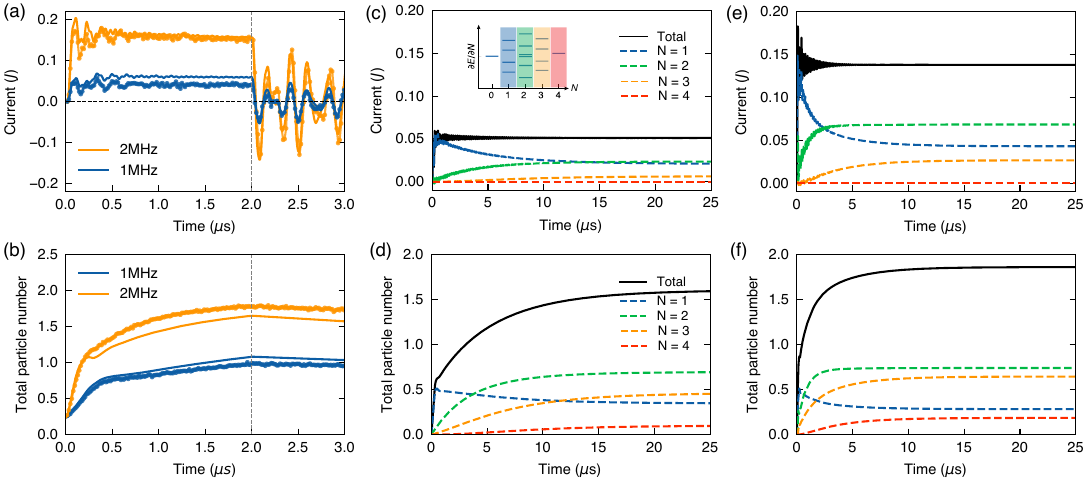}
    \caption{Current dynamics in bath-coupled lattice. (a) Numerical simulation (line) of the data (dots) in Fig.\,\ref{fig:fig4}c. The baths are on during the first 2$\mu$s and off afterward. (b) Same experiment as (a), showing the measured and calculated total particle number in the lattice. (c-f) Numerical simulation of the dynamics towards the non-equilibrium steady state. Black solid lines are the total current $\ev{j}$ and total particle number in the lattice (same as in (a,b)); colored dotted lines are contributions from each particle number manifold labeled by $N$.
    }
    \label{fig:SI_bath-curr}
\end{figure}

In Fig.\,\ref{fig:fig4}b, we have claimed that the peaks in the current versus bath-detuning data correspond to particle transport involving single- and two-particle eigenstates. In Fig.\,\ref{fig:SI_bath-spec}(a) we illustrate the single-particle processes where particles are added to the $N=1$ eigenstates by the particle source and removed from the lattice by the particle drain. The energies of these $N=0\rightarrow 1$ transitions are shown in Fig.\,\ref{fig:SI_bath-spec}(c) as grey vertical dashed lines (same as black arrow locations in Fig.\,\ref{fig:fig4}b), and they align well with the observed peaks in current for weak bath-coupling $g_\text{S,D} = 2\pi\times 1$MHz. When calculating the many-body spectrum, we added a finite next-nearest-neighbor tunneling $J_{\text{NNN}}=2\pi\times 0.55$\,MHz in addition to the measured nearest-neighbor tunneling $J=2\pi\times 5.8$\,MHz at this lattice location. This $J_{\text{NNN}}$ is consistent with finite element simulation of the device, and allows us to match the lattice spectrum more accurately to the measured current and current dynamics. 
Recall from above argument, population in one stationary eigenstate is not sufficient to generate current, nor is it sufficient to populate several eigenstates incoherently; we rely on coherent superpositions to have a finite current. When the bath frequency is tuned near one of the $N=1$ states (which gets populated resonantly), other $N=1$ states have to be off-resonantly populated as well. The different $N=1$ states are populated with some coherence between them, as a result of the intricate interplay between the incoherent particle baths and the coherent lattice Hamiltonian.

With larger bath-coupling $g_\text{S,D} =2\pi\times 2$MHz, the same single-particle processes contribute to the observed current, in addition to new two-particle processes.
In \ref{fig:SI_bath-spec}(b), we illustrate the two-particle processes where the particle baths add and remove two particles, each at half the energy of the transition between the $N=0$ ground state and one of the $N=2$ eigenstates. In (c), the red dotted lines indicate these transitions and show good agreement with peaks observed in the data.
Some of the $N=2$ states are more easily excited because of the presence of a near-resonant $N=1$ eigenstate at the required detuning, e.g., the second highest energy states in $N=2$. Other $N=2$ states can be harder to excite because the required bath detuning is off-resonant with all $N=1$ eigenstates, e.g. the highest energy $N=2$ state. In the latter case, the two-particle eigenstate is excited via a virtual population of the off-resonant $N=1$ states.

\subsection{Current dynamics in Fig.\,\ref{fig:fig4}c}

In Fig.\,\ref{fig:SI_bath-curr}(a), we replot the measured current dynamics in Fig.\,\ref{fig:fig4}c (dots) and show good agreement with numerical simulation (lines). We included the estimated $J_\text{NNN}$ in these Lindblad master equation simulations. The current dynamics at short times and steady-state value depend sensitively on the exact parameters of the system, e.g., small residual lattice disorder. 
When the current reaches a finite steady-state value, the many-body state contains coherent superpositions of different eigenstates. When the baths are turned off, the many-body state evolves in the closed lattice, with the different eigenstate components evolving at different energies. Therefore the current undergoes coherent oscillations. The amplitude of the oscillations will gradually decrease as a result of dephasing; here, the decay constant will be given by the effective many-body dephasing time of $\sim 10\mu$s. 

In Fig.\,\ref{fig:SI_bath-curr}(b), we show additional experimental data for the measured total particle number in the lattice for the same experiments in (a) and the results of numerical simulation. While the current through the lattice reaches the steady-state value within $1\,\mu$s, the total particle number continues to increase slowly until the baths are turned off at $2\,\mu$s. The bath-coupled lattice needs more time to reach the many-body non-equilibrium steady state (NESS), and that time scale is \textit{not} reflected in the current dynamics $\ev{j(t)}$. We briefly discuss the dynamics towards the NESS in the next section. After the bath is turned off, the total particle number in the lattice will decrease due to single-particle relaxation with a decay constant of $T_1 \sim 30\mu$s.

\subsection{Dynamics towards the NESS} 

To better understand the dynamics of the system towards the NESS, we simulate the experiments in Fig.\,\ref{fig:fig4}c but keep both baths on at all times for a longer duration of $25\mu$s. In Fig.\,\ref{fig:SI_bath-curr}(c-f), we plot the simulated current and particle number for bath-coupling rate $g_\text{S,D} = 2\pi\times 1$MHz (c,d) and $g_\text{S,D} = 2\pi\times 2$MHz (e,f). For the current, we calculate both the current contributions from each particle number manifold $\ev*{\hat j_N}$ (dotted color lines) and the total current expectation value $\ev*{j} = \sum_N \ev*{\hat j_N}$ (solid back line). Similarly for the particle number plots, we calculate both the number of particles in each particle number manifold (color dotted lines) and the total particle number in the lattice (solid black line).

The results for $g_\text{S,D} = 2\pi\times 1$MHz are shown in (c, d). The total current $\ev{j}$ settled near the steady-state value within $1\mu$s, but the currents from each particle number manifold $\ev*{\hat j_N}$ only reached their steady-state values after $\sim 20\mu$s when the system reaches its NESS. In the first few $\mu$s after the baths are on, the current is dominated by the contribution from $N=1$. The bath detuning here is aligned with one of the single-particle eigenstates, resonantly populating this state. All other eigenstates are driven off-resonantly and/or via multi-particle processes, therefore being populated at much slower rates. Hence, the initial current is dominated by the $N=1$ contribution, coming from the coherent superposition between one resonantly driven single-particle state and other states in $N=1$.
At longer times, the current contributions from $N>2$ gradually increase while the $N=1$ contribution decreases, but the total current remains roughly the same. This change in current contribution is also seen from the population in each particle manifold (d). In the final NESS, there is a finite population in the $N=4$ Mott state, but it does not contribute to the current: current is generated from superpositions within a particle number manifold, but there is only one eigenstate with $N=4$ in our case.
For $g_\text{S,D} = 2\pi\times 2$MHz at this same bath detuning (e, f), the current and population dynamics show qualitatively similar behavior towards the NESS. The larger bath-lattice coupling enables higher particle number eigenstates to contribute more in the particle transport, and allows faster relaxation towards the NESS. Here we see more current contributions from $N=2$ and $N=3$ manifolds in the NESS.

It remains an open question why the total current expectation value settles close to a steady-state value much faster than the underlying many-body evolution toward the NESS. We will leave for future work to explore the detailed properties of the NESS and its dependence on the baths' spectral properties, lattice interactions, and decoherence effects. 


\begin{thebibliography}{70}%
\makeatletter
\providecommand \@ifxundefined [1]{%
 \@ifx{#1\undefined}
}%
\providecommand \@ifnum [1]{%
 \ifnum #1\expandafter \@firstoftwo
 \else \expandafter \@secondoftwo
 \fi
}%
\providecommand \@ifx [1]{%
 \ifx #1\expandafter \@firstoftwo
 \else \expandafter \@secondoftwo
 \fi
}%
\providecommand \natexlab [1]{#1}%
\providecommand \enquote  [1]{``#1''}%
\providecommand \bibnamefont  [1]{#1}%
\providecommand \bibfnamefont [1]{#1}%
\providecommand \citenamefont [1]{#1}%
\providecommand \href@noop [0]{\@secondoftwo}%
\providecommand \href [0]{\begingroup \@sanitize@url \@href}%
\providecommand \@href[1]{\@@startlink{#1}\@@href}%
\providecommand \@@href[1]{\endgroup#1\@@endlink}%
\providecommand \@sanitize@url [0]{\catcode `\\12\catcode `\$12\catcode `\&12\catcode `\#12\catcode `\^12\catcode `\_12\catcode `\%12\relax}%
\providecommand \@@startlink[1]{}%
\providecommand \@@endlink[0]{}%
\providecommand \url  [0]{\begingroup\@sanitize@url \@url }%
\providecommand \@url [1]{\endgroup\@href {#1}{\urlprefix }}%
\providecommand \urlprefix  [0]{URL }%
\providecommand \Eprint [0]{\href }%
\providecommand \doibase [0]{https://doi.org/}%
\providecommand \selectlanguage [0]{\@gobble}%
\providecommand \bibinfo  [0]{\@secondoftwo}%
\providecommand \bibfield  [0]{\@secondoftwo}%
\providecommand \translation [1]{[#1]}%
\providecommand \BibitemOpen [0]{}%
\providecommand \bibitemStop [0]{}%
\providecommand \bibitemNoStop [0]{.\EOS\space}%
\providecommand \EOS [0]{\spacefactor3000\relax}%
\providecommand \BibitemShut  [1]{\csname bibitem#1\endcsname}%
\let\auto@bib@innerbib\@empty
\bibitem [{\citenamefont {Sachdev}(2011)}]{Sachdev2011-mn}%
  \BibitemOpen
  \bibfield  {author} {\bibinfo {author} {\bibfnamefont {S.}~\bibnamefont {Sachdev}},\ }\href {https://www.cambridge.org/core/books/quantum-phase-transitions/33C1C81500346005E54C1DE4223E5562} {\emph {\bibinfo {title} {Quantum Phase Transitions}}}\ (\bibinfo  {publisher} {Cambridge University Press},\ \bibinfo {year} {2011})\BibitemShut {NoStop}%
\bibitem [{\citenamefont {Blanter}\ and\ \citenamefont {B{\"u}ttiker}(2000)}]{Blanter2000-ah}%
  \BibitemOpen
  \bibfield  {author} {\bibinfo {author} {\bibfnamefont {Y.~M.}\ \bibnamefont {Blanter}}\ and\ \bibinfo {author} {\bibfnamefont {M.}~\bibnamefont {B{\"u}ttiker}},\ }\href {https://www.sciencedirect.com/science/article/pii/S0370157399001234} {\bibfield  {journal} {\bibinfo  {journal} {Phys. Rep.}\ }\textbf {\bibinfo {volume} {336}},\ \bibinfo {pages} {1} (\bibinfo {year} {2000})}\BibitemShut {NoStop}%
\bibitem [{\citenamefont {Georgescu}\ \emph {et~al.}(2014)\citenamefont {Georgescu}, \citenamefont {Ashhab},\ and\ \citenamefont {Nori}}]{Georgescu2014-mt}%
  \BibitemOpen
  \bibfield  {author} {\bibinfo {author} {\bibfnamefont {I.~M.}\ \bibnamefont {Georgescu}}, \bibinfo {author} {\bibfnamefont {S.}~\bibnamefont {Ashhab}},\ and\ \bibinfo {author} {\bibfnamefont {F.}~\bibnamefont {Nori}},\ }\href {https://link.aps.org/doi/10.1103/RevModPhys.86.153} {\bibfield  {journal} {\bibinfo  {journal} {Rev. Mod. Phys.}\ }\textbf {\bibinfo {volume} {86}},\ \bibinfo {pages} {153} (\bibinfo {year} {2014})}\BibitemShut {NoStop}%
\bibitem [{\citenamefont {Altman}\ \emph {et~al.}(2021)\citenamefont {Altman}, \citenamefont {Brown}, \citenamefont {Carleo}, \citenamefont {Carr}, \citenamefont {Demler}, \citenamefont {Chin}, \citenamefont {DeMarco}, \citenamefont {Economou}, \citenamefont {Eriksson}, \citenamefont {Fu}, \citenamefont {Greiner}, \citenamefont {Hazzard}, \citenamefont {Hulet}, \citenamefont {Koll{\'a}r}, \citenamefont {Lev}, \citenamefont {Lukin}, \citenamefont {Ma}, \citenamefont {Mi}, \citenamefont {Misra}, \citenamefont {Monroe}, \citenamefont {Murch}, \citenamefont {Nazario}, \citenamefont {Ni}, \citenamefont {Potter}, \citenamefont {Roushan}, \citenamefont {Saffman}, \citenamefont {Schleier-Smith}, \citenamefont {Siddiqi}, \citenamefont {Simmonds}, \citenamefont {Singh}, \citenamefont {Spielman}, \citenamefont {Temme}, \citenamefont {Weiss}, \citenamefont {Vu{\v c}kovi{\'c}}, \citenamefont {Vuleti{\'c}}, \citenamefont {Ye},\ and\ \citenamefont {Zwierlein}}]{Altman2021-yw}%
  \BibitemOpen
  \bibfield  {author} {\bibinfo {author} {\bibfnamefont {E.}~\bibnamefont {Altman}}, \bibinfo {author} {\bibfnamefont {K.~R.}\ \bibnamefont {Brown}}, \bibinfo {author} {\bibfnamefont {G.}~\bibnamefont {Carleo}}, \bibinfo {author} {\bibfnamefont {L.~D.}\ \bibnamefont {Carr}}, \bibinfo {author} {\bibfnamefont {E.}~\bibnamefont {Demler}}, \bibinfo {author} {\bibfnamefont {C.}~\bibnamefont {Chin}}, \bibinfo {author} {\bibfnamefont {B.}~\bibnamefont {DeMarco}}, \bibinfo {author} {\bibfnamefont {S.~E.}\ \bibnamefont {Economou}}, \bibinfo {author} {\bibfnamefont {M.~A.}\ \bibnamefont {Eriksson}}, \bibinfo {author} {\bibfnamefont {K.-M.~C.}\ \bibnamefont {Fu}}, \bibinfo {author} {\bibfnamefont {\textit{et al.}}},\ }\href {https://link.aps.org/doi/10.1103/PRXQuantum.2.017003} {\bibfield  {journal} {\bibinfo  {journal} {PRX Quantum}\ }\textbf {\bibinfo {volume} {2}},\ \bibinfo {pages} {017003} (\bibinfo {year} {2021})}\BibitemShut {NoStop}%
\bibitem [{\citenamefont {Daley}\ \emph {et~al.}(2022)\citenamefont {Daley}, \citenamefont {Bloch}, \citenamefont {Kokail}, \citenamefont {Flannigan}, \citenamefont {Pearson}, \citenamefont {Troyer},\ and\ \citenamefont {Zoller}}]{Daley2022-xm}%
  \BibitemOpen
  \bibfield  {author} {\bibinfo {author} {\bibfnamefont {A.~J.}\ \bibnamefont {Daley}}, \bibinfo {author} {\bibfnamefont {I.}~\bibnamefont {Bloch}}, \bibinfo {author} {\bibfnamefont {C.}~\bibnamefont {Kokail}}, \bibinfo {author} {\bibfnamefont {S.}~\bibnamefont {Flannigan}}, \bibinfo {author} {\bibfnamefont {N.}~\bibnamefont {Pearson}}, \bibinfo {author} {\bibfnamefont {M.}~\bibnamefont {Troyer}},\ and\ \bibinfo {author} {\bibfnamefont {P.}~\bibnamefont {Zoller}},\ }\href {http://dx.doi.org/10.1038/s41586-022-04940-6} {\bibfield  {journal} {\bibinfo  {journal} {Nature}\ }\textbf {\bibinfo {volume} {607}},\ \bibinfo {pages} {667} (\bibinfo {year} {2022})}\BibitemShut {NoStop}%
\bibitem [{\citenamefont {Choi}\ \emph {et~al.}(2016)\citenamefont {Choi}, \citenamefont {Hild}, \citenamefont {Zeiher}, \citenamefont {Schau{\ss}}, \citenamefont {Rubio-Abadal}, \citenamefont {Yefsah}, \citenamefont {Khemani}, \citenamefont {Huse}, \citenamefont {Bloch},\ and\ \citenamefont {Gross}}]{Choi2016-ej}%
  \BibitemOpen
  \bibfield  {author} {\bibinfo {author} {\bibfnamefont {J.-Y.}\ \bibnamefont {Choi}}, \bibinfo {author} {\bibfnamefont {S.}~\bibnamefont {Hild}}, \bibinfo {author} {\bibfnamefont {J.}~\bibnamefont {Zeiher}}, \bibinfo {author} {\bibfnamefont {P.}~\bibnamefont {Schau{\ss}}}, \bibinfo {author} {\bibfnamefont {A.}~\bibnamefont {Rubio-Abadal}}, \bibinfo {author} {\bibfnamefont {T.}~\bibnamefont {Yefsah}}, \bibinfo {author} {\bibfnamefont {V.}~\bibnamefont {Khemani}}, \bibinfo {author} {\bibfnamefont {D.~A.}\ \bibnamefont {Huse}}, \bibinfo {author} {\bibfnamefont {I.}~\bibnamefont {Bloch}},\ and\ \bibinfo {author} {\bibfnamefont {C.}~\bibnamefont {Gross}},\ }\href {http://dx.doi.org/10.1126/science.aaf8834} {\bibfield  {journal} {\bibinfo  {journal} {Science}\ }\textbf {\bibinfo {volume} {352}},\ \bibinfo {pages} {1547} (\bibinfo {year} {2016})}\BibitemShut {NoStop}%
\bibitem [{\citenamefont {Brown}\ \emph {et~al.}(2019)\citenamefont {Brown}, \citenamefont {Mitra}, \citenamefont {Guardado-Sanchez}, \citenamefont {Nourafkan}, \citenamefont {Reymbaut}, \citenamefont {H{\'e}bert}, \citenamefont {Bergeron}, \citenamefont {Tremblay}, \citenamefont {Kokalj}, \citenamefont {Huse}, \citenamefont {Schau{\ss}},\ and\ \citenamefont {Bakr}}]{Brown2019-nx}%
  \BibitemOpen
  \bibfield  {author} {\bibinfo {author} {\bibfnamefont {P.~T.}\ \bibnamefont {Brown}}, \bibinfo {author} {\bibfnamefont {D.}~\bibnamefont {Mitra}}, \bibinfo {author} {\bibfnamefont {E.}~\bibnamefont {Guardado-Sanchez}}, \bibinfo {author} {\bibfnamefont {R.}~\bibnamefont {Nourafkan}}, \bibinfo {author} {\bibfnamefont {A.}~\bibnamefont {Reymbaut}}, \bibinfo {author} {\bibfnamefont {C.-D.}\ \bibnamefont {H{\'e}bert}}, \bibinfo {author} {\bibfnamefont {S.}~\bibnamefont {Bergeron}}, \bibinfo {author} {\bibfnamefont {A.-M.~S.}\ \bibnamefont {Tremblay}}, \bibinfo {author} {\bibfnamefont {J.}~\bibnamefont {Kokalj}}, \bibinfo {author} {\bibfnamefont {D.~A.}\ \bibnamefont {Huse}}, \bibinfo {author} {\bibfnamefont {P.}~\bibnamefont {Schau{\ss}}},\ and\ \bibinfo {author} {\bibfnamefont {W.~S.}\ \bibnamefont {Bakr}},\ }\href {http://dx.doi.org/10.1126/science.aat4134} {\bibfield  {journal} {\bibinfo  {journal} {Science}\ }\textbf {\bibinfo {volume} {363}},\ \bibinfo {pages} {379} (\bibinfo {year} {2019})}\BibitemShut
  {NoStop}%
\bibitem [{\citenamefont {Anderson}\ \emph {et~al.}(2019)\citenamefont {Anderson}, \citenamefont {Wang}, \citenamefont {Xu}, \citenamefont {Venu}, \citenamefont {Trotzky}, \citenamefont {Chevy},\ and\ \citenamefont {Thywissen}}]{Anderson2019-xy}%
  \BibitemOpen
  \bibfield  {author} {\bibinfo {author} {\bibfnamefont {R.}~\bibnamefont {Anderson}}, \bibinfo {author} {\bibfnamefont {F.}~\bibnamefont {Wang}}, \bibinfo {author} {\bibfnamefont {P.}~\bibnamefont {Xu}}, \bibinfo {author} {\bibfnamefont {V.}~\bibnamefont {Venu}}, \bibinfo {author} {\bibfnamefont {S.}~\bibnamefont {Trotzky}}, \bibinfo {author} {\bibfnamefont {F.}~\bibnamefont {Chevy}},\ and\ \bibinfo {author} {\bibfnamefont {J.~H.}\ \bibnamefont {Thywissen}},\ }\href {http://dx.doi.org/10.1103/PhysRevLett.122.153602} {\bibfield  {journal} {\bibinfo  {journal} {Phys. Rev. Lett.}\ }\textbf {\bibinfo {volume} {122}},\ \bibinfo {pages} {153602} (\bibinfo {year} {2019})}\BibitemShut {NoStop}%
\bibitem [{\citenamefont {Brantut}\ \emph {et~al.}(2012)\citenamefont {Brantut}, \citenamefont {Meineke}, \citenamefont {Stadler}, \citenamefont {Krinner},\ and\ \citenamefont {Esslinger}}]{Brantut2012-gz}%
  \BibitemOpen
  \bibfield  {author} {\bibinfo {author} {\bibfnamefont {J.-P.}\ \bibnamefont {Brantut}}, \bibinfo {author} {\bibfnamefont {J.}~\bibnamefont {Meineke}}, \bibinfo {author} {\bibfnamefont {D.}~\bibnamefont {Stadler}}, \bibinfo {author} {\bibfnamefont {S.}~\bibnamefont {Krinner}},\ and\ \bibinfo {author} {\bibfnamefont {T.}~\bibnamefont {Esslinger}},\ }\href {http://dx.doi.org/10.1126/science.1223175} {\bibfield  {journal} {\bibinfo  {journal} {Science}\ }\textbf {\bibinfo {volume} {337}},\ \bibinfo {pages} {1069} (\bibinfo {year} {2012})}\BibitemShut {NoStop}%
\bibitem [{\citenamefont {H{\"a}usler}\ \emph {et~al.}(2017)\citenamefont {H{\"a}usler}, \citenamefont {Nakajima}, \citenamefont {Lebrat}, \citenamefont {Husmann}, \citenamefont {Krinner}, \citenamefont {Esslinger},\ and\ \citenamefont {Brantut}}]{Hausler2017-ag}%
  \BibitemOpen
  \bibfield  {author} {\bibinfo {author} {\bibfnamefont {S.}~\bibnamefont {H{\"a}usler}}, \bibinfo {author} {\bibfnamefont {S.}~\bibnamefont {Nakajima}}, \bibinfo {author} {\bibfnamefont {M.}~\bibnamefont {Lebrat}}, \bibinfo {author} {\bibfnamefont {D.}~\bibnamefont {Husmann}}, \bibinfo {author} {\bibfnamefont {S.}~\bibnamefont {Krinner}}, \bibinfo {author} {\bibfnamefont {T.}~\bibnamefont {Esslinger}},\ and\ \bibinfo {author} {\bibfnamefont {J.-P.}\ \bibnamefont {Brantut}},\ }\href {http://dx.doi.org/10.1103/PhysRevLett.119.030403} {\bibfield  {journal} {\bibinfo  {journal} {Phys. Rev. Lett.}\ }\textbf {\bibinfo {volume} {119}},\ \bibinfo {pages} {030403} (\bibinfo {year} {2017})}\BibitemShut {NoStop}%
\bibitem [{\citenamefont {Noh}\ and\ \citenamefont {Angelakis}(2017)}]{Noh2017-ow}%
  \BibitemOpen
  \bibfield  {author} {\bibinfo {author} {\bibfnamefont {C.}~\bibnamefont {Noh}}\ and\ \bibinfo {author} {\bibfnamefont {D.~G.}\ \bibnamefont {Angelakis}},\ }\href {http://dx.doi.org/10.1088/0034-4885/80/1/016401} {\bibfield  {journal} {\bibinfo  {journal} {Rep. Prog. Phys.}\ }\textbf {\bibinfo {volume} {80}},\ \bibinfo {pages} {016401} (\bibinfo {year} {2017})}\BibitemShut {NoStop}%
\bibitem [{\citenamefont {Hartmann}(2016)}]{Hartmann2016-tl}%
  \BibitemOpen
  \bibfield  {author} {\bibinfo {author} {\bibfnamefont {M.~J.}\ \bibnamefont {Hartmann}},\ }\href {https://iopscience.iop.org/article/10.1088/2040-8978/18/10/104005/meta} {\bibfield  {journal} {\bibinfo  {journal} {J. Opt.}\ }\textbf {\bibinfo {volume} {18}},\ \bibinfo {pages} {104005} (\bibinfo {year} {2016})}\BibitemShut {NoStop}%
\bibitem [{\citenamefont {Gu}\ \emph {et~al.}(2017)\citenamefont {Gu}, \citenamefont {Kockum}, \citenamefont {Miranowicz}, \citenamefont {Liu},\ and\ \citenamefont {Nori}}]{Gu2017-iu}%
  \BibitemOpen
  \bibfield  {author} {\bibinfo {author} {\bibfnamefont {X.}~\bibnamefont {Gu}}, \bibinfo {author} {\bibfnamefont {A.~F.}\ \bibnamefont {Kockum}}, \bibinfo {author} {\bibfnamefont {A.}~\bibnamefont {Miranowicz}}, \bibinfo {author} {\bibfnamefont {Y.-X.}\ \bibnamefont {Liu}},\ and\ \bibinfo {author} {\bibfnamefont {F.}~\bibnamefont {Nori}},\ }\href {https://www.sciencedirect.com/science/article/pii/S0370157317303290} {\bibfield  {journal} {\bibinfo  {journal} {Phys. Rep.}\ }\textbf {\bibinfo {volume} {718-719}},\ \bibinfo {pages} {1} (\bibinfo {year} {2017})}\BibitemShut {NoStop}%
\bibitem [{\citenamefont {Carusotto}\ \emph {et~al.}(2020)\citenamefont {Carusotto}, \citenamefont {Houck}, \citenamefont {Koll{\'a}r}, \citenamefont {Roushan}, \citenamefont {Schuster},\ and\ \citenamefont {Simon}}]{Carusotto2020-ct}%
  \BibitemOpen
  \bibfield  {author} {\bibinfo {author} {\bibfnamefont {I.}~\bibnamefont {Carusotto}}, \bibinfo {author} {\bibfnamefont {A.~A.}\ \bibnamefont {Houck}}, \bibinfo {author} {\bibfnamefont {A.~J.}\ \bibnamefont {Koll{\'a}r}}, \bibinfo {author} {\bibfnamefont {P.}~\bibnamefont {Roushan}}, \bibinfo {author} {\bibfnamefont {D.~I.}\ \bibnamefont {Schuster}},\ and\ \bibinfo {author} {\bibfnamefont {J.}~\bibnamefont {Simon}},\ }\href {https://www.nature.com/articles/s41567-020-0815-y} {\bibfield  {journal} {\bibinfo  {journal} {Nat. Phys.}\ }\textbf {\bibinfo {volume} {16}},\ \bibinfo {pages} {268} (\bibinfo {year} {2020})}\BibitemShut {NoStop}%
\bibitem [{\citenamefont {Hacohen-Gourgy}\ \emph {et~al.}(2015)\citenamefont {Hacohen-Gourgy}, \citenamefont {Ramasesh}, \citenamefont {De~Grandi}, \citenamefont {Siddiqi},\ and\ \citenamefont {Girvin}}]{Hacohen-Gourgy2015-zc}%
  \BibitemOpen
  \bibfield  {author} {\bibinfo {author} {\bibfnamefont {S.}~\bibnamefont {Hacohen-Gourgy}}, \bibinfo {author} {\bibfnamefont {V.~V.}\ \bibnamefont {Ramasesh}}, \bibinfo {author} {\bibfnamefont {C.}~\bibnamefont {De~Grandi}}, \bibinfo {author} {\bibfnamefont {I.}~\bibnamefont {Siddiqi}},\ and\ \bibinfo {author} {\bibfnamefont {S.~M.}\ \bibnamefont {Girvin}},\ }\href {http://dx.doi.org/10.1103/PhysRevLett.115.240501} {\bibfield  {journal} {\bibinfo  {journal} {Phys. Rev. Lett.}\ }\textbf {\bibinfo {volume} {115}},\ \bibinfo {pages} {240501} (\bibinfo {year} {2015})}\BibitemShut {NoStop}%
\bibitem [{\citenamefont {Collodo}\ \emph {et~al.}(2019)\citenamefont {Collodo}, \citenamefont {Poto{\v c}nik}, \citenamefont {Gasparinetti}, \citenamefont {Besse}, \citenamefont {Pechal}, \citenamefont {Sameti}, \citenamefont {Hartmann}, \citenamefont {Wallraff},\ and\ \citenamefont {Eichler}}]{Collodo2019-oa}%
  \BibitemOpen
  \bibfield  {author} {\bibinfo {author} {\bibfnamefont {M.~C.}\ \bibnamefont {Collodo}}, \bibinfo {author} {\bibfnamefont {A.}~\bibnamefont {Poto{\v c}nik}}, \bibinfo {author} {\bibfnamefont {S.}~\bibnamefont {Gasparinetti}}, \bibinfo {author} {\bibfnamefont {J.-C.}\ \bibnamefont {Besse}}, \bibinfo {author} {\bibfnamefont {M.}~\bibnamefont {Pechal}}, \bibinfo {author} {\bibfnamefont {M.}~\bibnamefont {Sameti}}, \bibinfo {author} {\bibfnamefont {M.~J.}\ \bibnamefont {Hartmann}}, \bibinfo {author} {\bibfnamefont {A.}~\bibnamefont {Wallraff}},\ and\ \bibinfo {author} {\bibfnamefont {C.}~\bibnamefont {Eichler}},\ }\href {http://dx.doi.org/10.1103/PhysRevLett.122.183601} {\bibfield  {journal} {\bibinfo  {journal} {Phys. Rev. Lett.}\ }\textbf {\bibinfo {volume} {122}},\ \bibinfo {pages} {183601} (\bibinfo {year} {2019})}\BibitemShut {NoStop}%
\bibitem [{\citenamefont {Ma}\ \emph {et~al.}(2019)\citenamefont {Ma}, \citenamefont {Saxberg}, \citenamefont {Owens}, \citenamefont {Leung}, \citenamefont {Lu}, \citenamefont {Simon},\ and\ \citenamefont {Schuster}}]{Ma2019-ye}%
  \BibitemOpen
  \bibfield  {author} {\bibinfo {author} {\bibfnamefont {R.}~\bibnamefont {Ma}}, \bibinfo {author} {\bibfnamefont {B.}~\bibnamefont {Saxberg}}, \bibinfo {author} {\bibfnamefont {C.}~\bibnamefont {Owens}}, \bibinfo {author} {\bibfnamefont {N.}~\bibnamefont {Leung}}, \bibinfo {author} {\bibfnamefont {Y.}~\bibnamefont {Lu}}, \bibinfo {author} {\bibfnamefont {J.}~\bibnamefont {Simon}},\ and\ \bibinfo {author} {\bibfnamefont {D.~I.}\ \bibnamefont {Schuster}},\ }\href {http://dx.doi.org/10.1038/s41586-019-0897-9} {\bibfield  {journal} {\bibinfo  {journal} {Nature}\ }\textbf {\bibinfo {volume} {566}},\ \bibinfo {pages} {51} (\bibinfo {year} {2019})}\BibitemShut {NoStop}%
\bibitem [{\citenamefont {Saxberg}\ \emph {et~al.}(2022)\citenamefont {Saxberg}, \citenamefont {Vrajitoarea}, \citenamefont {Roberts}, \citenamefont {Panetta}, \citenamefont {Simon},\ and\ \citenamefont {Schuster}}]{Saxberg2022-tt}%
  \BibitemOpen
  \bibfield  {author} {\bibinfo {author} {\bibfnamefont {B.}~\bibnamefont {Saxberg}}, \bibinfo {author} {\bibfnamefont {A.}~\bibnamefont {Vrajitoarea}}, \bibinfo {author} {\bibfnamefont {G.}~\bibnamefont {Roberts}}, \bibinfo {author} {\bibfnamefont {M.~G.}\ \bibnamefont {Panetta}}, \bibinfo {author} {\bibfnamefont {J.}~\bibnamefont {Simon}},\ and\ \bibinfo {author} {\bibfnamefont {D.~I.}\ \bibnamefont {Schuster}},\ }\href {http://dx.doi.org/10.1038/s41586-022-05357-x} {\bibfield  {journal} {\bibinfo  {journal} {Nature}\ }\textbf {\bibinfo {volume} {612}},\ \bibinfo {pages} {435} (\bibinfo {year} {2022})}\BibitemShut {NoStop}%
\bibitem [{\citenamefont {Morvan}\ \emph {et~al.}(2022)\citenamefont {Morvan}, \citenamefont {Andersen}, \citenamefont {Mi}, \citenamefont {Neill}, \citenamefont {Petukhov}, \citenamefont {Kechedzhi}, \citenamefont {Abanin}, \citenamefont {Michailidis}, \citenamefont {Acharya}, \citenamefont {Arute}, \citenamefont {Arya}, \citenamefont {Asfaw}, \citenamefont {Atalaya}, \citenamefont {Bardin}, \citenamefont {Basso}, \citenamefont {Bengtsson}, \citenamefont {Bortoli}, \citenamefont {Bourassa}, \citenamefont {Bovaird}, \citenamefont {Brill}, \citenamefont {Broughton}, \citenamefont {Buckley}, \citenamefont {Buell}, \citenamefont {Burger}, \citenamefont {Burkett}, \citenamefont {Bushnell}, \citenamefont {Chen}, \citenamefont {Chiaro}, \citenamefont {Collins}, \citenamefont {Conner}, \citenamefont {Courtney}, \citenamefont {Crook}, \citenamefont {Curtin}, \citenamefont {Debroy}, \citenamefont {Del Toro~Barba}, \citenamefont {Demura}, \citenamefont {Dunsworth}, \citenamefont {Eppens}, \citenamefont {Erickson},
  \citenamefont {Faoro}, \citenamefont {Farhi}, \citenamefont {Fatemi}, \citenamefont {Flores~Burgos}, \citenamefont {Forati}, \citenamefont {Fowler}, \citenamefont {Foxen}, \citenamefont {Giang}, \citenamefont {Gidney}, \citenamefont {Gilboa}, \citenamefont {Giustina}, \citenamefont {Grajales~Dau}, \citenamefont {Gross}, \citenamefont {Habegger}, \citenamefont {Hamilton}, \citenamefont {Harrigan}, \citenamefont {Harrington}, \citenamefont {Hoffmann}, \citenamefont {Hong}, \citenamefont {Huang}, \citenamefont {Huff}, \citenamefont {Huggins}, \citenamefont {Isakov}, \citenamefont {Iveland}, \citenamefont {Jeffrey}, \citenamefont {Jiang}, \citenamefont {Jones}, \citenamefont {Juhas}, \citenamefont {Kafri}, \citenamefont {Khattar}, \citenamefont {Khezri}, \citenamefont {Kieferov{\'a}}, \citenamefont {Kim}, \citenamefont {Kitaev}, \citenamefont {Klimov}, \citenamefont {Klots}, \citenamefont {Korotkov}, \citenamefont {Kostritsa}, \citenamefont {Kreikebaum}, \citenamefont {Landhuis}, \citenamefont {Laptev},
  \citenamefont {Lau}, \citenamefont {Laws}, \citenamefont {Lee}, \citenamefont {Lee}, \citenamefont {Lester}, \citenamefont {Lill}, \citenamefont {Liu}, \citenamefont {Locharla}, \citenamefont {Malone}, \citenamefont {Martin}, \citenamefont {McClean}, \citenamefont {McEwen}, \citenamefont {Meurer~Costa}, \citenamefont {Miao}, \citenamefont {Mohseni}, \citenamefont {Montazeri}, \citenamefont {Mount}, \citenamefont {Mruczkiewicz}, \citenamefont {Naaman}, \citenamefont {Neeley}, \citenamefont {Nersisyan}, \citenamefont {Newman}, \citenamefont {Nguyen}, \citenamefont {Nguyen}, \citenamefont {Niu}, \citenamefont {O'Brien}, \citenamefont {Olenewa}, \citenamefont {Opremcak}, \citenamefont {Potter}, \citenamefont {Quintana}, \citenamefont {Rubin}, \citenamefont {Saei}, \citenamefont {Sank}, \citenamefont {Sankaragomathi}, \citenamefont {Satzinger}, \citenamefont {Schurkus}, \citenamefont {Schuster}, \citenamefont {Shearn}, \citenamefont {Shorter}, \citenamefont {Shvarts}, \citenamefont {Skruzny}, \citenamefont
  {Smith}, \citenamefont {Strain}, \citenamefont {Sterling}, \citenamefont {Su}, \citenamefont {Szalay}, \citenamefont {Torres}, \citenamefont {Vidal}, \citenamefont {Villalonga}, \citenamefont {Vollgraff-Heidweiller}, \citenamefont {White}, \citenamefont {Xing}, \citenamefont {Yao}, \citenamefont {Yeh}, \citenamefont {Yoo}, \citenamefont {Zalcman}, \citenamefont {Zhang}, \citenamefont {Zhu}, \citenamefont {Neven}, \citenamefont {Bacon}, \citenamefont {Hilton}, \citenamefont {Lucero}, \citenamefont {Babbush}, \citenamefont {Boixo}, \citenamefont {Megrant}, \citenamefont {Kelly}, \citenamefont {Chen}, \citenamefont {Smelyanskiy}, \citenamefont {Aleiner}, \citenamefont {Ioffe},\ and\ \citenamefont {Roushan}}]{Morvan2022-am}%
  \BibitemOpen
  \bibfield  {author} {\bibinfo {author} {\bibfnamefont {A.}~\bibnamefont {Morvan}}, \bibinfo {author} {\bibfnamefont {T.~I.}\ \bibnamefont {Andersen}}, \bibinfo {author} {\bibfnamefont {X.}~\bibnamefont {Mi}}, \bibinfo {author} {\bibfnamefont {C.}~\bibnamefont {Neill}}, \bibinfo {author} {\bibfnamefont {A.}~\bibnamefont {Petukhov}}, \bibinfo {author} {\bibfnamefont {K.}~\bibnamefont {Kechedzhi}}, \bibinfo {author} {\bibfnamefont {D.~A.}\ \bibnamefont {Abanin}}, \bibinfo {author} {\bibfnamefont {A.}~\bibnamefont {Michailidis}}, \bibinfo {author} {\bibfnamefont {R.}~\bibnamefont {Acharya}}, \bibinfo {author} {\bibfnamefont {F.}~\bibnamefont {Arute}}, \bibinfo {author} {\bibfnamefont {\textit{et al.}}},\ }\href {http://dx.doi.org/10.1038/s41586-022-05348-y} {\bibfield  {journal} {\bibinfo  {journal} {Nature}\ }\textbf {\bibinfo {volume} {612}},\ \bibinfo {pages} {240} (\bibinfo {year} {2022})}\BibitemShut {NoStop}%
\bibitem [{\citenamefont {Zhang}\ \emph {et~al.}(2023)\citenamefont {Zhang}, \citenamefont {Kim}, \citenamefont {Mark}, \citenamefont {Choi},\ and\ \citenamefont {Painter}}]{Zhang2023-fg}%
  \BibitemOpen
  \bibfield  {author} {\bibinfo {author} {\bibfnamefont {X.}~\bibnamefont {Zhang}}, \bibinfo {author} {\bibfnamefont {E.}~\bibnamefont {Kim}}, \bibinfo {author} {\bibfnamefont {D.~K.}\ \bibnamefont {Mark}}, \bibinfo {author} {\bibfnamefont {S.}~\bibnamefont {Choi}},\ and\ \bibinfo {author} {\bibfnamefont {O.}~\bibnamefont {Painter}},\ }\href {http://dx.doi.org/10.1126/science.ade7651} {\bibfield  {journal} {\bibinfo  {journal} {Science}\ }\textbf {\bibinfo {volume} {379}},\ \bibinfo {pages} {278} (\bibinfo {year} {2023})}\BibitemShut {NoStop}%
\bibitem [{\citenamefont {Roushan}\ \emph {et~al.}(2017)\citenamefont {Roushan}, \citenamefont {Neill}, \citenamefont {Tangpanitanon}, \citenamefont {Bastidas}, \citenamefont {Megrant}, \citenamefont {Barends}, \citenamefont {Chen}, \citenamefont {Chen}, \citenamefont {Chiaro}, \citenamefont {Dunsworth}, \citenamefont {Fowler}, \citenamefont {Foxen}, \citenamefont {Giustina}, \citenamefont {Jeffrey}, \citenamefont {Kelly}, \citenamefont {Lucero}, \citenamefont {Mutus}, \citenamefont {Neeley}, \citenamefont {Quintana}, \citenamefont {Sank}, \citenamefont {Vainsencher}, \citenamefont {Wenner}, \citenamefont {White}, \citenamefont {Neven}, \citenamefont {Angelakis},\ and\ \citenamefont {Martinis}}]{Roushan2017-hh}%
  \BibitemOpen
  \bibfield  {author} {\bibinfo {author} {\bibfnamefont {P.}~\bibnamefont {Roushan}}, \bibinfo {author} {\bibfnamefont {C.}~\bibnamefont {Neill}}, \bibinfo {author} {\bibfnamefont {J.}~\bibnamefont {Tangpanitanon}}, \bibinfo {author} {\bibfnamefont {V.~M.}\ \bibnamefont {Bastidas}}, \bibinfo {author} {\bibfnamefont {A.}~\bibnamefont {Megrant}}, \bibinfo {author} {\bibfnamefont {R.}~\bibnamefont {Barends}}, \bibinfo {author} {\bibfnamefont {Y.}~\bibnamefont {Chen}}, \bibinfo {author} {\bibfnamefont {Z.}~\bibnamefont {Chen}}, \bibinfo {author} {\bibfnamefont {B.}~\bibnamefont {Chiaro}}, \bibinfo {author} {\bibfnamefont {A.}~\bibnamefont {Dunsworth}}, \bibinfo {author} {\bibfnamefont {\textit{et al.}}},\ }\href {http://dx.doi.org/10.1126/science.aao1401} {\bibfield  {journal} {\bibinfo  {journal} {Science}\ }\textbf {\bibinfo {volume} {358}},\ \bibinfo {pages} {1175} (\bibinfo {year} {2017})}\BibitemShut {NoStop}%
\bibitem [{\citenamefont {Guo}\ \emph {et~al.}(2021)\citenamefont {Guo}, \citenamefont {Cheng}, \citenamefont {Li}, \citenamefont {Xu}, \citenamefont {Zhang}, \citenamefont {Wang}, \citenamefont {Song}, \citenamefont {Liu}, \citenamefont {Ren}, \citenamefont {Dong}, \citenamefont {Mondaini},\ and\ \citenamefont {Wang}}]{Guo2021-jy}%
  \BibitemOpen
  \bibfield  {author} {\bibinfo {author} {\bibfnamefont {Q.}~\bibnamefont {Guo}}, \bibinfo {author} {\bibfnamefont {C.}~\bibnamefont {Cheng}}, \bibinfo {author} {\bibfnamefont {H.}~\bibnamefont {Li}}, \bibinfo {author} {\bibfnamefont {S.}~\bibnamefont {Xu}}, \bibinfo {author} {\bibfnamefont {P.}~\bibnamefont {Zhang}}, \bibinfo {author} {\bibfnamefont {Z.}~\bibnamefont {Wang}}, \bibinfo {author} {\bibfnamefont {C.}~\bibnamefont {Song}}, \bibinfo {author} {\bibfnamefont {W.}~\bibnamefont {Liu}}, \bibinfo {author} {\bibfnamefont {W.}~\bibnamefont {Ren}}, \bibinfo {author} {\bibfnamefont {H.}~\bibnamefont {Dong}}, \bibinfo {author} {\bibfnamefont {R.}~\bibnamefont {Mondaini}},\ and\ \bibinfo {author} {\bibfnamefont {H.}~\bibnamefont {Wang}},\ }\href {http://dx.doi.org/10.1103/PhysRevLett.127.240502} {\bibfield  {journal} {\bibinfo  {journal} {Phys. Rev. Lett.}\ }\textbf {\bibinfo {volume} {127}},\ \bibinfo {pages} {240502} (\bibinfo {year} {2021})}\BibitemShut {NoStop}%
\bibitem [{\citenamefont {Guo}\ \emph {et~al.}(2020)\citenamefont {Guo}, \citenamefont {Cheng}, \citenamefont {Sun}, \citenamefont {Song}, \citenamefont {Li}, \citenamefont {Wang}, \citenamefont {Ren}, \citenamefont {Dong}, \citenamefont {Zheng}, \citenamefont {Zhang}, \citenamefont {Mondaini}, \citenamefont {Fan},\ and\ \citenamefont {Wang}}]{Guo2020-eo}%
  \BibitemOpen
  \bibfield  {author} {\bibinfo {author} {\bibfnamefont {Q.}~\bibnamefont {Guo}}, \bibinfo {author} {\bibfnamefont {C.}~\bibnamefont {Cheng}}, \bibinfo {author} {\bibfnamefont {Z.-H.}\ \bibnamefont {Sun}}, \bibinfo {author} {\bibfnamefont {Z.}~\bibnamefont {Song}}, \bibinfo {author} {\bibfnamefont {H.}~\bibnamefont {Li}}, \bibinfo {author} {\bibfnamefont {Z.}~\bibnamefont {Wang}}, \bibinfo {author} {\bibfnamefont {W.}~\bibnamefont {Ren}}, \bibinfo {author} {\bibfnamefont {H.}~\bibnamefont {Dong}}, \bibinfo {author} {\bibfnamefont {D.}~\bibnamefont {Zheng}}, \bibinfo {author} {\bibfnamefont {Y.-R.}\ \bibnamefont {Zhang}}, \bibinfo {author} {\bibfnamefont {R.}~\bibnamefont {Mondaini}}, \bibinfo {author} {\bibfnamefont {H.}~\bibnamefont {Fan}},\ and\ \bibinfo {author} {\bibfnamefont {H.}~\bibnamefont {Wang}},\ }\href {https://www.nature.com/articles/s41567-020-1035-1} {\bibfield  {journal} {\bibinfo  {journal} {Nat. Phys.}\ }\textbf {\bibinfo {volume} {17}},\ \bibinfo {pages} {234} (\bibinfo {year}
  {2020})}\BibitemShut {NoStop}%
\bibitem [{\citenamefont {Braum{\"u}ller}\ \emph {et~al.}(2021)\citenamefont {Braum{\"u}ller}, \citenamefont {Karamlou}, \citenamefont {Yanay}, \citenamefont {Kannan}, \citenamefont {Kim}, \citenamefont {Kjaergaard}, \citenamefont {Melville}, \citenamefont {Niedzielski}, \citenamefont {Sung}, \citenamefont {Veps{\"a}l{\"a}inen}, \citenamefont {Winik}, \citenamefont {Yoder}, \citenamefont {Orlando}, \citenamefont {Gustavsson}, \citenamefont {Tahan},\ and\ \citenamefont {Oliver}}]{Braumuller2021-nj}%
  \BibitemOpen
  \bibfield  {author} {\bibinfo {author} {\bibfnamefont {J.}~\bibnamefont {Braum{\"u}ller}}, \bibinfo {author} {\bibfnamefont {A.~H.}\ \bibnamefont {Karamlou}}, \bibinfo {author} {\bibfnamefont {Y.}~\bibnamefont {Yanay}}, \bibinfo {author} {\bibfnamefont {B.}~\bibnamefont {Kannan}}, \bibinfo {author} {\bibfnamefont {D.}~\bibnamefont {Kim}}, \bibinfo {author} {\bibfnamefont {M.}~\bibnamefont {Kjaergaard}}, \bibinfo {author} {\bibfnamefont {A.}~\bibnamefont {Melville}}, \bibinfo {author} {\bibfnamefont {B.~M.}\ \bibnamefont {Niedzielski}}, \bibinfo {author} {\bibfnamefont {Y.}~\bibnamefont {Sung}}, \bibinfo {author} {\bibfnamefont {A.}~\bibnamefont {Veps{\"a}l{\"a}inen}}, \bibinfo {author} {\bibfnamefont {R.}~\bibnamefont {Winik}}, \bibinfo {author} {\bibfnamefont {J.~L.}\ \bibnamefont {Yoder}}, \bibinfo {author} {\bibfnamefont {T.~P.}\ \bibnamefont {Orlando}}, \bibinfo {author} {\bibfnamefont {S.}~\bibnamefont {Gustavsson}}, \bibinfo {author} {\bibfnamefont {C.}~\bibnamefont {Tahan}},\ and\ \bibinfo {author}
  {\bibfnamefont {W.~D.}\ \bibnamefont {Oliver}},\ }\href {https://www.nature.com/articles/s41567-021-01430-w} {\bibfield  {journal} {\bibinfo  {journal} {Nat. Phys.}\ }\textbf {\bibinfo {volume} {18}},\ \bibinfo {pages} {172} (\bibinfo {year} {2021})}\BibitemShut {NoStop}%
\bibitem [{\citenamefont {Zhang}\ \emph {et~al.}(2022)\citenamefont {Zhang}, \citenamefont {Dong}, \citenamefont {Gao}, \citenamefont {Zhao}, \citenamefont {Hao}, \citenamefont {Desaules}, \citenamefont {Guo}, \citenamefont {Chen}, \citenamefont {Deng}, \citenamefont {Liu}, \citenamefont {Ren}, \citenamefont {Yao}, \citenamefont {Zhang}, \citenamefont {Xu}, \citenamefont {Wang}, \citenamefont {Jin}, \citenamefont {Zhu}, \citenamefont {Zhang}, \citenamefont {Li}, \citenamefont {Song}, \citenamefont {Wang}, \citenamefont {Liu}, \citenamefont {Papi{\'c}}, \citenamefont {Ying}, \citenamefont {Wang},\ and\ \citenamefont {Lai}}]{Zhang2022-ts}%
  \BibitemOpen
  \bibfield  {author} {\bibinfo {author} {\bibfnamefont {P.}~\bibnamefont {Zhang}}, \bibinfo {author} {\bibfnamefont {H.}~\bibnamefont {Dong}}, \bibinfo {author} {\bibfnamefont {Y.}~\bibnamefont {Gao}}, \bibinfo {author} {\bibfnamefont {L.}~\bibnamefont {Zhao}}, \bibinfo {author} {\bibfnamefont {J.}~\bibnamefont {Hao}}, \bibinfo {author} {\bibfnamefont {J.-Y.}\ \bibnamefont {Desaules}}, \bibinfo {author} {\bibfnamefont {Q.}~\bibnamefont {Guo}}, \bibinfo {author} {\bibfnamefont {J.}~\bibnamefont {Chen}}, \bibinfo {author} {\bibfnamefont {J.}~\bibnamefont {Deng}}, \bibinfo {author} {\bibfnamefont {B.}~\bibnamefont {Liu}}, \bibinfo {author} {\bibfnamefont {W.}~\bibnamefont {Ren}}, \bibinfo {author} {\bibfnamefont {Y.}~\bibnamefont {Yao}}, \bibinfo {author} {\bibfnamefont {X.}~\bibnamefont {Zhang}}, \bibinfo {author} {\bibfnamefont {S.}~\bibnamefont {Xu}}, \bibinfo {author} {\bibfnamefont {K.}~\bibnamefont {Wang}}, \bibinfo {author} {\bibfnamefont {F.}~\bibnamefont {Jin}}, \bibinfo {author} {\bibfnamefont
  {X.}~\bibnamefont {Zhu}}, \bibinfo {author} {\bibfnamefont {B.}~\bibnamefont {Zhang}}, \bibinfo {author} {\bibfnamefont {H.}~\bibnamefont {Li}}, \bibinfo {author} {\bibfnamefont {C.}~\bibnamefont {Song}}, \bibinfo {author} {\bibfnamefont {Z.}~\bibnamefont {Wang}}, \bibinfo {author} {\bibfnamefont {F.}~\bibnamefont {Liu}}, \bibinfo {author} {\bibfnamefont {Z.}~\bibnamefont {Papi{\'c}}}, \bibinfo {author} {\bibfnamefont {L.}~\bibnamefont {Ying}}, \bibinfo {author} {\bibfnamefont {H.}~\bibnamefont {Wang}},\ and\ \bibinfo {author} {\bibfnamefont {Y.-C.}\ \bibnamefont {Lai}},\ }\href {https://www.nature.com/articles/s41567-022-01784-9} {\bibfield  {journal} {\bibinfo  {journal} {Nat. Phys.}\ }\textbf {\bibinfo {volume} {19}},\ \bibinfo {pages} {120} (\bibinfo {year} {2022})}\BibitemShut {NoStop}%
\bibitem [{\citenamefont {Karamlou}\ \emph {et~al.}(2024)\citenamefont {Karamlou}, \citenamefont {Rosen}, \citenamefont {Muschinske}, \citenamefont {Barrett}, \citenamefont {Di~Paolo}, \citenamefont {Ding}, \citenamefont {Harrington}, \citenamefont {Hays}, \citenamefont {Das}, \citenamefont {Kim}, \citenamefont {Niedzielski}, \citenamefont {Schuldt}, \citenamefont {Serniak}, \citenamefont {Schwartz}, \citenamefont {Yoder}, \citenamefont {Gustavsson}, \citenamefont {Yanay}, \citenamefont {Grover},\ and\ \citenamefont {Oliver}}]{Karamlou2023HCB}%
  \BibitemOpen
  \bibfield  {author} {\bibinfo {author} {\bibfnamefont {A.~H.}\ \bibnamefont {Karamlou}}, \bibinfo {author} {\bibfnamefont {I.~T.}\ \bibnamefont {Rosen}}, \bibinfo {author} {\bibfnamefont {S.~E.}\ \bibnamefont {Muschinske}}, \bibinfo {author} {\bibfnamefont {C.~N.}\ \bibnamefont {Barrett}}, \bibinfo {author} {\bibfnamefont {A.}~\bibnamefont {Di~Paolo}}, \bibinfo {author} {\bibfnamefont {L.}~\bibnamefont {Ding}}, \bibinfo {author} {\bibfnamefont {P.~M.}\ \bibnamefont {Harrington}}, \bibinfo {author} {\bibfnamefont {M.}~\bibnamefont {Hays}}, \bibinfo {author} {\bibfnamefont {R.}~\bibnamefont {Das}}, \bibinfo {author} {\bibfnamefont {D.~K.}\ \bibnamefont {Kim}}, \bibinfo {author} {\bibfnamefont {\textit{et al.}}},\ }\href {http://dx.doi.org/10.1038/s41586-024-07325-z} {\bibfield  {journal} {\bibinfo  {journal} {Nature}\ }\textbf {\bibinfo {volume} {629}},\ \bibinfo {pages} {561} (\bibinfo {year} {2024})}\BibitemShut {NoStop}%
\bibitem [{\citenamefont {Koll{\'a}r}\ \emph {et~al.}(2019)\citenamefont {Koll{\'a}r}, \citenamefont {Fitzpatrick},\ and\ \citenamefont {Houck}}]{Kollar2019-vn}%
  \BibitemOpen
  \bibfield  {author} {\bibinfo {author} {\bibfnamefont {A.~J.}\ \bibnamefont {Koll{\'a}r}}, \bibinfo {author} {\bibfnamefont {M.}~\bibnamefont {Fitzpatrick}},\ and\ \bibinfo {author} {\bibfnamefont {A.~A.}\ \bibnamefont {Houck}},\ }\href {http://dx.doi.org/10.1038/s41586-019-1348-3} {\bibfield  {journal} {\bibinfo  {journal} {Nature}\ }\textbf {\bibinfo {volume} {571}},\ \bibinfo {pages} {45} (\bibinfo {year} {2019})}\BibitemShut {NoStop}%
\bibitem [{\citenamefont {Martinez}\ \emph {et~al.}(2023)\citenamefont {Martinez}, \citenamefont {Chiu}, \citenamefont {Smitham},\ and\ \citenamefont {Houck}}]{Martinez2023-fh}%
  \BibitemOpen
  \bibfield  {author} {\bibinfo {author} {\bibfnamefont {J.~G.~C.}\ \bibnamefont {Martinez}}, \bibinfo {author} {\bibfnamefont {C.~S.}\ \bibnamefont {Chiu}}, \bibinfo {author} {\bibfnamefont {B.~M.}\ \bibnamefont {Smitham}},\ and\ \bibinfo {author} {\bibfnamefont {A.~A.}\ \bibnamefont {Houck}},\ }\href {http://dx.doi.org/10.1126/sciadv.adj7195} {\bibfield  {journal} {\bibinfo  {journal} {Sci Adv}\ }\textbf {\bibinfo {volume} {9}},\ \bibinfo {pages} {eadj7195} (\bibinfo {year} {2023})}\BibitemShut {NoStop}%
\bibitem [{\citenamefont {Roushan}\ \emph {et~al.}(2016)\citenamefont {Roushan}, \citenamefont {Neill}, \citenamefont {Megrant}, \citenamefont {Chen}, \citenamefont {Babbush}, \citenamefont {Barends}, \citenamefont {Campbell}, \citenamefont {Chen}, \citenamefont {Chiaro}, \citenamefont {Dunsworth}, \citenamefont {Fowler}, \citenamefont {Jeffrey}, \citenamefont {Kelly}, \citenamefont {Lucero}, \citenamefont {Mutus}, \citenamefont {O'Malley}, \citenamefont {Neeley}, \citenamefont {Quintana}, \citenamefont {Sank}, \citenamefont {Vainsencher}, \citenamefont {Wenner}, \citenamefont {White}, \citenamefont {Kapit}, \citenamefont {Neven},\ and\ \citenamefont {Martinis}}]{Roushan2016-nr}%
  \BibitemOpen
  \bibfield  {author} {\bibinfo {author} {\bibfnamefont {P.}~\bibnamefont {Roushan}}, \bibinfo {author} {\bibfnamefont {C.}~\bibnamefont {Neill}}, \bibinfo {author} {\bibfnamefont {A.}~\bibnamefont {Megrant}}, \bibinfo {author} {\bibfnamefont {Y.}~\bibnamefont {Chen}}, \bibinfo {author} {\bibfnamefont {R.}~\bibnamefont {Babbush}}, \bibinfo {author} {\bibfnamefont {R.}~\bibnamefont {Barends}}, \bibinfo {author} {\bibfnamefont {B.}~\bibnamefont {Campbell}}, \bibinfo {author} {\bibfnamefont {Z.}~\bibnamefont {Chen}}, \bibinfo {author} {\bibfnamefont {B.}~\bibnamefont {Chiaro}}, \bibinfo {author} {\bibfnamefont {A.}~\bibnamefont {Dunsworth}}, \bibinfo {author} {\bibfnamefont {A.}~\bibnamefont {Fowler}}, \bibinfo {author} {\bibfnamefont {E.}~\bibnamefont {Jeffrey}}, \bibinfo {author} {\bibfnamefont {J.}~\bibnamefont {Kelly}}, \bibinfo {author} {\bibfnamefont {E.}~\bibnamefont {Lucero}}, \bibinfo {author} {\bibfnamefont {J.}~\bibnamefont {Mutus}}, \bibinfo {author} {\bibfnamefont {P.~J.~J.}\ \bibnamefont
  {O'Malley}}, \bibinfo {author} {\bibfnamefont {M.}~\bibnamefont {Neeley}}, \bibinfo {author} {\bibfnamefont {C.}~\bibnamefont {Quintana}}, \bibinfo {author} {\bibfnamefont {D.}~\bibnamefont {Sank}}, \bibinfo {author} {\bibfnamefont {A.}~\bibnamefont {Vainsencher}}, \bibinfo {author} {\bibfnamefont {J.}~\bibnamefont {Wenner}}, \bibinfo {author} {\bibfnamefont {T.}~\bibnamefont {White}}, \bibinfo {author} {\bibfnamefont {E.}~\bibnamefont {Kapit}}, \bibinfo {author} {\bibfnamefont {H.}~\bibnamefont {Neven}},\ and\ \bibinfo {author} {\bibfnamefont {J.}~\bibnamefont {Martinis}},\ }\href {https://www.nature.com/articles/nphys3930} {\bibfield  {journal} {\bibinfo  {journal} {Nat. Phys.}\ }\textbf {\bibinfo {volume} {13}},\ \bibinfo {pages} {146} (\bibinfo {year} {2016})}\BibitemShut {NoStop}%
\bibitem [{\citenamefont {Leib}\ and\ \citenamefont {Hartmann}(2010)}]{Leib2010-ph}%
  \BibitemOpen
  \bibfield  {author} {\bibinfo {author} {\bibfnamefont {M.}~\bibnamefont {Leib}}\ and\ \bibinfo {author} {\bibfnamefont {M.~J.}\ \bibnamefont {Hartmann}},\ }\href {https://iopscience.iop.org/article/10.1088/1367-2630/12/9/093031/meta} {\bibfield  {journal} {\bibinfo  {journal} {New J. Phys.}\ }\textbf {\bibinfo {volume} {12}},\ \bibinfo {pages} {093031} (\bibinfo {year} {2010})}\BibitemShut {NoStop}%
\bibitem [{\citenamefont {Biella}\ \emph {et~al.}(2015)\citenamefont {Biella}, \citenamefont {Mazza}, \citenamefont {Carusotto}, \citenamefont {Rossini},\ and\ \citenamefont {Fazio}}]{Biella2015-xg}%
  \BibitemOpen
  \bibfield  {author} {\bibinfo {author} {\bibfnamefont {A.}~\bibnamefont {Biella}}, \bibinfo {author} {\bibfnamefont {L.}~\bibnamefont {Mazza}}, \bibinfo {author} {\bibfnamefont {I.}~\bibnamefont {Carusotto}}, \bibinfo {author} {\bibfnamefont {D.}~\bibnamefont {Rossini}},\ and\ \bibinfo {author} {\bibfnamefont {R.}~\bibnamefont {Fazio}},\ }\href {https://link.aps.org/doi/10.1103/PhysRevA.91.053815} {\bibfield  {journal} {\bibinfo  {journal} {Phys. Rev. A}\ }\textbf {\bibinfo {volume} {91}},\ \bibinfo {pages} {053815} (\bibinfo {year} {2015})}\BibitemShut {NoStop}%
\bibitem [{\citenamefont {Mertz}\ \emph {et~al.}(2016)\citenamefont {Mertz}, \citenamefont {Vasi{\'c}}, \citenamefont {Hartmann},\ and\ \citenamefont {Hofstetter}}]{Mertz2016-ad}%
  \BibitemOpen
  \bibfield  {author} {\bibinfo {author} {\bibfnamefont {T.}~\bibnamefont {Mertz}}, \bibinfo {author} {\bibfnamefont {I.}~\bibnamefont {Vasi{\'c}}}, \bibinfo {author} {\bibfnamefont {M.~J.}\ \bibnamefont {Hartmann}},\ and\ \bibinfo {author} {\bibfnamefont {W.}~\bibnamefont {Hofstetter}},\ }\href {https://link.aps.org/doi/10.1103/PhysRevA.94.013809} {\bibfield  {journal} {\bibinfo  {journal} {Phys. Rev. A}\ }\textbf {\bibinfo {volume} {94}},\ \bibinfo {pages} {013809} (\bibinfo {year} {2016})}\BibitemShut {NoStop}%
\bibitem [{\citenamefont {Bychek}\ \emph {et~al.}(2020)\citenamefont {Bychek}, \citenamefont {Muraev}, \citenamefont {Maksimov},\ and\ \citenamefont {Kolovsky}}]{Bychek2020-bl}%
  \BibitemOpen
  \bibfield  {author} {\bibinfo {author} {\bibfnamefont {A.~A.}\ \bibnamefont {Bychek}}, \bibinfo {author} {\bibfnamefont {P.~S.}\ \bibnamefont {Muraev}}, \bibinfo {author} {\bibfnamefont {D.~N.}\ \bibnamefont {Maksimov}},\ and\ \bibinfo {author} {\bibfnamefont {A.~R.}\ \bibnamefont {Kolovsky}},\ }\href {http://dx.doi.org/10.1103/PhysRevE.101.012208} {\bibfield  {journal} {\bibinfo  {journal} {Phys Rev E}\ }\textbf {\bibinfo {volume} {101}},\ \bibinfo {pages} {012208} (\bibinfo {year} {2020})}\BibitemShut {NoStop}%
\bibitem [{\citenamefont {Fitzpatrick}\ \emph {et~al.}(2017)\citenamefont {Fitzpatrick}, \citenamefont {Sundaresan}, \citenamefont {Li}, \citenamefont {Koch},\ and\ \citenamefont {Houck}}]{Fitzpatrick2017-sy}%
  \BibitemOpen
  \bibfield  {author} {\bibinfo {author} {\bibfnamefont {M.}~\bibnamefont {Fitzpatrick}}, \bibinfo {author} {\bibfnamefont {N.~M.}\ \bibnamefont {Sundaresan}}, \bibinfo {author} {\bibfnamefont {A.~C.~Y.}\ \bibnamefont {Li}}, \bibinfo {author} {\bibfnamefont {J.}~\bibnamefont {Koch}},\ and\ \bibinfo {author} {\bibfnamefont {A.~A.}\ \bibnamefont {Houck}},\ }\href {https://link.aps.org/doi/10.1103/PhysRevX.7.011016} {\bibfield  {journal} {\bibinfo  {journal} {Phys. Rev. X}\ }\textbf {\bibinfo {volume} {7}},\ \bibinfo {pages} {011016} (\bibinfo {year} {2017})}\BibitemShut {NoStop}%
\bibitem [{\citenamefont {Fedorov}\ \emph {et~al.}(2021)\citenamefont {Fedorov}, \citenamefont {Remizov}, \citenamefont {Shapiro}, \citenamefont {Pogosov}, \citenamefont {Egorova}, \citenamefont {Tsitsilin}, \citenamefont {Andronik}, \citenamefont {Dobronosova}, \citenamefont {Rodionov}, \citenamefont {Astafiev},\ and\ \citenamefont {Ustinov}}]{Fedorov2021-sr}%
  \BibitemOpen
  \bibfield  {author} {\bibinfo {author} {\bibfnamefont {G.~P.}\ \bibnamefont {Fedorov}}, \bibinfo {author} {\bibfnamefont {S.~V.}\ \bibnamefont {Remizov}}, \bibinfo {author} {\bibfnamefont {D.~S.}\ \bibnamefont {Shapiro}}, \bibinfo {author} {\bibfnamefont {W.~V.}\ \bibnamefont {Pogosov}}, \bibinfo {author} {\bibfnamefont {E.}~\bibnamefont {Egorova}}, \bibinfo {author} {\bibfnamefont {I.}~\bibnamefont {Tsitsilin}}, \bibinfo {author} {\bibfnamefont {M.}~\bibnamefont {Andronik}}, \bibinfo {author} {\bibfnamefont {A.~A.}\ \bibnamefont {Dobronosova}}, \bibinfo {author} {\bibfnamefont {I.~A.}\ \bibnamefont {Rodionov}}, \bibinfo {author} {\bibfnamefont {O.~V.}\ \bibnamefont {Astafiev}},\ and\ \bibinfo {author} {\bibfnamefont {A.~V.}\ \bibnamefont {Ustinov}},\ }\href {http://dx.doi.org/10.1103/PhysRevLett.126.180503} {\bibfield  {journal} {\bibinfo  {journal} {Phys. Rev. Lett.}\ }\textbf {\bibinfo {volume} {126}},\ \bibinfo {pages} {180503} (\bibinfo {year} {2021})}\BibitemShut {NoStop}%
\bibitem [{\citenamefont {Vrajitoarea}\ \emph {et~al.}(2024)\citenamefont {Vrajitoarea}, \citenamefont {Belyansky}, \citenamefont {Lundgren}, \citenamefont {Whitsitt}, \citenamefont {Gorshkov},\ and\ \citenamefont {Houck}}]{vrajitoarea2024ultrastrong}%
  \BibitemOpen
  \bibfield  {author} {\bibinfo {author} {\bibfnamefont {A.}~\bibnamefont {Vrajitoarea}}, \bibinfo {author} {\bibfnamefont {R.}~\bibnamefont {Belyansky}}, \bibinfo {author} {\bibfnamefont {R.}~\bibnamefont {Lundgren}}, \bibinfo {author} {\bibfnamefont {S.}~\bibnamefont {Whitsitt}}, \bibinfo {author} {\bibfnamefont {A.~V.}\ \bibnamefont {Gorshkov}},\ and\ \bibinfo {author} {\bibfnamefont {A.~A.}\ \bibnamefont {Houck}},\ }\href@noop {} {\bibinfo {title} {Ultrastrong light-matter interaction in a multimode photonic crystal}} (\bibinfo {year} {2024}),\ \Eprint {https://arxiv.org/abs/2209.14972} {arXiv:2209.14972 [quant-ph]} \BibitemShut {NoStop}%
\bibitem [{\citenamefont {Ke{\ss}ler}\ and\ \citenamefont {Marquardt}(2014)}]{Kesler2014-gv}%
  \BibitemOpen
  \bibfield  {author} {\bibinfo {author} {\bibfnamefont {S.}~\bibnamefont {Ke{\ss}ler}}\ and\ \bibinfo {author} {\bibfnamefont {F.}~\bibnamefont {Marquardt}},\ }\href {https://link.aps.org/doi/10.1103/PhysRevA.89.061601} {\bibfield  {journal} {\bibinfo  {journal} {Phys. Rev. A}\ }\textbf {\bibinfo {volume} {89}},\ \bibinfo {pages} {061601} (\bibinfo {year} {2014})}\BibitemShut {NoStop}%
\bibitem [{\citenamefont {Atala}\ \emph {et~al.}(2014)\citenamefont {Atala}, \citenamefont {Aidelsburger}, \citenamefont {Lohse}, \citenamefont {Barreiro}, \citenamefont {Paredes},\ and\ \citenamefont {Bloch}}]{Atala2014-tw}%
  \BibitemOpen
  \bibfield  {author} {\bibinfo {author} {\bibfnamefont {M.}~\bibnamefont {Atala}}, \bibinfo {author} {\bibfnamefont {M.}~\bibnamefont {Aidelsburger}}, \bibinfo {author} {\bibfnamefont {M.}~\bibnamefont {Lohse}}, \bibinfo {author} {\bibfnamefont {J.~T.}\ \bibnamefont {Barreiro}}, \bibinfo {author} {\bibfnamefont {B.}~\bibnamefont {Paredes}},\ and\ \bibinfo {author} {\bibfnamefont {I.}~\bibnamefont {Bloch}},\ }\href {https://www.nature.com/articles/nphys2998} {\bibfield  {journal} {\bibinfo  {journal} {Nat. Phys.}\ }\textbf {\bibinfo {volume} {10}},\ \bibinfo {pages} {588} (\bibinfo {year} {2014})}\BibitemShut {NoStop}%
\bibitem [{\citenamefont {Impertro}\ \emph {et~al.}(2023)\citenamefont {Impertro}, \citenamefont {Karch}, \citenamefont {Wienand}, \citenamefont {Huh}, \citenamefont {Schweizer}, \citenamefont {Bloch},\ and\ \citenamefont {Aidelsburger}}]{Impertro2023current}%
  \BibitemOpen
  \bibfield  {author} {\bibinfo {author} {\bibfnamefont {A.}~\bibnamefont {Impertro}}, \bibinfo {author} {\bibfnamefont {S.}~\bibnamefont {Karch}}, \bibinfo {author} {\bibfnamefont {J.~F.}\ \bibnamefont {Wienand}}, \bibinfo {author} {\bibfnamefont {S.}~\bibnamefont {Huh}}, \bibinfo {author} {\bibfnamefont {C.}~\bibnamefont {Schweizer}}, \bibinfo {author} {\bibfnamefont {I.}~\bibnamefont {Bloch}},\ and\ \bibinfo {author} {\bibfnamefont {M.}~\bibnamefont {Aidelsburger}},\ }\href@noop {} {\bibinfo {title} {Local readout and control of current and kinetic energy operators in optical lattices}} (\bibinfo {year} {2023}),\ \Eprint {https://arxiv.org/abs/2312.13268} {arXiv:2312.13268 [cond-mat.quant-gas]} \BibitemShut {NoStop}%
\bibitem [{Note1()}]{Note1}%
  \BibitemOpen
  \bibinfo {note} {See Supplemental Material for device characterization, measurement setup, numerical modeling, and additional analysis of the data, which includes Refs.\protect \,\cite {Rol2020-mr, Johnson2011-jg, Geerlings2013-kl, Pedregosa2011-rf, roberts2023manybody}.}\BibitemShut {Stop}%
\bibitem [{\citenamefont {Koch}\ \emph {et~al.}(2007)\citenamefont {Koch}, \citenamefont {Yu}, \citenamefont {Gambetta}, \citenamefont {Houck}, \citenamefont {Schuster}, \citenamefont {Majer}, \citenamefont {Blais}, \citenamefont {Devoret}, \citenamefont {Girvin},\ and\ \citenamefont {Schoelkopf}}]{Koch2007-cz}%
  \BibitemOpen
  \bibfield  {author} {\bibinfo {author} {\bibfnamefont {J.}~\bibnamefont {Koch}}, \bibinfo {author} {\bibfnamefont {T.~M.}\ \bibnamefont {Yu}}, \bibinfo {author} {\bibfnamefont {J.}~\bibnamefont {Gambetta}}, \bibinfo {author} {\bibfnamefont {A.~A.}\ \bibnamefont {Houck}}, \bibinfo {author} {\bibfnamefont {D.~I.}\ \bibnamefont {Schuster}}, \bibinfo {author} {\bibfnamefont {J.}~\bibnamefont {Majer}}, \bibinfo {author} {\bibfnamefont {A.}~\bibnamefont {Blais}}, \bibinfo {author} {\bibfnamefont {M.~H.}\ \bibnamefont {Devoret}}, \bibinfo {author} {\bibfnamefont {S.~M.}\ \bibnamefont {Girvin}},\ and\ \bibinfo {author} {\bibfnamefont {R.~J.}\ \bibnamefont {Schoelkopf}},\ }\href {https://link.aps.org/doi/10.1103/PhysRevA.76.042319} {\bibfield  {journal} {\bibinfo  {journal} {Phys. Rev. A}\ }\textbf {\bibinfo {volume} {76}},\ \bibinfo {pages} {042319} (\bibinfo {year} {2007})}\BibitemShut {NoStop}%
\bibitem [{\citenamefont {Bakr}\ \emph {et~al.}(2010)\citenamefont {Bakr}, \citenamefont {Peng}, \citenamefont {Tai}, \citenamefont {Ma}, \citenamefont {Simon}, \citenamefont {Gillen}, \citenamefont {F{\"o}lling}, \citenamefont {Pollet},\ and\ \citenamefont {Greiner}}]{Bakr2010-jm}%
  \BibitemOpen
  \bibfield  {author} {\bibinfo {author} {\bibfnamefont {W.~S.}\ \bibnamefont {Bakr}}, \bibinfo {author} {\bibfnamefont {A.}~\bibnamefont {Peng}}, \bibinfo {author} {\bibfnamefont {M.~E.}\ \bibnamefont {Tai}}, \bibinfo {author} {\bibfnamefont {R.}~\bibnamefont {Ma}}, \bibinfo {author} {\bibfnamefont {J.}~\bibnamefont {Simon}}, \bibinfo {author} {\bibfnamefont {J.~I.}\ \bibnamefont {Gillen}}, \bibinfo {author} {\bibfnamefont {S.}~\bibnamefont {F{\"o}lling}}, \bibinfo {author} {\bibfnamefont {L.}~\bibnamefont {Pollet}},\ and\ \bibinfo {author} {\bibfnamefont {M.}~\bibnamefont {Greiner}},\ }\href {http://dx.doi.org/10.1126/science.1192368} {\bibfield  {journal} {\bibinfo  {journal} {Science}\ }\textbf {\bibinfo {volume} {329}},\ \bibinfo {pages} {547} (\bibinfo {year} {2010})}\BibitemShut {NoStop}%
\bibitem [{\citenamefont {Umucal{\i}lar}\ and\ \citenamefont {Carusotto}(2017)}]{Umucalilar2017-ak}%
  \BibitemOpen
  \bibfield  {author} {\bibinfo {author} {\bibfnamefont {R.~O.}\ \bibnamefont {Umucal{\i}lar}}\ and\ \bibinfo {author} {\bibfnamefont {I.}~\bibnamefont {Carusotto}},\ }\href {https://link.aps.org/doi/10.1103/PhysRevA.96.053808} {\bibfield  {journal} {\bibinfo  {journal} {Phys. Rev. A}\ }\textbf {\bibinfo {volume} {96}},\ \bibinfo {pages} {053808} (\bibinfo {year} {2017})}\BibitemShut {NoStop}%
\bibitem [{\citenamefont {Hafezi}\ \emph {et~al.}(2015)\citenamefont {Hafezi}, \citenamefont {Adhikari},\ and\ \citenamefont {Taylor}}]{Hafezi2015-mw}%
  \BibitemOpen
  \bibfield  {author} {\bibinfo {author} {\bibfnamefont {M.}~\bibnamefont {Hafezi}}, \bibinfo {author} {\bibfnamefont {P.}~\bibnamefont {Adhikari}},\ and\ \bibinfo {author} {\bibfnamefont {J.~M.}\ \bibnamefont {Taylor}},\ }\href {https://link.aps.org/doi/10.1103/PhysRevB.92.174305} {\bibfield  {journal} {\bibinfo  {journal} {Phys. Rev. B Condens. Matter}\ }\textbf {\bibinfo {volume} {92}},\ \bibinfo {pages} {174305} (\bibinfo {year} {2015})}\BibitemShut {NoStop}%
\bibitem [{\citenamefont {Lebreuilly}\ and\ \citenamefont {Carusotto}(2018)}]{Lebreuilly2018-bs}%
  \BibitemOpen
  \bibfield  {author} {\bibinfo {author} {\bibfnamefont {J.}~\bibnamefont {Lebreuilly}}\ and\ \bibinfo {author} {\bibfnamefont {I.}~\bibnamefont {Carusotto}},\ }\href {https://doi.org/10.1016/j.crhy.2018.07.001} {\bibfield  {journal} {\bibinfo  {journal} {C. R. Phys.}\ }\textbf {\bibinfo {volume} {19}},\ \bibinfo {pages} {433} (\bibinfo {year} {2018})}\BibitemShut {NoStop}%
\bibitem [{\citenamefont {Ma}\ \emph {et~al.}(2017)\citenamefont {Ma}, \citenamefont {Owens}, \citenamefont {Houck}, \citenamefont {Schuster},\ and\ \citenamefont {Simon}}]{Ma2017-vw}%
  \BibitemOpen
  \bibfield  {author} {\bibinfo {author} {\bibfnamefont {R.}~\bibnamefont {Ma}}, \bibinfo {author} {\bibfnamefont {C.}~\bibnamefont {Owens}}, \bibinfo {author} {\bibfnamefont {A.}~\bibnamefont {Houck}}, \bibinfo {author} {\bibfnamefont {D.~I.}\ \bibnamefont {Schuster}},\ and\ \bibinfo {author} {\bibfnamefont {J.}~\bibnamefont {Simon}},\ }\href {https://link.aps.org/doi/10.1103/PhysRevA.95.043811} {\bibfield  {journal} {\bibinfo  {journal} {Phys. Rev. A}\ }\textbf {\bibinfo {volume} {95}},\ \bibinfo {pages} {043811} (\bibinfo {year} {2017})}\BibitemShut {NoStop}%
\bibitem [{\citenamefont {Mi}\ \emph {et~al.}(2024)\citenamefont {Mi}, \citenamefont {Michailidis}, \citenamefont {Shabani}, \citenamefont {Miao}, \citenamefont {Klimov}, \citenamefont {Lloyd}, \citenamefont {Rosenberg}, \citenamefont {Acharya}, \citenamefont {Aleiner}, \citenamefont {Andersen}, \citenamefont {Ansmann}, \citenamefont {Arute}, \citenamefont {Arya}, \citenamefont {Asfaw}, \citenamefont {Atalaya}, \citenamefont {Bardin}, \citenamefont {Bengtsson}, \citenamefont {Bortoli}, \citenamefont {Bourassa}, \citenamefont {Bovaird}, \citenamefont {Brill}, \citenamefont {Broughton}, \citenamefont {Buckley}, \citenamefont {Buell}, \citenamefont {Burger}, \citenamefont {Burkett}, \citenamefont {Bushnell}, \citenamefont {Chen}, \citenamefont {Chiaro}, \citenamefont {Chik}, \citenamefont {Chou}, \citenamefont {Cogan}, \citenamefont {Collins}, \citenamefont {Conner}, \citenamefont {Courtney}, \citenamefont {Crook}, \citenamefont {Curtin}, \citenamefont {Dau}, \citenamefont {Debroy}, \citenamefont {Del
  Toro~Barba}, \citenamefont {Demura}, \citenamefont {Di~Paolo}, \citenamefont {Drozdov}, \citenamefont {Dunsworth}, \citenamefont {Erickson}, \citenamefont {Faoro}, \citenamefont {Farhi}, \citenamefont {Fatemi}, \citenamefont {Ferreira}, \citenamefont {Burgos}, \citenamefont {Forati}, \citenamefont {Fowler}, \citenamefont {Foxen}, \citenamefont {Genois}, \citenamefont {Giang}, \citenamefont {Gidney}, \citenamefont {Gilboa}, \citenamefont {Giustina}, \citenamefont {Gosula}, \citenamefont {Gross}, \citenamefont {Habegger}, \citenamefont {Hamilton}, \citenamefont {Hansen}, \citenamefont {Harrigan}, \citenamefont {Harrington}, \citenamefont {Heu}, \citenamefont {Hoffmann}, \citenamefont {Hong}, \citenamefont {Huang}, \citenamefont {Huff}, \citenamefont {Huggins}, \citenamefont {Ioffe}, \citenamefont {Isakov}, \citenamefont {Iveland}, \citenamefont {Jeffrey}, \citenamefont {Jiang}, \citenamefont {Jones}, \citenamefont {Juhas}, \citenamefont {Kafri}, \citenamefont {Kechedzhi}, \citenamefont {Khattar},
  \citenamefont {Khezri}, \citenamefont {Kieferov{\'a}}, \citenamefont {Kim}, \citenamefont {Kitaev}, \citenamefont {Klots}, \citenamefont {Korotkov}, \citenamefont {Kostritsa}, \citenamefont {Kreikebaum}, \citenamefont {Landhuis}, \citenamefont {Laptev}, \citenamefont {Lau}, \citenamefont {Laws}, \citenamefont {Lee}, \citenamefont {Lee}, \citenamefont {Lensky}, \citenamefont {Lester}, \citenamefont {Lill}, \citenamefont {Liu}, \citenamefont {Locharla}, \citenamefont {Malone}, \citenamefont {Martin}, \citenamefont {McClean}, \citenamefont {McEwen}, \citenamefont {Mieszala}, \citenamefont {Montazeri}, \citenamefont {Morvan}, \citenamefont {Movassagh}, \citenamefont {Mruczkiewicz}, \citenamefont {Neeley}, \citenamefont {Neill}, \citenamefont {Nersisyan}, \citenamefont {Newman}, \citenamefont {Ng}, \citenamefont {Nguyen}, \citenamefont {Nguyen}, \citenamefont {Niu}, \citenamefont {O'Brien}, \citenamefont {Opremcak}, \citenamefont {Petukhov}, \citenamefont {Potter}, \citenamefont {Pryadko}, \citenamefont
  {Quintana}, \citenamefont {Rocque}, \citenamefont {Rubin}, \citenamefont {Saei}, \citenamefont {Sank}, \citenamefont {Sankaragomathi}, \citenamefont {Satzinger}, \citenamefont {Schurkus}, \citenamefont {Schuster}, \citenamefont {Shearn}, \citenamefont {Shorter}, \citenamefont {Shutty}, \citenamefont {Shvarts}, \citenamefont {Skruzny}, \citenamefont {Smith}, \citenamefont {Somma}, \citenamefont {Sterling}, \citenamefont {Strain}, \citenamefont {Szalay}, \citenamefont {Torres}, \citenamefont {Vidal}, \citenamefont {Villalonga}, \citenamefont {Heidweiller}, \citenamefont {White}, \citenamefont {Woo}, \citenamefont {Xing}, \citenamefont {Yao}, \citenamefont {Yeh}, \citenamefont {Yoo}, \citenamefont {Young}, \citenamefont {Zalcman}, \citenamefont {Zhang}, \citenamefont {Zhu}, \citenamefont {Zobrist}, \citenamefont {Neven}, \citenamefont {Babbush}, \citenamefont {Bacon}, \citenamefont {Boixo}, \citenamefont {Hilton}, \citenamefont {Lucero}, \citenamefont {Megrant}, \citenamefont {Kelly}, \citenamefont {Chen},
  \citenamefont {Roushan}, \citenamefont {Smelyanskiy},\ and\ \citenamefont {Abanin}}]{Mi2024-in}%
  \BibitemOpen
  \bibfield  {author} {\bibinfo {author} {\bibfnamefont {X.}~\bibnamefont {Mi}}, \bibinfo {author} {\bibfnamefont {A.~A.}\ \bibnamefont {Michailidis}}, \bibinfo {author} {\bibfnamefont {S.}~\bibnamefont {Shabani}}, \bibinfo {author} {\bibfnamefont {K.~C.}\ \bibnamefont {Miao}}, \bibinfo {author} {\bibfnamefont {P.~V.}\ \bibnamefont {Klimov}}, \bibinfo {author} {\bibfnamefont {J.}~\bibnamefont {Lloyd}}, \bibinfo {author} {\bibfnamefont {E.}~\bibnamefont {Rosenberg}}, \bibinfo {author} {\bibfnamefont {R.}~\bibnamefont {Acharya}}, \bibinfo {author} {\bibfnamefont {I.}~\bibnamefont {Aleiner}}, \bibinfo {author} {\bibfnamefont {T.~I.}\ \bibnamefont {Andersen}}, \bibinfo {author} {\bibfnamefont {\textit{et al.}}},\ }\href {http://dx.doi.org/10.1126/science.adh9932} {\bibfield  {journal} {\bibinfo  {journal} {Science}\ }\textbf {\bibinfo {volume} {383}},\ \bibinfo {pages} {1332} (\bibinfo {year} {2024})}\BibitemShut {NoStop}%
\bibitem [{\citenamefont {de~L{\'e}s{\'e}leuc}\ \emph {et~al.}(2019)\citenamefont {de~L{\'e}s{\'e}leuc}, \citenamefont {Lienhard}, \citenamefont {Scholl}, \citenamefont {Barredo}, \citenamefont {Weber}, \citenamefont {Lang}, \citenamefont {B{\"u}chler}, \citenamefont {Lahaye},\ and\ \citenamefont {Browaeys}}]{De_Leseleuc2019-ux}%
  \BibitemOpen
  \bibfield  {author} {\bibinfo {author} {\bibfnamefont {S.}~\bibnamefont {de~L{\'e}s{\'e}leuc}}, \bibinfo {author} {\bibfnamefont {V.}~\bibnamefont {Lienhard}}, \bibinfo {author} {\bibfnamefont {P.}~\bibnamefont {Scholl}}, \bibinfo {author} {\bibfnamefont {D.}~\bibnamefont {Barredo}}, \bibinfo {author} {\bibfnamefont {S.}~\bibnamefont {Weber}}, \bibinfo {author} {\bibfnamefont {N.}~\bibnamefont {Lang}}, \bibinfo {author} {\bibfnamefont {H.~P.}\ \bibnamefont {B{\"u}chler}}, \bibinfo {author} {\bibfnamefont {T.}~\bibnamefont {Lahaye}},\ and\ \bibinfo {author} {\bibfnamefont {A.}~\bibnamefont {Browaeys}},\ }\href {http://dx.doi.org/10.1126/science.aav9105} {\bibfield  {journal} {\bibinfo  {journal} {Science}\ }\textbf {\bibinfo {volume} {365}},\ \bibinfo {pages} {775} (\bibinfo {year} {2019})}\BibitemShut {NoStop}%
\bibitem [{\citenamefont {Landi}\ \emph {et~al.}(2022)\citenamefont {Landi}, \citenamefont {Poletti},\ and\ \citenamefont {Schaller}}]{Landi2022-ga}%
  \BibitemOpen
  \bibfield  {author} {\bibinfo {author} {\bibfnamefont {G.~T.}\ \bibnamefont {Landi}}, \bibinfo {author} {\bibfnamefont {D.}~\bibnamefont {Poletti}},\ and\ \bibinfo {author} {\bibfnamefont {G.}~\bibnamefont {Schaller}},\ }\href {https://link.aps.org/doi/10.1103/RevModPhys.94.045006} {\bibfield  {journal} {\bibinfo  {journal} {Rev. Mod. Phys.}\ }\textbf {\bibinfo {volume} {94}},\ \bibinfo {pages} {045006} (\bibinfo {year} {2022})}\BibitemShut {NoStop}%
\bibitem [{\citenamefont {Bertini}\ \emph {et~al.}(2021)\citenamefont {Bertini}, \citenamefont {Heidrich-Meisner}, \citenamefont {Karrasch}, \citenamefont {Prosen}, \citenamefont {Steinigeweg},\ and\ \citenamefont {{\v Z}nidari{\v c}}}]{Bertini2021-lo}%
  \BibitemOpen
  \bibfield  {author} {\bibinfo {author} {\bibfnamefont {B.}~\bibnamefont {Bertini}}, \bibinfo {author} {\bibfnamefont {F.}~\bibnamefont {Heidrich-Meisner}}, \bibinfo {author} {\bibfnamefont {C.}~\bibnamefont {Karrasch}}, \bibinfo {author} {\bibfnamefont {T.}~\bibnamefont {Prosen}}, \bibinfo {author} {\bibfnamefont {R.}~\bibnamefont {Steinigeweg}},\ and\ \bibinfo {author} {\bibfnamefont {M.}~\bibnamefont {{\v Z}nidari{\v c}}},\ }\href {https://link.aps.org/doi/10.1103/RevModPhys.93.025003} {\bibfield  {journal} {\bibinfo  {journal} {Rev. Mod. Phys.}\ }\textbf {\bibinfo {volume} {93}},\ \bibinfo {pages} {025003} (\bibinfo {year} {2021})}\BibitemShut {NoStop}%
\bibitem [{\citenamefont {Beaudoin}\ \emph {et~al.}(2012)\citenamefont {Beaudoin}, \citenamefont {da~Silva}, \citenamefont {Dutton},\ and\ \citenamefont {Blais}}]{Beaudoin2012-xh}%
  \BibitemOpen
  \bibfield  {author} {\bibinfo {author} {\bibfnamefont {F.}~\bibnamefont {Beaudoin}}, \bibinfo {author} {\bibfnamefont {M.~P.}\ \bibnamefont {da~Silva}}, \bibinfo {author} {\bibfnamefont {Z.}~\bibnamefont {Dutton}},\ and\ \bibinfo {author} {\bibfnamefont {A.}~\bibnamefont {Blais}},\ }\href {https://link.aps.org/doi/10.1103/PhysRevA.86.022305} {\bibfield  {journal} {\bibinfo  {journal} {Phys. Rev. A}\ }\textbf {\bibinfo {volume} {86}},\ \bibinfo {pages} {022305} (\bibinfo {year} {2012})}\BibitemShut {NoStop}%
\bibitem [{\citenamefont {Strand}\ \emph {et~al.}(2013)\citenamefont {Strand}, \citenamefont {Ware}, \citenamefont {Beaudoin}, \citenamefont {Ohki}, \citenamefont {Johnson}, \citenamefont {Blais},\ and\ \citenamefont {Plourde}}]{Strand2013-wt}%
  \BibitemOpen
  \bibfield  {author} {\bibinfo {author} {\bibfnamefont {J.~D.}\ \bibnamefont {Strand}}, \bibinfo {author} {\bibfnamefont {M.}~\bibnamefont {Ware}}, \bibinfo {author} {\bibfnamefont {F.}~\bibnamefont {Beaudoin}}, \bibinfo {author} {\bibfnamefont {T.~A.}\ \bibnamefont {Ohki}}, \bibinfo {author} {\bibfnamefont {B.~R.}\ \bibnamefont {Johnson}}, \bibinfo {author} {\bibfnamefont {A.}~\bibnamefont {Blais}},\ and\ \bibinfo {author} {\bibfnamefont {B.~L.~T.}\ \bibnamefont {Plourde}},\ }\href {https://link.aps.org/doi/10.1103/PhysRevB.87.220505} {\bibfield  {journal} {\bibinfo  {journal} {Phys. Rev. B Condens. Matter}\ }\textbf {\bibinfo {volume} {87}},\ \bibinfo {pages} {220505} (\bibinfo {year} {2013})}\BibitemShut {NoStop}%
\bibitem [{\citenamefont {Lu}\ \emph {et~al.}(2017)\citenamefont {Lu}, \citenamefont {Chakram}, \citenamefont {Leung}, \citenamefont {Earnest}, \citenamefont {Naik}, \citenamefont {Huang}, \citenamefont {Groszkowski}, \citenamefont {Kapit}, \citenamefont {Koch},\ and\ \citenamefont {Schuster}}]{Lu2017-uq}%
  \BibitemOpen
  \bibfield  {author} {\bibinfo {author} {\bibfnamefont {Y.}~\bibnamefont {Lu}}, \bibinfo {author} {\bibfnamefont {S.}~\bibnamefont {Chakram}}, \bibinfo {author} {\bibfnamefont {N.}~\bibnamefont {Leung}}, \bibinfo {author} {\bibfnamefont {N.}~\bibnamefont {Earnest}}, \bibinfo {author} {\bibfnamefont {R.~K.}\ \bibnamefont {Naik}}, \bibinfo {author} {\bibfnamefont {Z.}~\bibnamefont {Huang}}, \bibinfo {author} {\bibfnamefont {P.}~\bibnamefont {Groszkowski}}, \bibinfo {author} {\bibfnamefont {E.}~\bibnamefont {Kapit}}, \bibinfo {author} {\bibfnamefont {J.}~\bibnamefont {Koch}},\ and\ \bibinfo {author} {\bibfnamefont {D.~I.}\ \bibnamefont {Schuster}},\ }\href {http://dx.doi.org/10.1103/PhysRevLett.119.150502} {\bibfield  {journal} {\bibinfo  {journal} {Phys. Rev. Lett.}\ }\textbf {\bibinfo {volume} {119}},\ \bibinfo {pages} {150502} (\bibinfo {year} {2017})}\BibitemShut {NoStop}%
\bibitem [{\citenamefont {Macklin}\ \emph {et~al.}(2015)\citenamefont {Macklin}, \citenamefont {O'Brien}, \citenamefont {Hover}, \citenamefont {Schwartz}, \citenamefont {Bolkhovsky}, \citenamefont {Zhang}, \citenamefont {Oliver},\ and\ \citenamefont {Siddiqi}}]{Macklin2015-pj}%
  \BibitemOpen
  \bibfield  {author} {\bibinfo {author} {\bibfnamefont {C.}~\bibnamefont {Macklin}}, \bibinfo {author} {\bibfnamefont {K.}~\bibnamefont {O'Brien}}, \bibinfo {author} {\bibfnamefont {D.}~\bibnamefont {Hover}}, \bibinfo {author} {\bibfnamefont {M.~E.}\ \bibnamefont {Schwartz}}, \bibinfo {author} {\bibfnamefont {V.}~\bibnamefont {Bolkhovsky}}, \bibinfo {author} {\bibfnamefont {X.}~\bibnamefont {Zhang}}, \bibinfo {author} {\bibfnamefont {W.~D.}\ \bibnamefont {Oliver}},\ and\ \bibinfo {author} {\bibfnamefont {I.}~\bibnamefont {Siddiqi}},\ }\href {http://dx.doi.org/10.1126/science.aaa8525} {\bibfield  {journal} {\bibinfo  {journal} {Science}\ }\textbf {\bibinfo {volume} {350}},\ \bibinfo {pages} {307} (\bibinfo {year} {2015})}\BibitemShut {NoStop}%
\bibitem [{\citenamefont {Naik}\ \emph {et~al.}(2017)\citenamefont {Naik}, \citenamefont {Leung}, \citenamefont {Chakram}, \citenamefont {Groszkowski}, \citenamefont {Lu}, \citenamefont {Earnest}, \citenamefont {McKay}, \citenamefont {Koch},\ and\ \citenamefont {Schuster}}]{Naik2017-pi}%
  \BibitemOpen
  \bibfield  {author} {\bibinfo {author} {\bibfnamefont {R.~K.}\ \bibnamefont {Naik}}, \bibinfo {author} {\bibfnamefont {N.}~\bibnamefont {Leung}}, \bibinfo {author} {\bibfnamefont {S.}~\bibnamefont {Chakram}}, \bibinfo {author} {\bibfnamefont {P.}~\bibnamefont {Groszkowski}}, \bibinfo {author} {\bibfnamefont {Y.}~\bibnamefont {Lu}}, \bibinfo {author} {\bibfnamefont {N.}~\bibnamefont {Earnest}}, \bibinfo {author} {\bibfnamefont {D.~C.}\ \bibnamefont {McKay}}, \bibinfo {author} {\bibfnamefont {J.}~\bibnamefont {Koch}},\ and\ \bibinfo {author} {\bibfnamefont {D.~I.}\ \bibnamefont {Schuster}},\ }\href {http://dx.doi.org/10.1038/s41467-017-02046-6} {\bibfield  {journal} {\bibinfo  {journal} {Nat. Commun.}\ }\textbf {\bibinfo {volume} {8}},\ \bibinfo {pages} {1904} (\bibinfo {year} {2017})}\BibitemShut {NoStop}%
\bibitem [{\citenamefont {Reagor}\ \emph {et~al.}(2018)\citenamefont {Reagor}, \citenamefont {Osborn}, \citenamefont {Tezak}, \citenamefont {Staley}, \citenamefont {Prawiroatmodjo}, \citenamefont {Scheer}, \citenamefont {Alidoust}, \citenamefont {Sete}, \citenamefont {Didier}, \citenamefont {da~Silva}, \citenamefont {Acala}, \citenamefont {Angeles}, \citenamefont {Bestwick}, \citenamefont {Block}, \citenamefont {Bloom}, \citenamefont {Bradley}, \citenamefont {Bui}, \citenamefont {Caldwell}, \citenamefont {Capelluto}, \citenamefont {Chilcott}, \citenamefont {Cordova}, \citenamefont {Crossman}, \citenamefont {Curtis}, \citenamefont {Deshpande}, \citenamefont {El~Bouayadi}, \citenamefont {Girshovich}, \citenamefont {Hong}, \citenamefont {Hudson}, \citenamefont {Karalekas}, \citenamefont {Kuang}, \citenamefont {Lenihan}, \citenamefont {Manenti}, \citenamefont {Manning}, \citenamefont {Marshall}, \citenamefont {Mohan}, \citenamefont {O'Brien}, \citenamefont {Otterbach}, \citenamefont {Papageorge}, \citenamefont
  {Paquette}, \citenamefont {Pelstring}, \citenamefont {Polloreno}, \citenamefont {Rawat}, \citenamefont {Ryan}, \citenamefont {Renzas}, \citenamefont {Rubin}, \citenamefont {Russel}, \citenamefont {Rust}, \citenamefont {Scarabelli}, \citenamefont {Selvanayagam}, \citenamefont {Sinclair}, \citenamefont {Smith}, \citenamefont {Suska}, \citenamefont {To}, \citenamefont {Vahidpour}, \citenamefont {Vodrahalli}, \citenamefont {Whyland}, \citenamefont {Yadav}, \citenamefont {Zeng},\ and\ \citenamefont {Rigetti}}]{Reagor2018-we}%
  \BibitemOpen
  \bibfield  {author} {\bibinfo {author} {\bibfnamefont {M.}~\bibnamefont {Reagor}}, \bibinfo {author} {\bibfnamefont {C.~B.}\ \bibnamefont {Osborn}}, \bibinfo {author} {\bibfnamefont {N.}~\bibnamefont {Tezak}}, \bibinfo {author} {\bibfnamefont {A.}~\bibnamefont {Staley}}, \bibinfo {author} {\bibfnamefont {G.}~\bibnamefont {Prawiroatmodjo}}, \bibinfo {author} {\bibfnamefont {M.}~\bibnamefont {Scheer}}, \bibinfo {author} {\bibfnamefont {N.}~\bibnamefont {Alidoust}}, \bibinfo {author} {\bibfnamefont {E.~A.}\ \bibnamefont {Sete}}, \bibinfo {author} {\bibfnamefont {N.}~\bibnamefont {Didier}}, \bibinfo {author} {\bibfnamefont {M.~P.}\ \bibnamefont {da~Silva}}, \bibinfo {author} {\bibfnamefont {\textit{et al.}}},\ }\href {http://dx.doi.org/10.1126/sciadv.aao3603} {\bibfield  {journal} {\bibinfo  {journal} {Sci Adv}\ }\textbf {\bibinfo {volume} {4}},\ \bibinfo {pages} {eaao3603} (\bibinfo {year} {2018})}\BibitemShut {NoStop}%
\bibitem [{\citenamefont {Kordas}\ \emph {et~al.}(2015)\citenamefont {Kordas}, \citenamefont {Witthaut}, \citenamefont {Buonsante}, \citenamefont {Vezzani}, \citenamefont {Burioni}, \citenamefont {Karanikas},\ and\ \citenamefont {Wimberger}}]{Kordas2015-ec}%
  \BibitemOpen
  \bibfield  {author} {\bibinfo {author} {\bibfnamefont {G.}~\bibnamefont {Kordas}}, \bibinfo {author} {\bibfnamefont {D.}~\bibnamefont {Witthaut}}, \bibinfo {author} {\bibfnamefont {P.}~\bibnamefont {Buonsante}}, \bibinfo {author} {\bibfnamefont {A.}~\bibnamefont {Vezzani}}, \bibinfo {author} {\bibfnamefont {R.}~\bibnamefont {Burioni}}, \bibinfo {author} {\bibfnamefont {A.~I.}\ \bibnamefont {Karanikas}},\ and\ \bibinfo {author} {\bibfnamefont {S.}~\bibnamefont {Wimberger}},\ }\href {http://dx.doi.org/10.1140/epjst/e2015-02528-2} {\bibfield  {journal} {\bibinfo  {journal} {Eur. Phys. J. Spec. Top.}\ }\textbf {\bibinfo {volume} {224}},\ \bibinfo {pages} {2127} (\bibinfo {year} {2015})}\BibitemShut {NoStop}%
\bibitem [{\citenamefont {Sieberer}\ \emph {et~al.}(2016)\citenamefont {Sieberer}, \citenamefont {Buchhold},\ and\ \citenamefont {Diehl}}]{Sieberer2016-rh}%
  \BibitemOpen
  \bibfield  {author} {\bibinfo {author} {\bibfnamefont {L.~M.}\ \bibnamefont {Sieberer}}, \bibinfo {author} {\bibfnamefont {M.}~\bibnamefont {Buchhold}},\ and\ \bibinfo {author} {\bibfnamefont {S.}~\bibnamefont {Diehl}},\ }\href {http://dx.doi.org/10.1088/0034-4885/79/9/096001} {\bibfield  {journal} {\bibinfo  {journal} {Rep. Prog. Phys.}\ }\textbf {\bibinfo {volume} {79}},\ \bibinfo {pages} {096001} (\bibinfo {year} {2016})}\BibitemShut {NoStop}%
\bibitem [{\citenamefont {Dutta}\ and\ \citenamefont {Cooper}(2021)}]{Dutta2021-pr}%
  \BibitemOpen
  \bibfield  {author} {\bibinfo {author} {\bibfnamefont {S.}~\bibnamefont {Dutta}}\ and\ \bibinfo {author} {\bibfnamefont {N.~R.}\ \bibnamefont {Cooper}},\ }\href {https://link.aps.org/doi/10.1103/PhysRevResearch.3.L012016} {\bibfield  {journal} {\bibinfo  {journal} {Phys. Rev. Res.}\ }\textbf {\bibinfo {volume} {3}},\ \bibinfo {pages} {L012016} (\bibinfo {year} {2021})}\BibitemShut {NoStop}%
\bibitem [{\citenamefont {Binder}\ \emph {et~al.}(2019)\citenamefont {Binder}, \citenamefont {Correa}, \citenamefont {Gogolin}, \citenamefont {Anders},\ and\ \citenamefont {Adesso}}]{Binder2019-qo}%
  \BibitemOpen
  \bibfield  {author} {\bibinfo {author} {\bibfnamefont {F.}~\bibnamefont {Binder}}, \bibinfo {author} {\bibfnamefont {L.~A.}\ \bibnamefont {Correa}}, \bibinfo {author} {\bibfnamefont {C.}~\bibnamefont {Gogolin}}, \bibinfo {author} {\bibfnamefont {J.}~\bibnamefont {Anders}},\ and\ \bibinfo {author} {\bibfnamefont {G.}~\bibnamefont {Adesso}},\ }\href {https://play.google.com/store/books/details?id=5uWPDwAAQBAJ} {\emph {\bibinfo {title} {Thermodynamics in the Quantum Regime: Fundamental Aspects and New Directions}}}\ (\bibinfo  {publisher} {Springer},\ \bibinfo {year} {2019})\BibitemShut {NoStop}%
\bibitem [{\citenamefont {Deffner}\ and\ \citenamefont {Campbell}(2019)}]{Deffner2019-ps}%
  \BibitemOpen
  \bibfield  {author} {\bibinfo {author} {\bibfnamefont {S.}~\bibnamefont {Deffner}}\ and\ \bibinfo {author} {\bibfnamefont {S.}~\bibnamefont {Campbell}},\ }\href {https://play.google.com/store/books/details?id=YdAuxwEACAAJ} {\emph {\bibinfo {title} {Quantum Thermodynamics: An Introduction to the Thermodynamics of Quantum Information}}}\ (\bibinfo  {publisher} {Morgan \& Claypool Publishers},\ \bibinfo {year} {2019})\BibitemShut {NoStop}%
\bibitem [{\citenamefont {Lenar{\v c}i{\v c}}\ \emph {et~al.}(2020)\citenamefont {Lenar{\v c}i{\v c}}, \citenamefont {Alberton}, \citenamefont {Rosch},\ and\ \citenamefont {Altman}}]{Lenarcic2020-hh}%
  \BibitemOpen
  \bibfield  {author} {\bibinfo {author} {\bibfnamefont {Z.}~\bibnamefont {Lenar{\v c}i{\v c}}}, \bibinfo {author} {\bibfnamefont {O.}~\bibnamefont {Alberton}}, \bibinfo {author} {\bibfnamefont {A.}~\bibnamefont {Rosch}},\ and\ \bibinfo {author} {\bibfnamefont {E.}~\bibnamefont {Altman}},\ }\href {http://dx.doi.org/10.1103/PhysRevLett.125.116601} {\bibfield  {journal} {\bibinfo  {journal} {Phys. Rev. Lett.}\ }\textbf {\bibinfo {volume} {125}},\ \bibinfo {pages} {116601} (\bibinfo {year} {2020})}\BibitemShut {NoStop}%
\bibitem [{\citenamefont {Klich}\ and\ \citenamefont {Levitov}(2009)}]{Klich2009-id}%
  \BibitemOpen
  \bibfield  {author} {\bibinfo {author} {\bibfnamefont {I.}~\bibnamefont {Klich}}\ and\ \bibinfo {author} {\bibfnamefont {L.}~\bibnamefont {Levitov}},\ }\href {http://dx.doi.org/10.1103/PhysRevLett.102.100502} {\bibfield  {journal} {\bibinfo  {journal} {Phys. Rev. Lett.}\ }\textbf {\bibinfo {volume} {102}},\ \bibinfo {pages} {100502} (\bibinfo {year} {2009})}\BibitemShut {NoStop}%
\bibitem [{\citenamefont {Anderson}\ \emph {et~al.}(2016)\citenamefont {Anderson}, \citenamefont {Ma}, \citenamefont {Owens}, \citenamefont {Schuster},\ and\ \citenamefont {Simon}}]{Anderson2016-sp}%
  \BibitemOpen
  \bibfield  {author} {\bibinfo {author} {\bibfnamefont {B.~M.}\ \bibnamefont {Anderson}}, \bibinfo {author} {\bibfnamefont {R.}~\bibnamefont {Ma}}, \bibinfo {author} {\bibfnamefont {C.}~\bibnamefont {Owens}}, \bibinfo {author} {\bibfnamefont {D.~I.}\ \bibnamefont {Schuster}},\ and\ \bibinfo {author} {\bibfnamefont {J.}~\bibnamefont {Simon}},\ }\href {https://link.aps.org/doi/10.1103/PhysRevX.6.041043} {\bibfield  {journal} {\bibinfo  {journal} {Phys. Rev. X}\ }\textbf {\bibinfo {volume} {6}},\ \bibinfo {pages} {041043} (\bibinfo {year} {2016})}\BibitemShut {NoStop}%
\bibitem [{\citenamefont {Owens}\ \emph {et~al.}(2022)\citenamefont {Owens}, \citenamefont {Panetta}, \citenamefont {Saxberg}, \citenamefont {Roberts}, \citenamefont {Chakram}, \citenamefont {Ma}, \citenamefont {Vrajitoarea}, \citenamefont {Simon},\ and\ \citenamefont {Schuster}}]{Owens2022-ua}%
  \BibitemOpen
  \bibfield  {author} {\bibinfo {author} {\bibfnamefont {J.~C.}\ \bibnamefont {Owens}}, \bibinfo {author} {\bibfnamefont {M.~G.}\ \bibnamefont {Panetta}}, \bibinfo {author} {\bibfnamefont {B.}~\bibnamefont {Saxberg}}, \bibinfo {author} {\bibfnamefont {G.}~\bibnamefont {Roberts}}, \bibinfo {author} {\bibfnamefont {S.}~\bibnamefont {Chakram}}, \bibinfo {author} {\bibfnamefont {R.}~\bibnamefont {Ma}}, \bibinfo {author} {\bibfnamefont {A.}~\bibnamefont {Vrajitoarea}}, \bibinfo {author} {\bibfnamefont {J.}~\bibnamefont {Simon}},\ and\ \bibinfo {author} {\bibfnamefont {D.~I.}\ \bibnamefont {Schuster}},\ }\href {https://www.nature.com/articles/s41567-022-01671-3} {\bibfield  {journal} {\bibinfo  {journal} {Nat. Phys.}\ }\textbf {\bibinfo {volume} {18}},\ \bibinfo {pages} {1048} (\bibinfo {year} {2022})}\BibitemShut {NoStop}%
\bibitem [{\citenamefont {Rol}\ \emph {et~al.}(2020)\citenamefont {Rol}, \citenamefont {Ciorciaro}, \citenamefont {Malinowski}, \citenamefont {Tarasinski}, \citenamefont {Sagastizabal}, \citenamefont {Bultink}, \citenamefont {Salathe}, \citenamefont {Haandbaek}, \citenamefont {Sedivy},\ and\ \citenamefont {DiCarlo}}]{Rol2020-mr}%
  \BibitemOpen
  \bibfield  {author} {\bibinfo {author} {\bibfnamefont {M.~A.}\ \bibnamefont {Rol}}, \bibinfo {author} {\bibfnamefont {L.}~\bibnamefont {Ciorciaro}}, \bibinfo {author} {\bibfnamefont {F.~K.}\ \bibnamefont {Malinowski}}, \bibinfo {author} {\bibfnamefont {B.~M.}\ \bibnamefont {Tarasinski}}, \bibinfo {author} {\bibfnamefont {R.~E.}\ \bibnamefont {Sagastizabal}}, \bibinfo {author} {\bibfnamefont {C.~C.}\ \bibnamefont {Bultink}}, \bibinfo {author} {\bibfnamefont {Y.}~\bibnamefont {Salathe}}, \bibinfo {author} {\bibfnamefont {N.}~\bibnamefont {Haandbaek}}, \bibinfo {author} {\bibfnamefont {J.}~\bibnamefont {Sedivy}},\ and\ \bibinfo {author} {\bibfnamefont {L.}~\bibnamefont {DiCarlo}},\ }\href {http://dx.doi.org/10.1063/1.5133894} {\bibfield  {journal} {\bibinfo  {journal} {Appl. Phys. Lett.}\ }\textbf {\bibinfo {volume} {116}},\ \bibinfo {pages} {054001} (\bibinfo {year} {2020})}\BibitemShut {NoStop}%
\bibitem [{\citenamefont {Johnson}(2011)}]{Johnson2011-jg}%
  \BibitemOpen
  \bibfield  {author} {\bibinfo {author} {\bibfnamefont {B.~R.}\ \bibnamefont {Johnson}},\ }\emph {\bibinfo {title} {Controlling Photons in Superconducting Electrical Circuits}},\ \href {https://www.proquest.com/dissertations-theses/controlling-photons-superconducting-electrical/docview/884221402/se-2} {Ph.D. thesis},\ \bibinfo  {school} {Yale University}, \bibinfo {address} {Ann Arbor, United States} (\bibinfo {year} {2011})\BibitemShut {NoStop}%
\bibitem [{\citenamefont {Geerlings}\ \emph {et~al.}(2013)\citenamefont {Geerlings}, \citenamefont {Leghtas}, \citenamefont {Pop}, \citenamefont {Shankar}, \citenamefont {Frunzio}, \citenamefont {Schoelkopf}, \citenamefont {Mirrahimi},\ and\ \citenamefont {Devoret}}]{Geerlings2013-kl}%
  \BibitemOpen
  \bibfield  {author} {\bibinfo {author} {\bibfnamefont {K.}~\bibnamefont {Geerlings}}, \bibinfo {author} {\bibfnamefont {Z.}~\bibnamefont {Leghtas}}, \bibinfo {author} {\bibfnamefont {I.~M.}\ \bibnamefont {Pop}}, \bibinfo {author} {\bibfnamefont {S.}~\bibnamefont {Shankar}}, \bibinfo {author} {\bibfnamefont {L.}~\bibnamefont {Frunzio}}, \bibinfo {author} {\bibfnamefont {R.~J.}\ \bibnamefont {Schoelkopf}}, \bibinfo {author} {\bibfnamefont {M.}~\bibnamefont {Mirrahimi}},\ and\ \bibinfo {author} {\bibfnamefont {M.~H.}\ \bibnamefont {Devoret}},\ }\href {http://dx.doi.org/10.1103/PhysRevLett.110.120501} {\bibfield  {journal} {\bibinfo  {journal} {Phys. Rev. Lett.}\ }\textbf {\bibinfo {volume} {110}},\ \bibinfo {pages} {120501} (\bibinfo {year} {2013})}\BibitemShut {NoStop}%
\bibitem [{\citenamefont {Pedregosa}\ \emph {et~al.}(2011)\citenamefont {Pedregosa}, \citenamefont {Varoquaux}, \citenamefont {Gramfort}, \citenamefont {Michel}, \citenamefont {Thirion}, \citenamefont {Grisel}, \citenamefont {Blondel}, \citenamefont {Prettenhofer}, \citenamefont {Weiss}, \citenamefont {Dubourg}, \citenamefont {Vanderplas}, \citenamefont {Passos}, \citenamefont {Cournapeau}, \citenamefont {Brucher}, \citenamefont {Perrot},\ and\ \citenamefont {Duchesnay}}]{Pedregosa2011-rf}%
  \BibitemOpen
  \bibfield  {author} {\bibinfo {author} {\bibfnamefont {F.}~\bibnamefont {Pedregosa}}, \bibinfo {author} {\bibfnamefont {G.}~\bibnamefont {Varoquaux}}, \bibinfo {author} {\bibfnamefont {A.}~\bibnamefont {Gramfort}}, \bibinfo {author} {\bibfnamefont {V.}~\bibnamefont {Michel}}, \bibinfo {author} {\bibfnamefont {B.}~\bibnamefont {Thirion}}, \bibinfo {author} {\bibfnamefont {O.}~\bibnamefont {Grisel}}, \bibinfo {author} {\bibfnamefont {M.}~\bibnamefont {Blondel}}, \bibinfo {author} {\bibfnamefont {P.}~\bibnamefont {Prettenhofer}}, \bibinfo {author} {\bibfnamefont {R.}~\bibnamefont {Weiss}}, \bibinfo {author} {\bibfnamefont {V.}~\bibnamefont {Dubourg}}, \bibinfo {author} {\bibfnamefont {J.}~\bibnamefont {Vanderplas}}, \bibinfo {author} {\bibfnamefont {A.}~\bibnamefont {Passos}}, \bibinfo {author} {\bibfnamefont {D.}~\bibnamefont {Cournapeau}}, \bibinfo {author} {\bibfnamefont {M.}~\bibnamefont {Brucher}}, \bibinfo {author} {\bibfnamefont {M.}~\bibnamefont {Perrot}},\ and\ \bibinfo {author} {\bibfnamefont
  {E.}~\bibnamefont {Duchesnay}},\ }\href {http://jmlr.org/papers/v12/pedregosa11a.html} {\bibfield  {journal} {\bibinfo  {journal} {J. Mach. Learn. Res.}\ }\textbf {\bibinfo {volume} {12}},\ \bibinfo {pages} {2825} (\bibinfo {year} {2011})}\BibitemShut {NoStop}%
\bibitem [{\citenamefont {Roberts}\ \emph {et~al.}(2023)\citenamefont {Roberts}, \citenamefont {Vrajitoarea}, \citenamefont {Saxberg}, \citenamefont {Panetta}, \citenamefont {Simon},\ and\ \citenamefont {Schuster}}]{roberts2023manybody}%
  \BibitemOpen
  \bibfield  {author} {\bibinfo {author} {\bibfnamefont {G.}~\bibnamefont {Roberts}}, \bibinfo {author} {\bibfnamefont {A.}~\bibnamefont {Vrajitoarea}}, \bibinfo {author} {\bibfnamefont {B.}~\bibnamefont {Saxberg}}, \bibinfo {author} {\bibfnamefont {M.~G.}\ \bibnamefont {Panetta}}, \bibinfo {author} {\bibfnamefont {J.}~\bibnamefont {Simon}},\ and\ \bibinfo {author} {\bibfnamefont {D.~I.}\ \bibnamefont {Schuster}},\ }\href@noop {} {\bibinfo {title} {Manybody interferometry of quantum fluids}} (\bibinfo {year} {2023}),\ \Eprint {https://arxiv.org/abs/2309.05727} {arXiv:2309.05727 [quant-ph]} \BibitemShut {NoStop}%
\end{thebibliography}
\end{document}